# Raman and IR spectra of graphdiyne nanoribbons

Patrick Serafini[1], Alberto Milani[1*], Matteo Tommasini[2], Chiara Castiglioni[2], Carlo S. Casari[1*]

[1]*Dipartimento di Energia, Politecnico di Milano, via Ponzio 34/3, Milano, Italy*
[2]*Dipartimento di Chimica, Materiali e Ing. Chimica, Politecnico di Milano, P.zza L. da Vinci 32, Milano, Italy*

*Corresponding authors: alberto.milani@polimi.it; carlo.casari@polimi.it

## Abstract

γ-graphdiyne is a 2D carbon structure beyond graphene: it is formed by sp and $sp^2$ carbon atoms organized as hexagonal rings connected by linear links, and it is predicted to be a semiconductor. The lateral confinement of γ-graphdiyne nanoribbons significantly affects the electronic and vibrational properties. By means of periodic Density Functional Theory (DFT) calculations we here investigate the electronic band structure, the Raman and IR spectra of γ-graphdiyne 2D crystal and related nanoribbons. We discuss the effect of the functional and basis set on the evaluation of the band gap, highlighting the reliability of hybrid functionals. By joining DFT calculations with a symmetry analysis, we assign in detail the IR and Raman spectra of γ-graphdiyne. On this basis we show the modulation of the gap in nanoribbons of increasing width and different edges (armchair, zigzag). We assess how confinement affects the Raman and IR spectra by comparing vibrational modes with phonons of the parent 2D crystal. Our symmetry-based classification allows identifying the marker bands sensitive to the edge structure and lateral confinement of nanoribbons of increasing width. These results show the effectiveness of vibrational spectroscopy for the characterization of such nanostructures.

## 1. Introduction

Carbon materials and their nanostructures played a relevant role in the science and technology of the last two decades: from fullerenes to carbon nanotubes, from polyconjugated polymers to graphene, the so-called "era of carbon allotropes" has been enlightened by groundbreaking results, paving the way to many interesting research topics [1]. In the last years, the interest of many scientists has been directed towards exotic forms of carbon, including systems based on sp-hybridized carbon (variously referred to as carbyne, carbon-atom wires, polyynes, cumulenes, …) or hybrid sp-sp$^2$ or sp-sp$^3$ nanostructures [2-6], focusing both on their fundamental properties and on their potential interest for applications.

Research on sp-carbon systems dates back to the end of XIX century and the first theoretical investigation of sp-sp$^2$ carbon systems (*e.g.*, graphyne GY and graphdiyne GDY) has been published in 1987 [7].The advent of graphene and related materials strongly renewed the interest in these peculiar systems, also from the experimental point of view [8-10]. In fact, GY and GDY can be considered as possible modification of graphene, where novel 2D crystals are formed by interconnecting sp$^2$-carbon hexagons with sp-carbon chains of different lengths, and show the occurrence of Dirac cones in their electronic structure [11, 12].

Bottom-up approaches have been successfully employed to produce sub-fragments of GDY of different topology and dimension. In particular, Haley's group has been very active in synthesizing a variety of molecules mimicking GDY [13-17]. More recently, different papers reported the preparation and characterization of extended 2D GDY sheets by means of organometallic synthesis techniques, showing promising routes to the production of these systems, even though significant efforts should still be put to further investigate and understand their properties [18-20]. Among these hybrid sp-sp$^2$ carbon nanostructures, the γ polymorph of GY and GDY is the system that attracted the widest attention both from the experimental and theoretical points of view.

In addition to the limited experimental work, a much larger amount of papers focused on the prediction of the properties of GDY (and related systems such as GDY nanoribbons, hereafter referred to as GDYNRs) mainly through Density Functional Theory (DFT) calculations [5,6,21-24].

Among the most important features of GDY, the electronic band structure and band gaphave been investigated in detail , including also the study of confinement effects in GDYNRs or defected systems. Due to the relevance of this family of hybrid sp-sp$^2$ carbon materials in current nanoscience and nanotechnology, effective characterization tools with a combined experimental/computational approach are mandatory for a detailed investigation of these systems.

Vibrational spectroscopy is extremely useful to characterize carbon-based materials and Raman in particular has been employed in the first experimental works reporting the preparation of GDY-based materials [25-29].

Despite the effectiveness of these spectroscopic techniques up to now and to our knowledge, only few investigations focused on the prediction of the vibrational properties of GDY [27-29], . Popov *et al*. computed the Raman spectrum of different polymorphs of GY [26], Zhang *et al*. gave an assignment of the Raman spectrum of GY and GDY and studied the effects of strain by Raman [27]. Wang *et al*. discussed the Raman spectra of different finite-size models of different sp-sp$^2$ carbon nanostructures [29] while IR active C≡C stretching band has been discussed in Ref. [28].

However, a basic description of the vibrational dynamics of GDY is lacking. Similarly, the effects of delocalization *vs.* confinement on the spectra of GDYNRs call for an investigation of the spectroscopic markers useful for the characterization of nanoribbons with different width and edge topology. Even if the experimental spectra of these material are not available so far, we believe that the theoretical study of their vibrational states can highlight several aspects related to the peculiar electronic and molecular structure of GDY and GDYNRs, and to the effectiveness of vibrational spectroscopy techniques for molecular recognition and characterization.

Here, we have computed the electronic structure of 2D-GDY by periodic DFT simulations (CRYSTAL code) adopting different functionals and basis sets. Thereafter, we have focused on the prediction and assignment of IR and Raman spectra. We have analyzed GDYNRs as a function of the nanoribbon width and edge-type (armchair or zigzag), disclosing the modulation of the band gap, and peculiar markers in IR and Raman that are sensitive to the lateral confinement and to the molecular structure of the edges.

## 2. Computational Details

Geometry optimization, band gap evaluation and prediction of IR and Raman spectra have been carried out for the two-dimensional (2D) crystal γ-GDY (also called (6,6,6)-graphdiyne and hereafter referred as 2D-GDY) and for GDYNRs, treated as one-dimensional (1D) crystals, by employing DFT simulations and applying Periodic Boundary Conditions in two and one dimensions respectively. The CRYSTAL17 code [30, 31] has been adopted to this aim, exploiting the possibility to employ hybrid exchange-correlation functionals together with Gaussian basis sets.

In the case of 2D-GDY, for which reference theoretical data are already available in the literature both considering band gaps [5, 6] and Raman spectra [27], LDA (VWN), GGA (PBE and BLYP) and hybrid functionals (PBE0, HSE06, B3LYP) have been considered with 6-31G(d,p) basis sets. For PBE0 and HSE06 only, also the more extended Ahlrichs VTZ + Polarization basis set [32] has been used. This allowed analyzing the reliability of different functionals, including hybrid ones, and the influence of basis set in the prediction of the band gap.

PBE0/6-31G(d,p) and HSE06/6-31G(d,p) levels of theory have been then chosen to predict the gap of the different GDYNRs of different width, while, among these, Raman spectra could be computed only at PBE0/6-31G(d,p) level in CRYSTAL17 since HSE06 is not yet implemented for Raman response. Our previous experience in the prediction of the Raman response of several sp-carbon based molecular systems shows that this functional provides a good agreement *vs*. experimental data [33-35]. According to the discussion reported in Ref. [36], the exponent of the diffuse sp orbitals in 6-31G(d) basis set of carbon atoms have been increased from 0.1687144 Bohr$^{-2}$ to 0.187 Bohr$^{-2}$ to avoid convergence problems in the SCF cycles, due to basis sets linear dependencies. For the same reason, when using the VTZ basis set, the exponent 0.12873135 Bohr$^{-2}$ of one the s function and the exponent 0.10084754 Bohr$^{-2}$ of one the p functions have been changed to 0.153 and 0.12 Bohr$^{-2}$ respectively.

Considering other parameters of the CRYSTAL simulations, the tolerance on integral screening (TOLINTEG parameters) have been fixed to 9,9,9,9,80, while the shrink parameters defining Monkhorst-Pack sampling points have been fixed to 100 for the calculation of the band structures and gaps and 50 for the vibrational analysis. In both case, full convergence on the total energy had been obtained.

Since, in absence of experimental data, the comparative analysis here presented concerns calculated spectra, it does not require the usually adopted scaling of the transition frequencies. The paper reports unscaled wavenumber values both in the discussion and in the Figures: a suitable scaling factor could be defined [37] whenever comparing these theoretical spectra with experimental ones.

## 3. Results and Discussion

To present the results coming from DFT calculations of 2D-GDY and of three series of nanoribbons, we refer to the nomenclature introduced in previous papers [5, 6, 21-23]. Armchair and zigzag are indicated by A(*n*)-GDYNR and Z(*n*)-GDYNR, respectively, where the index *n* refers to the width of GDYNR, which is an integer value for A($n$)-GDYNR ($n$ = 1, 2, … 5) and integer or half integer value in the case of Z($n$)-GDYNR ($n = 1, 3/2, … 5$), as depicted in Figure 1. Such GDYNRs models can be classified according to the proper line-symmetry groups, isomorphs to $D_{2h}$ group in the case of A($n$)-GDYNRs and Z($n$)-GDYNRs with half integer $n$. Z($n$)-GDYNRs with integer $n$, belong to the $C_{2v}$ group.
For all GDYNRs, hydrogen atoms have been added at the edges as end-groups, in agreement with most of the previous computational works [5, 6, 21-23]. This choice is reasonable also from the experimental point of view, based on the finite GDY fragments that have been synthetized in the past [17]. To assess the influence of the edge type on both electronic and vibrational properties, we have considered also the A(1) and Z(1) systems, which correspond to co-polymers formed by alternated di-acetylenic and phenyl-rings units connected respectively in para- and meta- positions [38]. These two limiting cases represent a reference system formed by just the edge of A($n$)- and Z($n$)-GDYNRs.

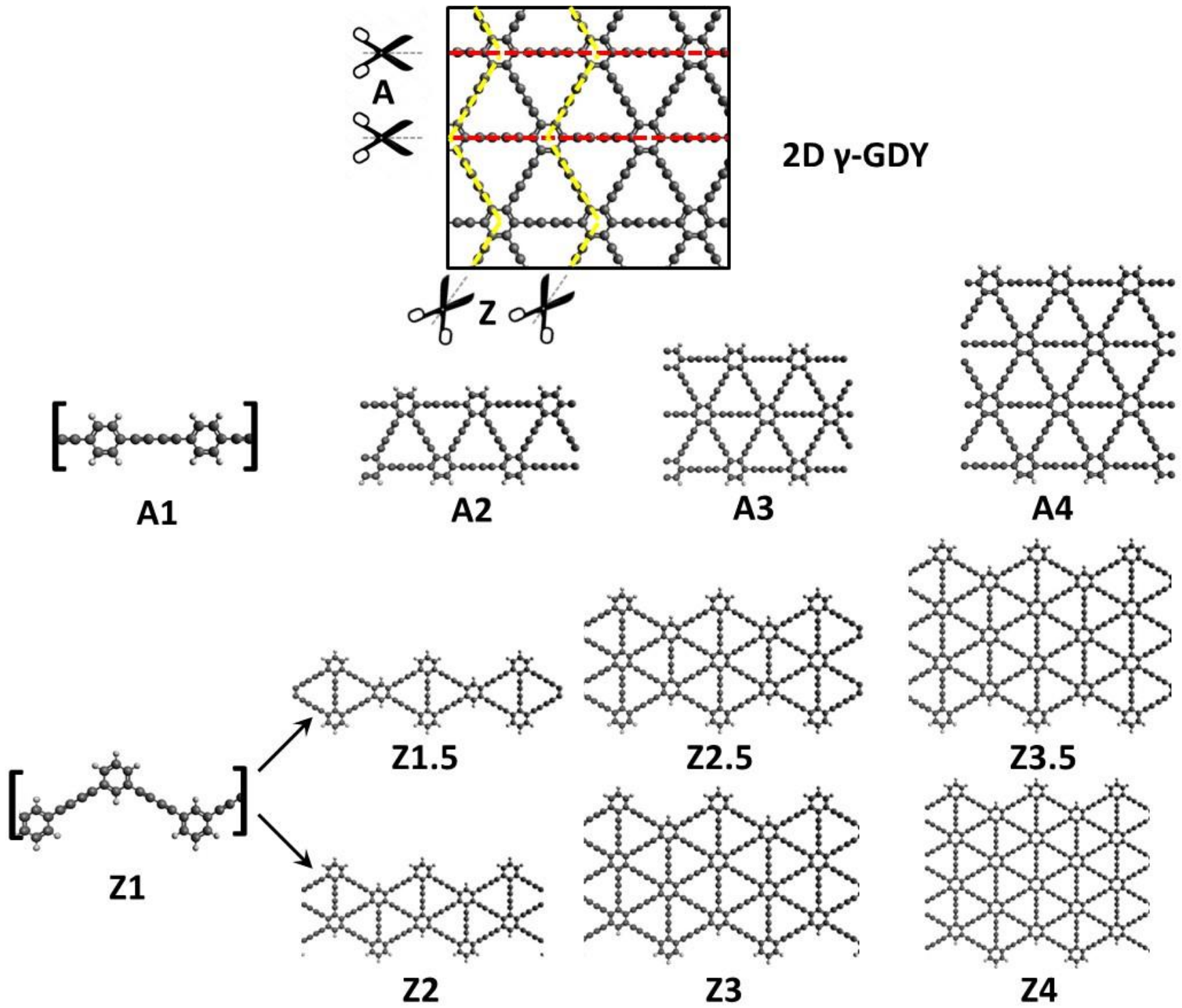

***Figure1****: Sketches of the armchair (An) and zigzag (Zn)-GDYNRs investigated in this work. n is the index representing the nanoribbon width.*

The lattice parameters and fractional coordinates of the optimized structure of all the systems here investigated are reported in the Supplementary Material [50], including the computed wavenumbers, IR intensities, Raman activities, and graphical representations of the normal modes of 2D-GDY and of the main GDYNR markers discussed in the text.

### *3.1 Electronic properties of 2D γ-graphdiyne and its nanoribbons*

As in the case of other carbon-based polyconjugated materials, the correct prediction of the gap suffers from the limitations of current DFT pure functionals in describing accurately π-electron delocalization, with a tendency to overestimate its effect, thus causing unrealistically low predictions of the band gap. For instance, the gap of 2D γ-GDY reported in the literature changes from about 0.4 eV (LDA), 0.5 eV (GGA) to 0.9 - 1.1 eV when employing hybrid functionals or the GW method. The latter values are considered more reliable [5, 6, 12, 24]. In fact, it is well-known that hybrid functionals significantly improve the accuracy in the description of π-electron delocalization, but their use is not convenient when employing plane waves and pseudopotential methods to carry out DFT calculations in periodic systems. A useful alternative is provided by Gaussian basis sets, which allow to use hybrid functionals at an acceptable computational cost: this approach is implemented in the CRYSTAL code [30-31] which offers the valuable opportunity to extend to infinite systems the same computational approach typical of finite molecular systems.

As illustrated in previous papers [5, 6], the gap in the band structure of γ-GDY is found at the Γ point (Figure 2).

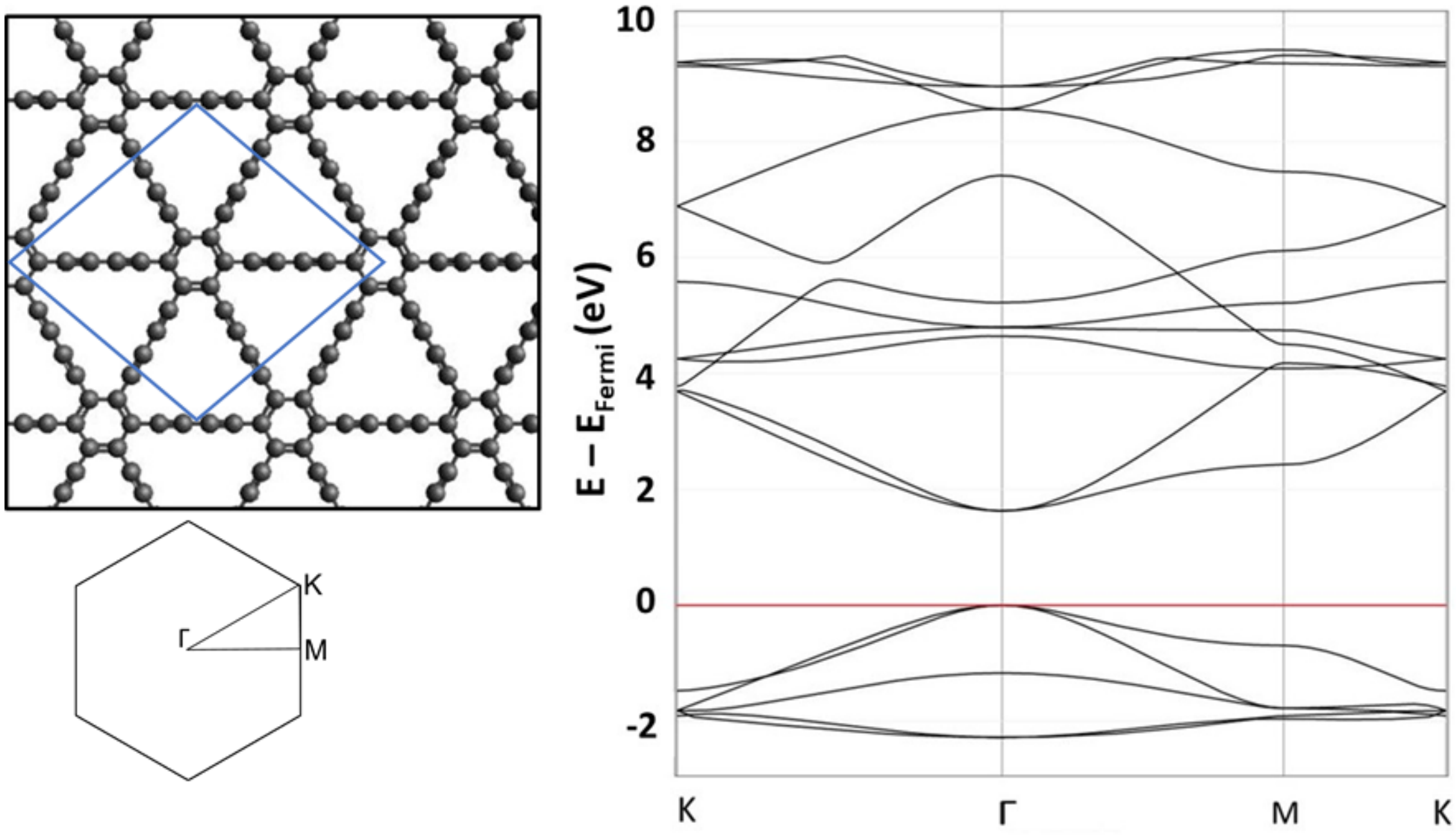

***Figure 2**: Electronic band structure of 2D-GDY (PBE0/6-31G(d) DFT calculation).*

We report in Table 1 the band gap computed by employing different LDA, GGA and hybrid functionals. For PBE0 and HSE06 we also provide the comparison of the results produced by a double-ζ basis set (6-31G(d)) *vs*. a triple-ζ basis set (VTZ). We note that the effect of the basis set is less important than the choice of the functional: indeed, both for PBE0 and HSE06, adopting the more accurate VTZ basis set gives in both cases a gap which is just about 0.05 eV larger than the values obtained with 6-31G(d) basis set. Therefore, we can suppose that the latter basis set is sufficiently accurate to model the properties of this system.

***Table 1:** Band gaps of 2D-GDY computed by employing different exchange-correlation functionals and gaussian basis sets (see text for details).*

| XC functional | Basis set | Band Gap (eV) |
|---|---|---|
| PBE0 | 6-31G(d) | 1.63 |
| | VTZ | 1.67 |
| HSE06 | 6-31G(d)<br>VTZ | 1.11<br>1.16 |
| B3LYP | 6-31G(d) | 1.40 |
| BLYP | 6-31G(d) | 0.48 |
| PBE | 6-31G(d) | 0.45 |
| LDA (VWN) | 6-31G(d) | 0.40 |

Considering LDA and GGA functionals (PBE, BLYP), we obtain band gaps of 0.4 (LDA), 0.45 (PBE) and 0.48 eV (BLYP), in full agreement with previous calculations. As expected, hybrid functionals provide higher values of the band gap: 1.6 eV (PBE0), 1.4 eV (B3LYP) and 1.1 eV (HSE06). The latter value agrees with the result of GW calculations, usually taken as an accurate reference value. This also agrees with the fact that HSE06 has been found to give a more accurate prediction of band gaps with respect to the parent PBE0 functional [39].

Considering zigzag and armchair GDYNR, in Figure 3 we plot the band gap of A(*n*)-GDYNRs and Z(*n*)-GDYNRs as a function of the *n* index ($n \leq 5$) and compared with the values found for 2D-GDY (*i.e.*, n → ∞).

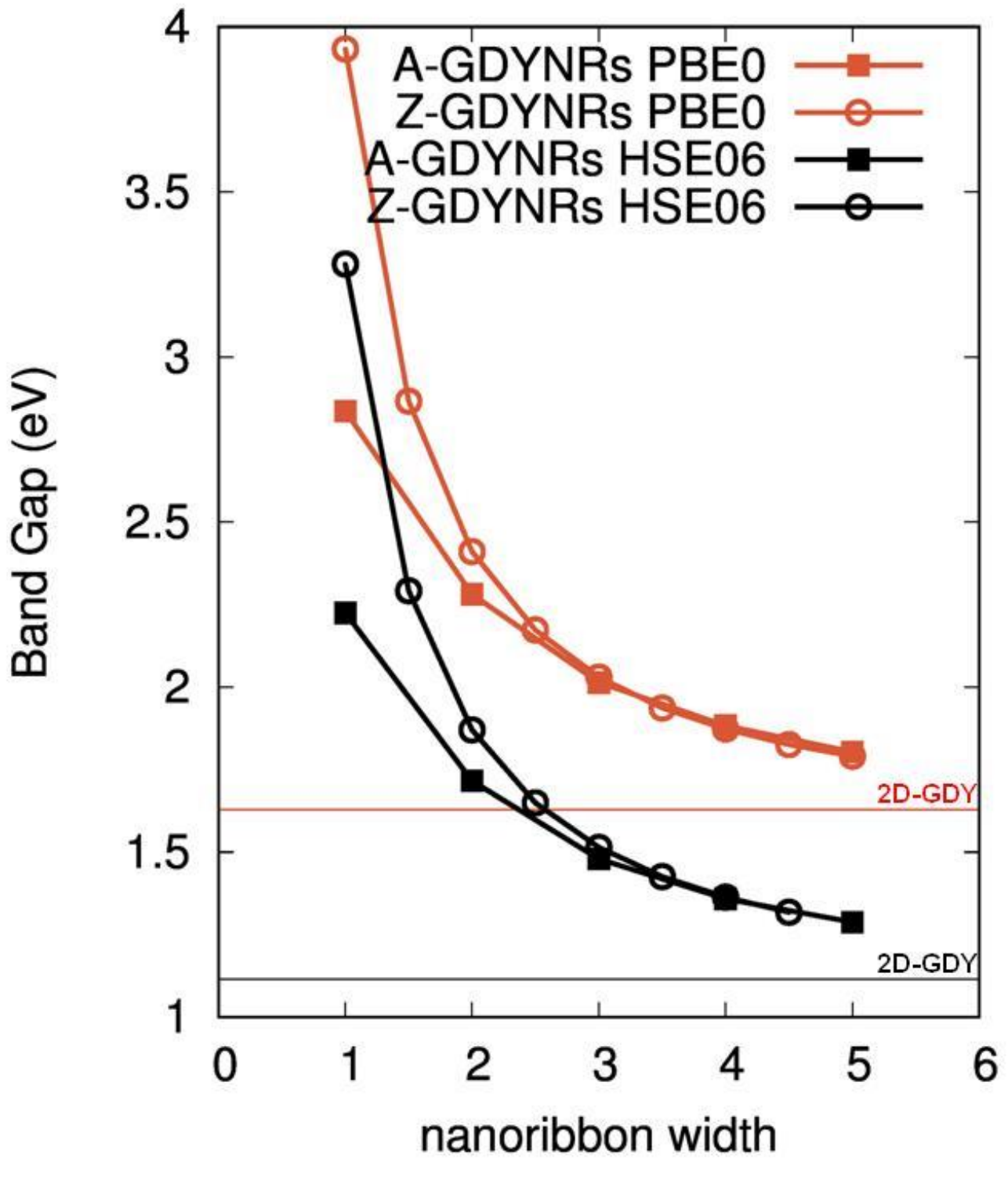

***Figure 3:*** *Band gap (eV) of armchair and zigzag GDYNR as a function of width index (n) as obtained by PBE0/6-31G(d) (in red) and HSE06/6-31G(d) (in black) calculations. Numerical values of these band gaps are reported in Table S2 of the Supplementary Material [50]. The value band gap of 2D-GDY is also indicated as a comparison by the horizontal lines.*

Several works report that the gap of confined GDY is larger than the gap of the infinitely extended 2D GDY and decreases for increasing size, towards the limit of 2D-GDY [5, 6, 21-23]. Confinement effects caused by the finite width of GDYNRs (*n*) modulate the gap, similar to the behavior of graphene nanoribbons with respect to the graphene crystal. As show in Figure 3 this modulation of the gap spans a range of about 1 eV (for $n \geq 2$ ). As expected, the values obtained with the HSE06 functional are systematically lower than those obtained with PBE0. However, both functionals essentially describe the same behavior.

Interestingly, while for $n > 3$ the trends are the same for both edges, for small widths ($n < 3$), Z(*n*)-GDYNRs have a larger gap with respect to A(*n*)-GDYNRs. The reason for this is evident if we consider the gap values obtained for the limiting case ($n = 1$) of the 1D polymers A(1) and Z(1) (Figure 1). Both polymers can be described as sp-carbon segments of four carbon atoms (*i.e.*, diacetylenic units) interconnected by phenyl rings. The connection on these rings can be done along three different conjugation paths, corresponding to para-, meta- and ortho- connections. As highlighted by Tahara et al. investigating the case of graphyne fragments [38], A(1) corresponds to a para-conjugated system where the delocalization of π-electrons along the system occurs more efficiently, thus promoting the lowest gap. Conversely, the meta-conjugation characteristic of Z(1) reduces π-electron delocalization, which justifies the higher band gap. As expected, the influence of the edge type is effective for small widths, but rapidly decreases for increasing width. For $n > 3$ the edge type (A- or Z-) has a very limited influence on the gap, even if the gap remains still quite far from the $n \rightarrow \infty$ limit of the 2D-GDY crystal.

### *3.2 Vibrational spectra and optically active phonons of 2D γ-GDY*

The phonons of 2D-GDY at the $\Gamma$ point can be classified according to the $D_{6h}$ point group, which is isomorphic to the P6/mmm layer group of the crystal. The structure of the representation of the $D_{6h}$ point group in the vibrational space is:

$$\Gamma^{vib} = 3A_{1g} + 2A_{2g} + 3B_{2g} + 3E_{1g} + 6E_{2g} + 2A_{2u} + 5E_{1u} + 3B_{1u} + 3B_{2u} + 3E_{2u}.$$

$A_{1g}$, $E_{1g}$ and $E_{2g}$ phonons are Raman active, whereas $A_{2u}$ and $E_{1u}$ are IR active; the rest of the irreducible representations ($A_{2g}$, $B_{2g}$, $B_{1u}$, $B_{2u}$, $E_{2u}$) are inactive. Therefore, 12 peaks (9 of which are double-degenerate), are expected in the first-order Raman spectrum, and 7 absorption bands (5 double-degenerate) in the IR spectrum. Furthermore, since the $D_{6h}$ point group possesses inversion symmetry, there is mutual exclusion between IR and Raman transitions. the Raman and IR spectra of 2D-GDY computed by DFT are reported in Figure 4.

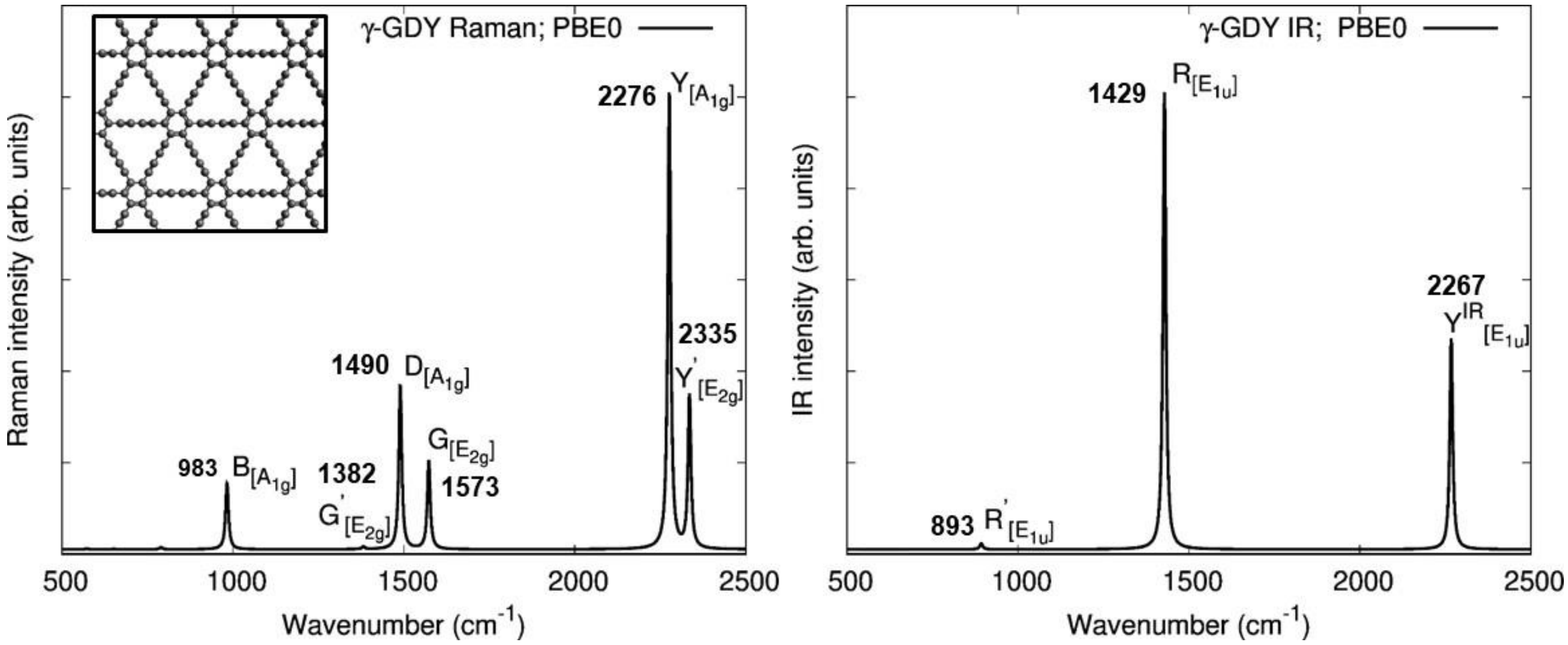

***Figure 4****: DFT computed (PBE0/6-31G(d)) Raman and IR spectra of 2D-GDY. The wavenumber values are not scaled.*

Our calculations provide a pattern of the Raman spectrum which is similar to the one determined by Zhang *et al.* adopting LDA and plane wave pseudopotential method [27]. The labels used in Figure 4 to identify the different Raman lines have been chosen analyzing the related vibrational eigenvectors, to suggest possible correlations with the vibrational transitions of related materials (*e.g.*, G and D Raman lines of graphene).
Table A of the Appendix (reported in the Supplementary Materials [50]) shows the basis set of the stretching coordinates and the symmetry-adapted vibrational coordinates defined accordingly. The high symmetry of 2D-GDY allows to draw several conclusions on the vibrational displacements of individual phonons at $\Gamma$, and to obtain information about allowed and forbidden dynamical couplings among vibrational coordinates. The behavior based on the analysis of symmetry coordinates is general, and independent from the level of theory adopted for spectra simulations.
The analysis presented here is limited to CC stretching internal coordinates: indeed, in 2D-GDY, CC stretching modes dominate the vibrational spectra above 1000 $cm^{-1}$, giving rise to the strongest Raman and IR transitions. Crystal redundancies and several local redundancies (*i.e.*, ring redundancies) [40, 41] involve both bending and stretching coordinates, thus intrinsically determining a coupling among bending and stretching displacements. However, we can obtain a sound description of the most relevant normal modes, limiting the analysis to their stretching component.

***3.2.1 CC stretching symmetry coordinates.***
The following internal coordinates belonging to the reference unit cell (Table A) allow describing the CC stretching subspace:
1) six equivalent CC stretching of the aromatic ring, $\{R_i; i = 1 \dots 6\}$
2) CC stretching of the bonds of the six polyyne arms. Each arm contributes to the reference cell with one half of its bonds, whose stretching coordinates are described by three sets, namely: $\{r_i; i = 1 \dots 6\}$, $\{T_i; i = 1 \dots 6\}$, and $\{P_i; i = 1 \dots 3\}$. $r_i$ refers to “single” bonds linked to the aromatic unit, $T_i$ are “triple” bonds stretching, and $P_i$ are the stretching coordinates of the central CC bonds, which carry inversion centers. Each $P_i$ is shared with an adjacent crystal cell, so that only three $P_i$ coordinates are associated to the reference cell.

To highlight correlations with the normal modes of a simple diphenyl-polyyne molecule it is very useful to adopt a description of the phonons using the displacement patterns involving the six polyyne arms (actually diynes, each consisting of 5 CC bonds). This description is obtained by introducing the group symmetry coordinates of the individual polyyne chains (Table A.2). Accordingly, we obtain the following sets of equivalent coordinates; every set contains the six group coordinates associated to each polyyne arm:

$\{r_i^+\}, \{r_i^-\}, \{T_i^+\}, \{T_i^-\}, \{P_i\}$

Thanks to translational symmetry, it is immediate to verify that the sets $\{r_i^+\}$, $\{T_i^+\}$, and $\{P_i\}$ are involved in phonons of “gerade” symmetry (Raman active). IR active “ungerade” phonons do involve $\{r_i^-\}$ and $\{T_i^-\}$. According to $D_{6h}$ symmetry, each set gives symmetry coordinates belonging to $A_{1g}$, $E_{2g}$, $B_{2u}$ and $E_{1u}$ species, as described below.

(a) the <u>$A_{1g}$ subspace</u> is spanned by 4 $A_g$ symmetry coordinates. These correspond to the in-phase combinations of the internal (or group) coordinates belonging to the different sets, which originate the symmetry coordinates $A_{1g}^{R}, A_{1g}^{r+}, A_{1g}^{T+}, A_{1g}^{P}$ defined in Table A.3. It is important to note that the 3 $A_{1g}$ Raman active phonons are linear combinations of these four $A_{1g}$ symmetry coordinates and the four dimensional $A_{1g}$ subspace contains a crystal redundancy (*i.e.,* the in-phase combination of the four $A_{1g}$ stretching coordinates, which describes a simultaneous stretching of all CC bonds belonging to each cell, and which would imply a change of the crystal volume).

(b) 4 phonons belonging to the $E_{2g}$ subspace can be described as linear combinations of the 4 $E_{2g}$ symmetry coordinates ($E_{2g}^{R}, E_{2g}^{r+}, E_{2g}^{T+}, E_{2g}^{P}$). The remaining 2 $E_{2g}$ phonons predicted for 2D-GDY can be accounted for by introducing bending coordinates.

(c) 3 phonons of the $E_{1u}$ subspace are linear combinations of the 3 symmetry coordinates $E_{1u}^{R}, E_{1u}^{r-}, E_{1u}^{T-}$ ($P_i$ coordinates cannot appear in *ungerade* species because each $P_i$ bond carries an inversion center). The remaining 2 $E_{1u}$ phonons are expected to involve mainly bending displacements.

(d) $B_{2u}$ is the last symmetry species involving stretching coordinates, and it is associated to phonons inactive in the vibrational spectra. It contains 3 phonons which are linear combinations the $B_{2u}$ symmetry coordinates $B_{2u}^{R}, B_{2u}^{r-}, B_{2u}^{T-}$. Based on the structure of the representation, we expect one further $B_{2u}$ phonon associated to bending displacements.

$A_{2u}$ IR-active phonons and $E_{1u}$ Raman-active phonons cannot be described by the chosen basis set made of just stretching coordinates: according to the character table, $A_{2u}$ and $E_{1u}$ phonons must involve out-of-plane displacements, such as those associated to out-of-plane bending and torsions. For this reason, such phonons are expected to lie in the low-wavenumber region of the IR spectrum (indeed, the $A_{2u}$ transitions are computed by CRYSTAL at 183 and 485 cm$^{-1}$ and the $E_{1g}$ transitions are computed at 137 and 474 cm$^{-1}$, showing in both cases a negligible intensity in the IR or Raman spectrum, see Supplementary Material [50])

DFT calculations supply the nuclear displacements of the phonons, which are used to draw the graphical representation of the vibrational patterns associated to the most relevant Raman and IR bands (see Supplementary Material [50]). The inspection of these graphics provides a description of such phonons as linear combinations of the symmetry coordinates introduced above.

### *3.2.2 Raman and IR assignment*

It is well-known that the marker bands of sp-hybridized CC bonds are found in the 2000-2500 cm$^{-1}$ range [2, 25]. Indeed, the Y and Y' phonons are respectively $A_{1g}$ and $E_{2g}$ collective CC stretching vibrations localized on the sp-carbon domains of GDY. The combination of the $r_i^{+}, T_i^{+}, P_i$ coordinates of each *i*-th polyyne arm follows the characteristic pattern shown by the Effective Conjugation Coordinate (ECC) mode of polyynes, described as a collective simultaneous C≡C stretching/C-C shrinking of the bonds forming the chain ($r_i^{+}$ and $P_i$ oscillate in phase, while $T_i^{+}$ vibrate out-of-phase). The properties of this mode have been widely discussed in the past [2, 25], extending the treatment of vibrational dynamics developed for polyconjugated polymers a few decades ago [42-44]. In the $A_{1g}$ Y mode (2276 cm$^{-1}$), all sp-carbon chains oscillate in-phase, according to the ECC coordinate. In the $E_{2g}$ Y' mode (2335 cm$^{-1}$) the ECC vibrational pattern is repeated on the 6 polyyne segments with different relative phases, according to $E_{2g}$ symmetry. The sizeable difference in wavenumber of these two phonons (59 cm$^{-1}$) indicates a non-negligible coupling among adjacent polyyne chains, caused by the interaction among the $\pi$ electrons of the phenyl unit and of the attached sp-chains. This is supported by the calculation of the vibrational structure of a di-phenyl polyyne consisting of a linear chain having 4 sp-hybridized carbon atoms capped with two phenyl groups (*i.e.*, the smallest structural unit defining GDY). The ECC mode of such model molecule is computed at 2368 $cm^{-1}$, much higher than the Y and Y' bands of GDY. This proves the occurrence in GDY of significant $\pi$ electron delocalization across phenyl groups (see Supplementary Material [50]).

The region below 1600 $cm^{-1}$ is commonly assigned to stretching vibrations of the aromatic rings. However, phonons involving $R_i$ coordinates, always show a remarkable coupling with "single" bonds stretchings of the polyyne arms ($r_i$ and $P_i$).

The phonons of $A_{1g}$ symmetry are associated to the following Raman bands:

- the D line (1490 $cm^{-1}$) is assigned to a phonon described as the $A_{1g}^{R}$ ring breathing coupled to $A_{1g}^{r+}$ coordinates, namely to the stretchings of the CC bonds linked to the aromatic moiety. The $A_{1g}^{R}$ and $A_{1g}^{r+}$ coordinates vibrate out of phase, with a pattern very similar to that of the phonon associated to the D line of graphite/graphene. The $A_{1g}^{P}$ coordinate is also involved, to a lesser extent.
- the B line (983 $cm^{-1}$), associated to the breathing mode of the phenyl units, out-of-phase coupled with the $A_{1g}^{P}$ coordinate. The $A_{1g}^{r+}$ coordinate is also involved, with a minor contribution.

  Symmetry selection rules for $A_{1g}$ phonons tell us that only the three diagonal elements of the Raman polar tensor, namely $\left(\alpha_{xx} = \alpha_{yy}, \alpha_{zz}\right)$ are non-vanishing. According to the calculations (see SI), all relevant Raman transitions listed above show a neligible out-of-plane $\alpha_{zz}$ component. So, the intensity of these Raman lines are determined by the value $|\alpha_{xx}|^2 = |\alpha_{yy}|^2$. This implies that for 2D-GDY sheets deposited on a substrate (*x, y* plane), Raman experiment in backscattering geometry at normal incidence (*z* direction) will provide the same intensity pattern, independently on the polarization of the incident and scattered light, because the symmetry of the Raman tensors does not allow assessing the orientation of the sheet in the x, y plane. A proof of the orientation of the sheet with respect to the substrate should be obtained with normal incidence and collection of the scattered photons at 90° (e.g. x direction). In this case $A_{1g}$ bands should vanish for z-polarization of the scattered photons.

<u>The $E_{2g}$ Raman transitions</u> are the following:

- the G mode (1573 $cm^{-1}$): it involves $E_{2g}^{R}$ coordinate of the phenyl units with in phase contribution by $E_{2g}^{r+}$ coordinate. The displacements pattern recalls that associated to the G line of graphene.
- the weak G' mode (1382 $cm^{-1}$); it also involves the $E_{2g}^{R}$ coordinate of the phenyl unit, as the G mode; the contribution from the polyyne chain involves $E_{2g}^{P}$ coordinate and, to a lesser extent, $E_{2g}^{r+}$.
- A very weak $E_{2g}$ bands at 790 $cm^{-1}$, corresponding to another different linear combination of $E_{2g}^{R}, E_{2g}^{r+}$ and $E_{2g}^{P}$ coordinates, with non-negligible bending contributions.
- two very weak bending phonons (573 $cm^{-1}$, 394 $cm^{-1}$).

Considering a degenerate pairs for each $E_{2g}$ phonons, the nonvanishing elements of the Raman tensor are $\alpha_{xx} = -\alpha_{yy}$ and $\alpha_{xy}$ respectively, with $|\alpha_{xx}|^2 = |\alpha_{xy}|^2$.

Accordingly, the polarization properties of the $E_{2g}$ bands of 2D-GDY sheets deposited on a substrate result to be the same as for $A_{1g}$ transitions: for backscattering geometry, one expects independence of the Raman intensity vs. the polarization direction within the xy plane.

In the IR spectrum, since the 2 <u>$A_{2u}$</u> modes practically show almost zero intensity (see Table S5 in the

Supplementary Material [50]) we essentially observe $E_{1u}$ phonons, namely:
- the $Y^{IR}$ band (2267 cm$^{-1}$) assigned to the out-of-phase combination of $E_{1u}^{r-}$ and $E_{1u}^{T-}$ coordinates; focusing on individual polyyne arms, it can be described as the ECC mode with a node, located on the central CC bond;
- the R band (1429 cm$^{-1}$), which is mainly due to CC stretching of the bonds in the phenyl units ($E_{1u}^{R}$) with out-of phase contribution of the $E_{1u}^{r-}$ coordinate;
- the very weak R' transition (893 cm$^{-1}$); it mainly involves the $E_{1u}^{R}$ coordinate (as the R band), coupled in-phase with the $E_{1u}^{r-}$ coordinate; a careful inspection of the vibrational patterns shows that a non-negligible contribution by bending displacements affects the R' phonon;
- two very weak bending phonons are computed at 419 cm$^{-1}$ and 171 cm$^{-1}$.

Considering a degenerate pair for each $E_{1u}$ phonon, the non-vanishing elements of the associated dipole derivatives are $\mu_x$ and $\mu_y$ respectively, with $|\mu_x|^2 = |\mu_y|^2$. The above relationship tells us that IR experiment with polarized light at normal incidence on the GDY plane cannot discriminate between oriented and un-oriented sheets in the plane.

The assignments here reported are the starting point for the discussion of the spectra of GDYNRs, and allow to discriminate among phonons characteristic of the bulk and phonons localized on the edges, which can be taken as markers of the spatial confinement.

### *3.3 Raman and IR spectra of GDYNRs: markers of confinement*

The detailed analysis of Raman and IR spectra of A- and Z- nanoribbons presented here aims at discussing the existence of marker bands that could be taken as signature of spatial confinement and/or edge type (A- or Z-).

#### *3.3.1 Raman spectra*

Figure 5 compares Raman spectra of A(*n*)- and Z(*n*)-GDYNR for increasing widths with the spectrum of 2D-GDY, discussed in the previous section. Starting from *n* = 2, both A(*n*)- and Z(*n*)-GDYNRs show strong bands which can be put in correspondence with the Raman bands of 2D-GDY. We discuss bands in the region of sp-carbon diacetilenic segments (2000-2300 cm$^{-1}$ range) and in the region of sp$^2$ carbon of the phenyls (1400-1600 cm$^{-1}$ range).

*2000-2500 cm$^{-1}$ region*. A strong doublet (≈ 2300 cm$^{-1}$) correlates with the Y and Y' lines of 2D-GDY. The strongest $A_{1g}$ band starts from 2307 cm$^{-1}$ (2305 cm$^{-1}$) in A(2)-GDYNR (Z(2)-GDYNR), and in larger ribbons converges to a unique wavenumber (2285 cm$^{-1}$, about 10 cm$^{-1}$ higher than Y line of 2D-GDY, and independent on edge topology – see Figure 5). The remarkable frequency dispersion (≈ 30 cm$^{-1}$) from *n* = 2 to *n* = ∞ (2D-GDY) suggests that this band is highly sensitive to π electron delocalization, thus indicating that there is a strong interplay between the π electrons of the polyyne arms and those of the aromatic rings. The effect of edge topology on the Y line can be appreciated only considering very thin ribbons (*n* = 1, 1.5).

The higher wavenumber component of such doublet correlates with the Y' $E_{2g}$ line of 2D-GDY, and shows a less pronounced dispersion (10 cm$^{-1}$ for A(*n*)-GDYNRs and 20 cm$^{-1}$ for Z(*n*)-GDYNRs) while it is more sensitive to the edge type.

In the case *n* = 1, the Y-Y' doublet merges in a single band predicted at 2344 cm$^{-1}$ and 2371 cm$^{-1}$ for A and Z species, respectively. This band corresponds to the ECC mode of the single unit of polyyne (one-unit cell) in the sp-carbon segment. Remarkably, the wavenumber of this mode in A(1)-GDYNR is significantly lower than the corresponding mode of the di-phenyl-polyyne model molecule (2368 $cm^{-1}$), while it almost coincides for Z(1)-GDYNR. This complies with the effect of conjugation between $\pi$ electrons of the polyyne chain and $\pi$ electrons of the aromatic ring being more effective in the case of the para-substitution (see the discussion above on the band gaps). A thorough discussion on the physical effects ruling this behavior can be found in Ref. [35].

The lowering of symmetry in GDYNRs with respect 2D-GDY concurs to the activation of different modes, close in frequency and mainly contributing to the two main Y and Y' bands. Sometimes, weak additional bands can be identified (*e.g.*, A(4), A(5), Z(2), Z(4)).

The Y-Y' doublet cannot be taken as characteristic marker of a ribbon, because of its clear relationship with the corresponding doublet of the 2D crystal. However, the remarkable redshift observed with increasing width for both families of GDYNRs (Figure 5) suggests that the position of Y and Y' could be a diagnostic tool regarding the width of GDYNRs, at least qualitatively.

*850-1600 $cm^{-1}$ region.* Starting from A(3)-GDYNR and Z(2)-GDYNR, we clearly identify three bands: the first is very close to the G line of 2D-GDY (at 1571 $cm^{-1}$), the second to the D band (at 1490 $cm^{-1}$ ), and the third to the B band (at 983 $cm^{-1}$). Several satellite lines also show up, givingrise to a rather rich spectrum. This feature is ascribed to the presence of CH bonds on the edges of the ribbons, whose characteristic bending frequencies lye in this spectral region.

By analyzing the vibrational eigenvectors of A(*n*)-GYNRs, we observe that the band at about 1570 $cm^{-1}$ corresponds to the G mode vibration (ring stretching) of the "bulk" regions, whereas the band at about 1630 $cm^{-1}$ is assigned to the same vibrational pattern, well localized on the edges and coupled to CH wagging vibration. A similar situation is found for the two lines at 1452 and 1499 $cm^{-1}$: the latter corresponds to the D line of 2D-GDY, and is associated to a "bulk" normal mode, whereas the former shows the same vibrational pattern, mainly localized on peripheral rings and coupled with CH wagging vibrations. Therefore, the lines at 1452 and 1630 $cm^{-1}$ can be taken as markers of armchair edges and, as expected, their relative intensity decreases with increasing the nanoribbon width *n*. These lines are rather intense and should allow a reliable identification of A(*n*)-GDYNRs. Remarkably, thanks to its $A_{1g}$ symmetry, the 1630 $cm^{-1}$ line shows a sizeable Raman intensity also for the largest A(*n*)-GDYNR here investigated (*n* = 5).

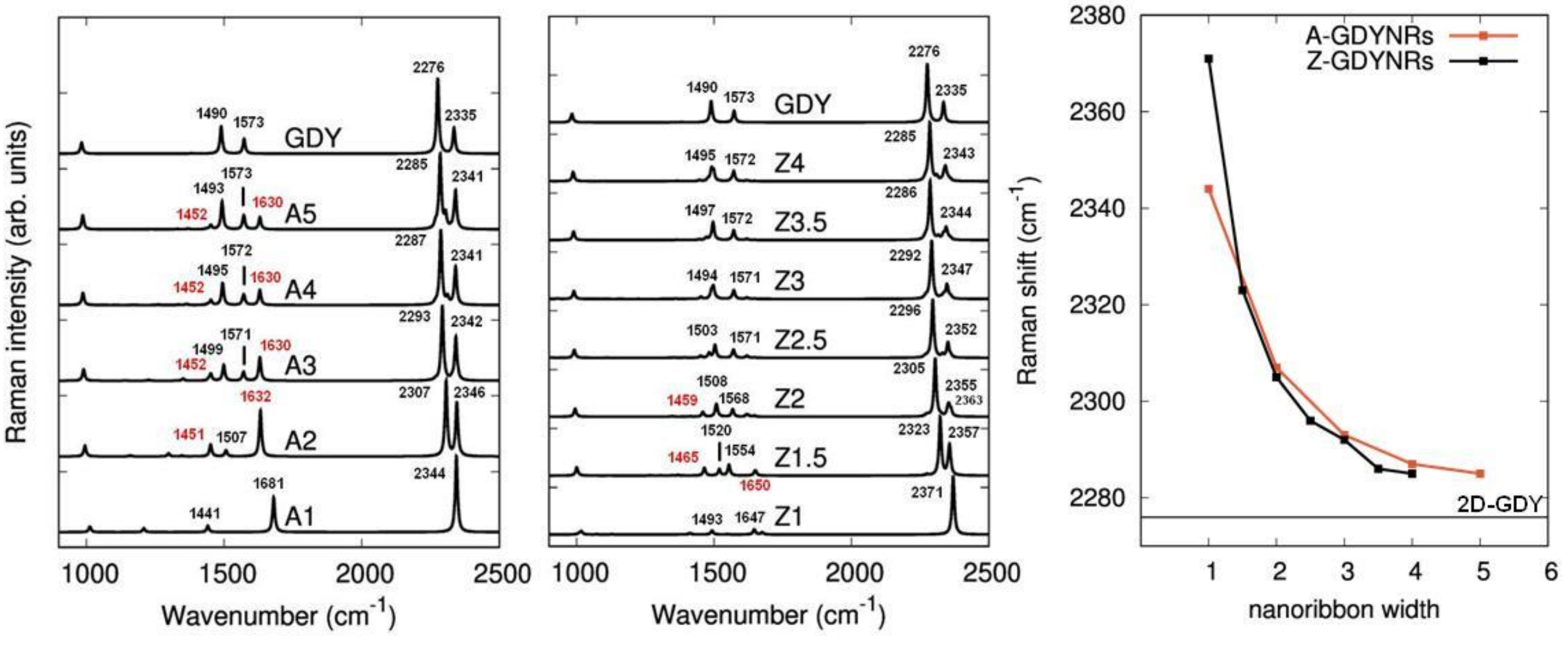

***Figure 5:*** *Comparison among the DFT computed (PBE0/6-31G(d)) Raman spectra of 2D-GDY, A(n)-GDYNRs (left) and Z(n)-GDYNRs (center) having increasing widths. The wavenumber values are not scaled. (right) Modulation of the DFT computed (PBE0/6-31G(d), unscaled values) wavenumber of the Raman Y bands predicted for armchair and zigzag GDYNRs as a function of their width. The wavenumber computed for 2D-GDY is also reported as a comparison.*

Considering Z(*n*)-GDYNRs, the inspection of Figure 5 shows that the edge type significantly affects the shape of the Raman spectrum. In the Raman spectra of the thinner Z(*n*)-GDYNRs we observe two bands (1650 cm$^{-1}$, 1460 cm$^{-1}$) that are the counterparts of the satellite bands observed for A(*n*)-GDYNRs. However, these lines are significantly weaker in Z(*n*)- than in A(*n*)-GDYNR, which would make more difficult the recognition of Z(*n*)-GDYNRs based on the Raman pattern.

### *3.3.2 IR spectra*

In Figure 6 we compare the IR spectra of A(*n*)- and Z(*n*)-GDYNRs with the spectra of the 1D infinite polymers (A(1) and Z(1)) and 2D-GDY.

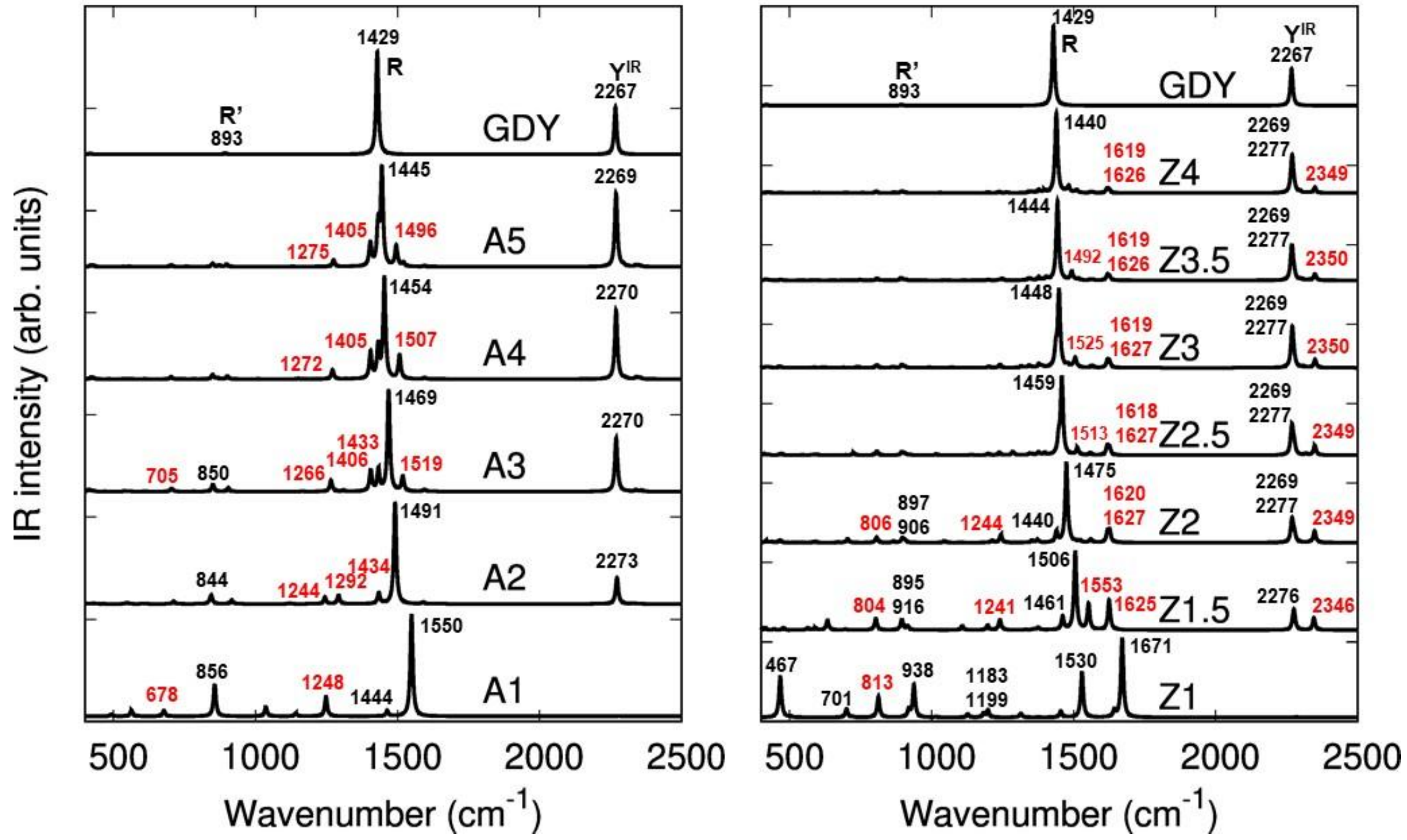

***Figure 6:*** *Comparison among the DFT computed (PBE0/6-31G(d)) IR spectra of 2D-GDY, A(n)-GDYNRs and Z(n)-GDYNRs having increasing widths. The wavenumber values are not scaled.*

*2000-2500 cm$^{-1}$ region*. For both edge types, the polymers A(1) and Z(1) do not show any IR band of appreciable intensity in the sp-carbon region above 2000 cm$^{-1}$. Starting from A(2) a band raises at about 2270 cm$^{-1}$, very close to the $Y^{IR}$ band of the 2D crystal. Similarly, in the IR spectrum of Z(*n*)-GDYNRs we find a rather strong $Y^{IR}$ band coming from two close phonons, computed at 2269 and 2277 cm$^{-1}$. Another IR peak at 2350 cm$^{-1}$ is clearly observed for Z(*n*)-GDYNRs, corresponding to the IR active ECC mode localized on the edges. On the opposite, only a very weak additional absorption band is present for A(*n*)-GDYNRs. This behavior makes this feature a potential marker of Z edges.

*850-1600 $cm^{-1}$ region.* The strong R band of 2D-GDY as well as the weak R' find a nice correspondence with two IR features of GYNRs, beginning with A(2) and Z(1.5).
The bands predicted at about 1270 $cm^{-1}$ and 700 $cm^{-1}$ in the larger A(*n*)-GDYNRs, even if weak, are significant markers of confinement. While many intense bands are predicted for Z(1), the bands below 1400 $cm^{-1}$ are weaker in Z(*n*)-GDYNRs: as in the Raman, markers of Z- nanoribbons are difficult to identify, also with IR spectroscopy in this spectral region. A good marker of Z(*n*)-GDYNRs could be the band at about 1620 $cm^{-1}$, which results from the convolution of two bands computed at 1619 and 1626 $cm^{-1}$.
In Figures 5 and 6 we highlight in red the more interesting IR and Raman markers of spatial confinement, which are often sensitive to the edge conformation. A brief discussion summarizing the spectroscopic marker bands allowing to distinguish between armchair and zigzag GDYNR is reported in the Supplementary Material [50].
In Supplementary Material [50] we provide also information about Raman tensor properties and dipole derivatives direction of some selected Raman and IR transitions of A(3). These data show that the remarkable ribbon anisotropy determine peculiar polarization properties of the Raman (and IR) transitions, which in turns make suitable the assessment of axial orientation, through polarized Raman and IR experiments.

### *3.3.3 Effects of the edge groups*

In addition to the spectral region above 1000 $cm^{-1}$, the IR spectra show another frequency range where significant features could be analyzed to discriminate between GDYNR with different edges. In particular, Ref. [45] showed that, similarly to CH stretching modes located in different intramolecular environments [46], also out-of-plane (opla) CH bending normal modes (600-900 $cm^{-1}$ range) are influenced by the molecular topology. In the case of polycyclic aromatic hydrocarbons (PAHs), out-of-plane CH bending (hereafter referred as opla) modes bring significant information about the edge: indeed, depending on the edge topology different "types" of vibrational patterns, associated to IR transitions, can be identified, each one at characteristic wavenumber. These vibrational patterns are sketched in Figure 7 and are described as collective out of plane displacements of one, two, or three adjacent CH bonds. In the literature, they have been named SOLO, DUO and TRIO [47].
In PAHs, TRIO modes are located at lower wavenumbers than SOLO modes while DUO vibrations are located between these two extremes. Moreover, since these modes are out-of-plane vibrations (orthogonal to the plane of π-conjugated atoms) and are completely localized on terminal CH bonds, they are not affected by π-electron delocalization. Therefore, the same behavior observed in PAHs, should be found also in the case of GDYNR.
We report in Figure 7 the analysis of the 750-950 $cm^{-1}$ range for 2D-GDY, A(3)- and Z(3)-GDYNR.

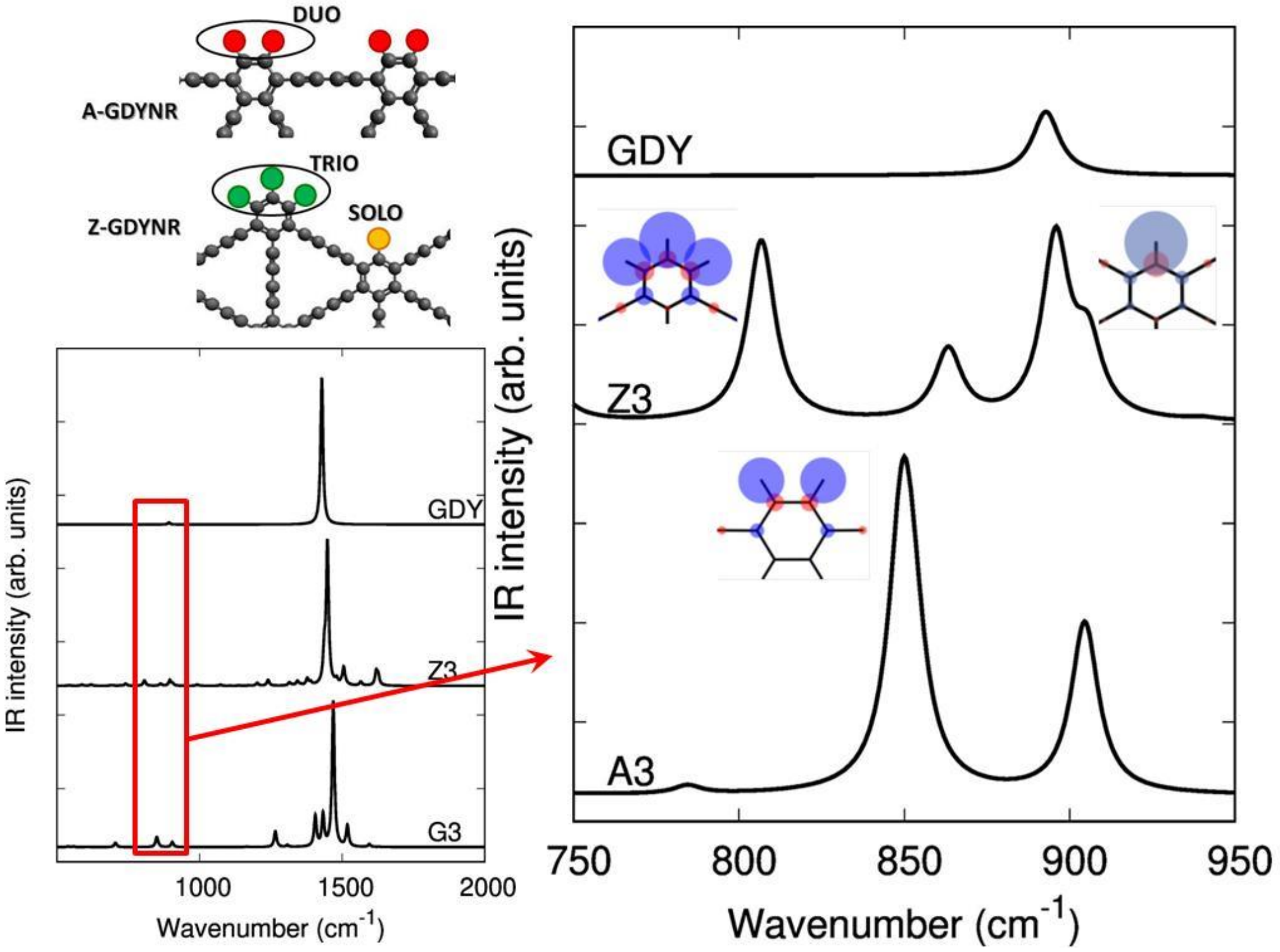

***Figure 7:*** *(top, left) Definition of SOLO, DUO and TRIO vibrational patterns associated to opla CH bending normal modes found in A- and Z-GDYNR. (right) Comparison among the DFT computed (PBE0/6-31G(d)) IR spectra of 2D-GDY, with A(3)- and Z(3)-GDYNR in the frequency range 750-950 cm$^{-1}$ where opla CH bending vibrations of the edges are found. Sketched of the OPLA CH normal modes describing the different bands are reported in figure: atomic displacements are indicated by red and purple circles with different colours indicating respectively opposite spatial phases of the motion. The amplitude of the displacement are proportional to the diameter of the circle. The wavenumber values are not scaled.*

Even if the opla CH bending vibrations give contributions to the IR spectra weaker than the bands above 1250 cm$^{-1}$ (see the bottom left panel of Figure 7), the bands between 750 and 950 cm$^{-1}$ are non-negligible and are potentially measurable. By analyzing more in details this region, we observe significant differences between A(3)- and Z(3)-GDYNR. A(3)-GDYNR shows two bands at 850 cm$^{-1}$ and 905 cm$^{-1}$: the latter is reminiscent of a band found in 2D-GDY and it is due to a bulk mode of carbons atoms. As expected, the band at 850 cm$^{-1}$ is an opla CH bending vibration, and it belongs to the DUO pattern. In the case of Z(*n*)-GDYNR, four IR bands are found (807, 863, 896 and 906 cm$^{-1}$). The bands at 863 cm$^{-1}$ and 896 cm$^{-1}$ are related to bulk modes of carbon domains, while the intense band at 807 cm$^{-1}$ and the shoulder at 906 cm$^{-1}$ are assigned to pure opla CH bending vibrations. In full agreement with the behavior of PAHs, the band at 807 cm$^{-1}$ is associated to TRIO moieties and the band at 906 cm$^{-1}$ to SOLO moieties. Based on these results, we infer that the analysis of IR spectra below 1000 cm$^{-1}$ would allow to detect useful markers able to clearly discriminate A(*n*)-GDYNR from Z(*n*)-GDYNR. We also note that a similar spectroscopic signature is predicted also for Raman-active opla CH bending vibrations. However, their Raman intensity is very weak and probably not measurable.

## 4. Summary and Conclusion

State-of-the-art periodic boundary conditions DFT calculations have been used here to compute the band gap and the vibrational response of 2D γ-GDY and related nanoribbons with different widths and edge-type.

We have obtained the following results:

1. The band gap of 2D-GDY (γ-GDY) computed at HSE06/6-31G(d) level of theory (hybrid functional, Gaussian basis set) provides a reliable band gap (1.11 eV), in very good agreement with the value of 1.1 eV obtained by using the GW method.
2. The trends of the band gap as a function of the GDYNRs width, for both armchair (A) and zigzag (Z) edge can be rationalized based on the different structures, and comply with the band gaps of the polymers A(1)-GDYNR and Z(1)-GDYNR, taken as the $n = 1$ width limit.
3. Raman and IR spectra of 2D-GDY have been computed, assigned and discussed in detail, showing a good agreement with available literature data. These spectra can be taken as a reference for the investigation of confined systems and for the interpretation of experimental measurements. Indeed, in a recent work [48], Raman spectroscopy has been applied to characterize GDY grown on graphene, obtaining a nice agreement with previous calculations [27].
4. We have computed the Raman and IR spectra of GDYNRs with different width and edge. We have unveiled Raman markers of A(*n*)-GDYNR at 1630 $cm^{-1}$ and 1450 $cm^{-1}$, whereas the weak bands computed at 700, 1270 and 1405 $cm^{-1}$ are IR markers. In the case of Z(*n*)-GDYNRs, Raman markers are too weak to be of experimental interest, while the most significant markers are found in the IR spectrum at about 1620 and 2350 $cm^{-1}$. The Raman bands above 2200 $cm^{-1}$ display an evident redshift for increasing GDYNRs width; hence their position can be useful for a qualitative assessment of the GDYNR width. Apart from minor details, no significant difference has been found in the spectroscopic response of integer *vs*. half-integer Z(*n*)-GDYNRs. Finally, IR-active CH out-of-plane modes have been found to have a characteristic dependence on the edge type, providing another significant marker for the characterization of GDYNR.

Our results are relevant as reference spectra of confined GDY-systems, suggesting that both Raman and IR techniques are useful tools for the identification of nanoribbons or to check for the existence of confinement phenomena in 2D-GDY. Our results indicate that it is possible to discriminate systems having different edges, suggesting the application of vibrational spectroscopy for the characterization of topology. As an example in this context, in a very recent paper [49], a combined experimental and computational approach has been used to characterize A1 polymer synthetised on Au surface, providing an interpretation of the bonding processes occurring in the synthesis.

**Acknowledgements**
Authors acknowledge funding from the European Research Council (ERC) under the European Union's Horizon 2020 research and innovation program ERC—Consolidator Grant (ERC CoG 2016 EspLORE grant agreement No. 724610, website: www.esplore.polimi.it)

# Supplementary Material:

## Raman and IR spectra of graphdiyne nanoribbons

Patrick Serafini[1], Alberto Milani[1*], Matteo Tommasini[2], Chiara Castiglioni[2], Carlo S. Casari[1*]

[1]*Dipartimento di Energia, Politecnico di Milano, via Ponzio 34/3, Milano, Italy*

[2]*Dipartimento di Chimica, Materiali e Ing. Chimica, Politecnico di Milano, P.zza L. da Vinci 32, Milano, Italy*

*Corresponding authors: alberto.milani@polimi.it; carlo.casari@polimi.it

# Appendix

***Table A.1**: Definition of the internal CC stretching coordinates of 2D-GDY fragments:*

| | | |
|---|---|---|
| 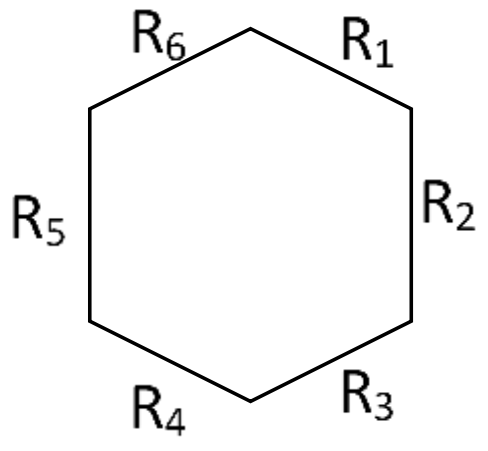 | 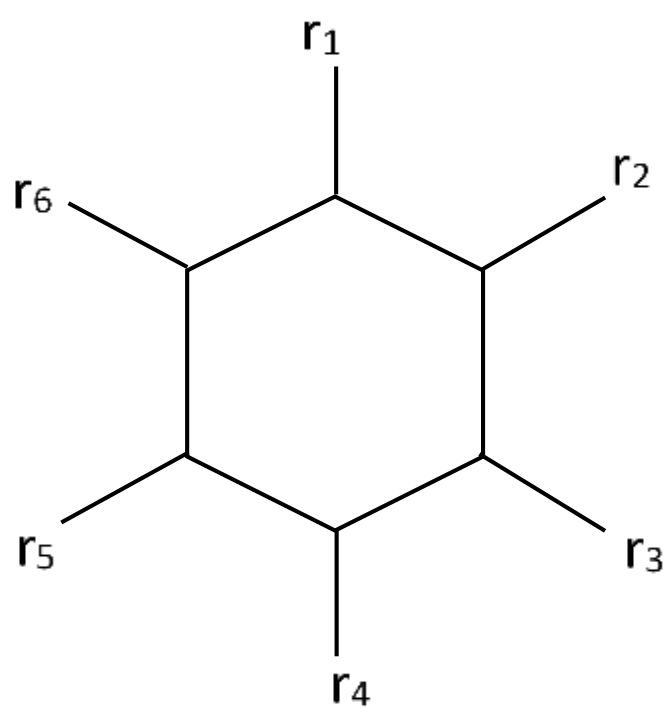 | 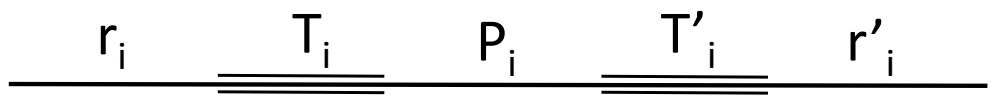 |
| CC stretching coordinates $\{R_i\}$ of the aromatic ring | CC stretching $\{r_i\}$ of the polyyne arms. $\{T_i\}$ and $\{P_i\}$ coordinates are defined consistently. | CC stretching coordinates of an individual i-th polyyne arm |

***Table A.2**: Group coordinates of the individual i-th polyyne arm*

| | |
|---|---|
| $r_i^+ = \frac{1}{\sqrt{2}}(r_i + r_i')$ | 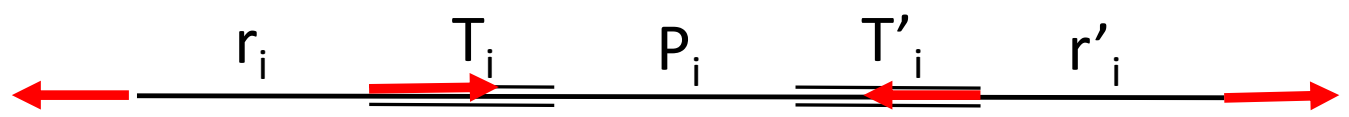 |
| $T_i^+ = \frac{1}{\sqrt{2}}(T_i + T_i')$ | 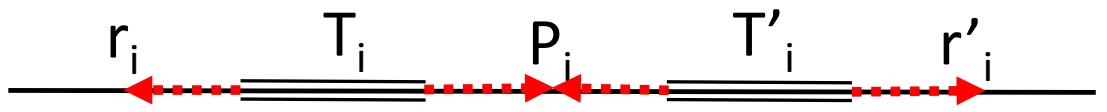 |
| $P_i$ | 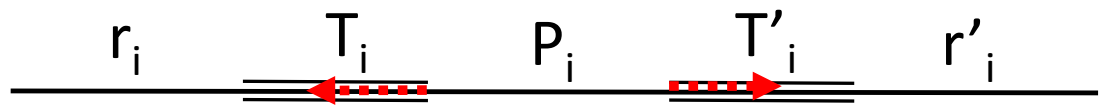 |
| $r_i^- = \frac{1}{\sqrt{2}}(r_i - r_i')$ | 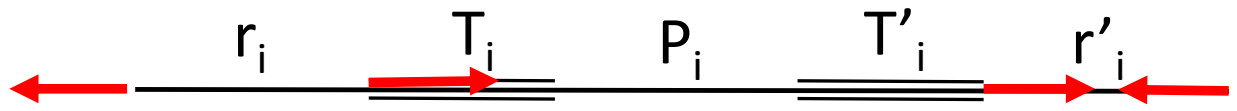 |
| $T_i^- = \frac{1}{\sqrt{2}}(T_i - T_i')$ | 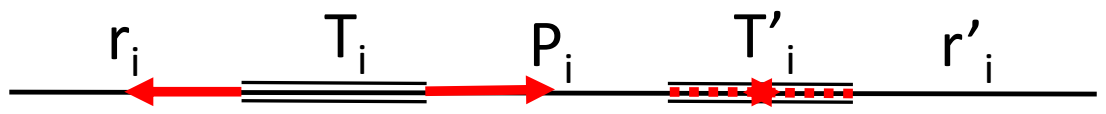 |

*Table A.3: CC stretching symmetry coordinates of 2D-GDY crystal. Last column: sketch of the stretching pattern associated to each individual symmetry coordinate; bonds which stretch: blue solid line, bonds which shrink: red, broken line.*

| Symmetry Species | Basis of internal coordinates of the unit cell (ring and polyyne arm) | Basis of group coordinates of six individual polyyne arms | stretching pattern |
|---|---|---|---|
| $A_{1g}$ | $A^R_{1g} = \frac{1}{\sqrt{6}}(R_1 + R_2 + R_3 + R_4 + R_5 + R_6)$ | | 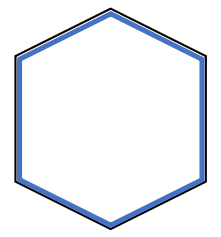 |
| | $A^r_{1g} = \frac{1}{\sqrt{6}}(r_1 + r_2 + r_3 + r_4 + r_5 + r_6)$ | $A^{r+}_{1g} = \frac{1}{\sqrt{6}}(r^+_1 + r^+_2 + r^+_3 + r^+_4 + r^+_5 + r^+_6)$ | 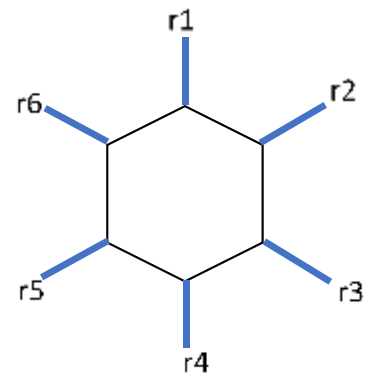  |
| | $A^T_{1g} = \frac{1}{\sqrt{6}}(T_1 + T_2 + T_3 + T_4 + T_5 + T_6)$ | $A^{T+}_{1g} = \frac{1}{\sqrt{6}}(T^+_1 + T^+_2 + T^+_3 + T^+_4 + T^+_5 + T^+_6)$ | the pattern can be deduced considering that of $A^r_{1g}$ |
| | $A^P_{1g} = \frac{1}{\sqrt{3}}(P_1 + P_2 + P_6)$ | $A^P_{1g} = \frac{1}{\sqrt{3}}(P_1 + P_2 + P_3 + P_4 + P_5 + P_6)$ | the pattern can be deduced considering that of $A^r_{1g}$ |
| $E_{2g}$ | $E^R_{2g}(Y) = \frac{1}{\sqrt{12}}(2R_2 + 2R_5 - R_1 - R_3 - R_4 - R_6)$<br>$E^R_{2g}(X) = \frac{1}{2}(R_1 - R_3 + R_4 - R_6)$ | | 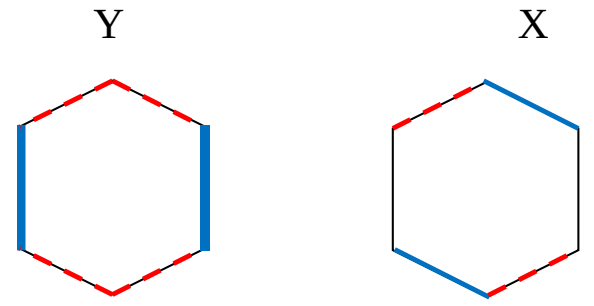  |
| | $E^r_{2g}(Y) = \frac{1}{\sqrt{12}}(2r_1 + 2r_4 - r_2 - r_3 - r_5 - r_6)$<br>$E^R_{2g}(X) = \frac{1}{2}(r_2 - r_3 + r_5 - r_6)$ | $E^{r+}_{2g}(Y) = \frac{1}{\sqrt{12}}(2r^+_1 + 2r^+_4 - r^+_2 - r^+_3 - r^+_5 - r^+_6)$<br>$E^{r+}_{2g}(X) = \frac{1}{2}(r^+_2 - r^+_3 + r^+_5 - r^+_6)$ | 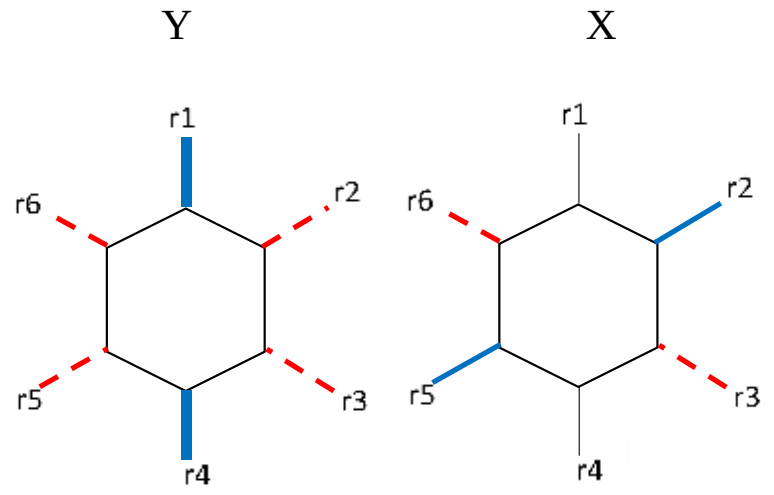  |
| | $E^T_{2g}(Y) = \frac{1}{\sqrt{12}}(2T_1 + 2T_4 - T_2 - T_3 - T_5 - T_6)$<br>$E^T_{2g}(X) = \frac{1}{2}(T_2 - T_3 + T_5 - T_6)$ | $E^{T+}_{2g}(Y) = \frac{1}{\sqrt{12}}(2T^+_1 + 2T^+_4 - T^+_2 - T^+_3 - T^+_5 - T^+_6)$<br>$E^{T+}_{2g}(X) = \frac{1}{2}(T^+_2 - T^+_3 + T^+_5 - T^+_6)$ | the pattern can be deduced considering that of $E^r_{2g}$ |
| | $E_{2g}^{P}(Y) = \frac{1}{\sqrt{6}}(2P_1 - P_2 - T_6)$ <br> $E_{2g}^{P}(X) = \frac{1}{2}(P_2 - P_6)$ | $E_{2g}^{P}(Y) = \frac{1}{\sqrt{12}}(2P_1 + 2P_4 - P_2 - P_3 - P_5 - P_6)$ <br> $E_{2g}^{P}(X) = \frac{1}{2}(P_2 - P_3 + P_5 - P_6)$ | the pattern can be deduced considering that of $E_{2g}^{r}$ |
| | $B_{2u}^{R} = \frac{1}{\sqrt{6}}(R_1 - R_2 + R_3 - R_4 + R_5 - R_6)$ | | 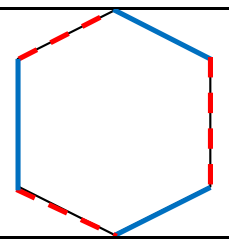 |
| $B_{2u}$ | $B_{2u}^{r} = \frac{1}{\sqrt{6}}(r_1 - r_2 + r_3 - r_4 + r_5 - r_6)$ | $B_{2u}^{r-} = \frac{1}{\sqrt{6}}(r_1^- - r_2^- + r_3^- - r_4^- + r_5^- - r_6^-)$ | 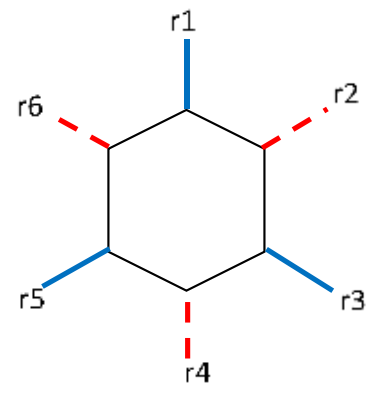  |
| | $B_{2u}^{T} = \frac{1}{\sqrt{6}}(T_1 - T_2 + T_3 - T_4 + T_5 - T_6)$ | $B_{2u}^{T-} = \frac{1}{\sqrt{6}}(T_1^- - T_2^- + T_3^- - T_4^- + T_5^- - T_6^-)$ | the pattern can be deduced considering that of $B_{2u}^{r}$ |
| | $E_{1u}^{R}(Y) = \frac{1}{2}(R_1 - R_3 - R_4 + R_6)$ <br> $E_{1u}^{R}(X) = \frac{1}{\sqrt{12}}(2R_2 - 2R_5 + R_1 + R_3 - R_4 - R_6)$ | | 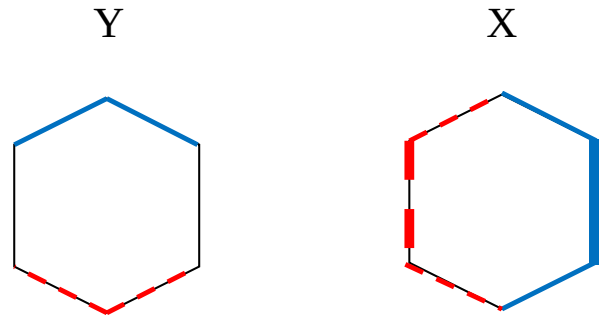  |
| $E_{1u}$ | $E_{1u}^{r}(Y) = \frac{1}{\sqrt{12}}(2r_1 - 2r_4 + r_2 - r_3 - r_5 + r_6)$ <br> $E_{1u}^{r}(X) = \frac{1}{2}(r_2 + r_3 - r_5 - r_6)$ | $E_{1u}^{r-}(Y) = \frac{1}{\sqrt{12}}(2r_1^- - 2r_4^- + r_2^- - r_3^- - r_5^- + r_6^-)$ <br> $E_{1u}^{r-}(X) = \frac{1}{2}(r_2^- + r_3^- - r_5^- - r_6^-)$ | 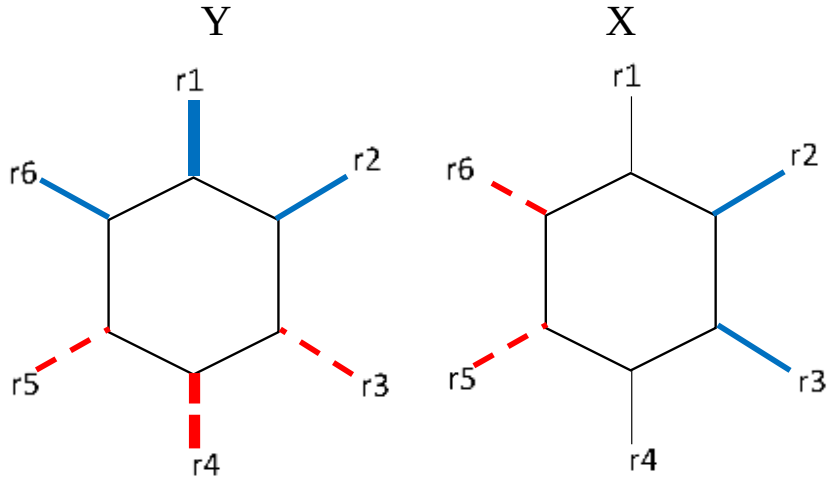  |
| | $E_{1u}^{T}(Y) = \frac{1}{\sqrt{12}}(2T_1 - 2T_4 + T_2 - T_3 - T_5 + T_6)$ <br> $E_{1u}^{T}(X) = \frac{1}{2}(T_2 + T_3 - T_5 - T_6)$ | $E_{1u}^{T-}(Y) = \frac{1}{\sqrt{12}}(2T_1^- - 2T_4^- + T_2^- - T_3^- - T_5^- + T_6^-)$ <br> $E_{1u}^{T-}(X) = \frac{1}{2}(T_2^- + T_3^- - T_5^- - T_6^-)$ | the pattern can be deduced considering that of $E_{1u}^{r}$ |

**Table S1:** DFT computed **[**PBE0/6-31G(d)] primitive cell parameters (in Angstrom) and fractional coordinates of 2D-GDY

space group :P 6/MMM

LATTICE PARAMETERS (ANGSTROMS AND DEGREES)

| A | B | C | ALPHA | BETA | GAMMA |
|---|---|---|---|---|---|
| 9.44081102 | 9.44081102 | 500.00000000 | 90.000000 | 90.000000 | 120.000000 |

*******************************************************************************

ATOMS IN THE ASYMMETRIC UNIT 3 - ATOMS IN THE UNIT CELL: 18

| ATOM | X/A | Y/B | Z(ANGSTROM) |
|---|---|---|---|
| C | 0.000000000000E+00 | 1.502058239380E-01 | 0.000000000000E+00 |
| C | 0.000000000000E+00 | -1.502058239380E-01 | 0.000000000000E+00 |
| C | -1.502058239380E-01 | -1.502058239380E-01 | 0.000000000000E+00 |
| C | 1.502058239380E-01 | 1.387778780781E-17 | 0.000000000000E+00 |
| C | -1.502058239380E-01 | -1.387778780781E-17 | 0.000000000000E+00 |
| C | 1.502058239380E-01 | 1.502058239380E-01 | 0.000000000000E+00 |
| C | -2.991003876810E-01 | -2.775557561563E-17 | 0.000000000000E+00 |
| C | 2.991003876810E-01 | 2.775557561563E-17 | 0.000000000000E+00 |
| C | -2.721815550437E-17 | -2.991003876810E-01 | 0.000000000000E+00 |
| C | 2.991003876810E-01 | 2.991003876810E-01 | 0.000000000000E+00 |
| C | -2.991003876810E-01 | -2.991003876810E-01 | 0.000000000000E+00 |
| C | 2.721815550437E-17 | 2.991003876810E-01 | 0.000000000000E+00 |
| C | -4.284415967800E-01 | 0.000000000000E+00 | 0.000000000000E+00 |
| C | 4.284415967800E-01 | 2.775557561563E-17 | 0.000000000000E+00 |
| C | 0.000000000000E+00 | -4.284415967800E-01 | 0.000000000000E+00 |
| C | 4.284415967800E-01 | 4.284415967800E-01 | 0.000000000000E+00 |
| C | -4.284415967800E-01 | -4.284415967800E-01 | 0.000000000000E+00 |
| C | 0.000000000000E+00 | 4.284415967800E-01 | 0.000000000000E+00 |

**Table S2:** DFT computed [PBE0/6-31G(d)] primitive cell parameters and fractional coordinates of A(n)-GDYNR of different widths. Space groups are all P1.

## A1

LATTICE PARAMETERS (ANGSTROMS AND DEGREES)

| A | B | C | ALPHA | BETA | GAMMA |
|---|---|---|---|---|---|
| 9.43619694 | 500.00000000 | 500.00000000 | 90.000000 | 90.000000 | 90.000000 |

*******************************************************************************

ATOMS IN THE ASYMMETRIC UNIT 14 - ATOMS IN THE UNIT CELL: 14

| ATOM | X/A | Y(ANGSTROM) | Z(ANGSTROM) |
|---|---|---|---|
| C | 2.212034929497E-01 | 1.211539588444E+00 | 2.654137930430E-04 |
| C | 7.475810552195E-02 | 1.211716427417E+00 | 2.994996334514E-04 |
| C | -4.238964086493E-01 | 3.698240912281E-06 | -1.026613190419E-06 |
| C | -2.801454905482E-01 | 4.872875894212E-06 | 2.999029709488E-06 |
| C | 4.467980866812E-01 | -1.098926222355E-04 | 3.219440647709E-06 |
| C | -1.508389950894E-01 | 8.677619826011E-05 | -1.204768866295E-05 |
| C | 2.968774159284E-01 | -6.511564883681E-05 | 6.230821559945E-05 |
| C | -9.331936697957E-04 | 3.509700334213E-05 | -4.328741689435E-05 |
| H | 2.793218922936E-01 | 2.148492927733E+00 | 3.932939710312E-04 |
| C | 2.211942712962E-01 | -1.211616314353E+00 | -2.611670848359E-04 |
| C | 7.474883748800E-02 | -1.211696375321E+00 | -3.405211587565E-04 |
| H | 1.668152338480E-02 | 2.148848961456E+00 | 5.262876417629E-04 |
| H | 2.793052858649E-01 | -2.148610565712E+00 | -6.032846907945E-04 |
| H | 1.666517654799E-02 | -2.148790085712E+00 | -5.616870721088E-04 |

## A2

LATTICE PARAMETERS (ANGSTROMS AND DEGREES)

| A | B | C | ALPHA | BETA | GAMMA |
|---|---|---|---|---|---|
| 9.43869055 | 500.00000000 | 500.00000000 | 90.000000 | 90.000000 | 90.000000 |

*******************************************************************************

ATOMS IN THE ASYMMETRIC UNIT 32 - ATOMS IN THE UNIT CELL: 32

| ATOM | X/A | Y(ANGSTROM) | Z(ANGSTROM) |
|---|---|---|---|
| C | -4.254000625681E-01 | 9.399187332857E+00 | 3.086078600736E-04 |
| C | -2.791343574068E-01 | 9.399124893117E+00 | -5.806113996774E-04 |
| H | -4.835436994160E-01 | 1.033551288806E+01 | 5.209927012823E-04 |
| C | 4.981486340743E-01 | 8.193882978524E+00 | 3.840369790689E-04 |
| H | -2.209808604232E-01 | 1.033539924091E+01 | -1.208455334460E-03 |
| C | -2.026834746840E-01 | 8.193803890603E+00 | -1.130373674865E-04 |
| C | -5.329187136983E-02 | 8.206608854863E+00 | 5.236628716020E-05 |
| C | -4.273942038309E-01 | 6.960837973132E+00 | 5.766815097722E-04 |
| C | 7.600739486134E-02 | 8.224823455280E+00 | -2.014476441927E-04 |
| C | -2.771448082145E-01 | 6.960725877736E+00 | 7.011788936198E-04 |
| C | 2.194547002114E-01 | 8.224888730501E+00 | -2.181637874389E-04 |
| C | 4.981534321064E-01 | 5.740238384991E+00 | 3.163516372525E-05 |
| C | 3.487554169970E-01 | 8.206645568051E+00 | -1.048552930959E-05 |
| C | 4.336187951974E-01 | 4.682818528881E+00 | -7.459053912106E-04 |

C -2.026875999680E-01 5.740183280177E+00 8.058154158979E-04
C 3.622452177501E-01 3.509764208636E+00 -1.390054891331E-03
C -1.378564439804E-01 4.684354044254E+00 7.602666949035E-04
C 2.977135061582E-01 2.452324431623E+00 -1.849872725450E-03
C 7.280333371417E-02 1.234865263202E+00 -8.595345136725E-04
C 2.230466825579E-01 1.232924030973E+00 -5.463181059646E-04
C -6.631891509260E-02 3.512168941979E+00 2.623627782419E-04
C -1.836669748748E-03 2.754171089640E-03 2.833479256309E-06
C -1.495102010349E-03 2.456330180625E+00 -8.681242652421E-04
C 2.973314492072E-01 -1.161951415020E-03 -5.465493040682E-04
C 7.443823853555E-02 -1.203545543557E+00 5.440960388187E-04
C 4.467219094107E-01 -1.556155531468E-02 -5.667888299237E-04
C -4.239741768687E-01 -3.034562941349E-02 -4.603384317417E-04
C 2.207079709870E-01 -1.205450337767E+00 6.702874792862E-04
C -2.805332216125E-01 -2.911924648773E-02 -3.005250356739E-04
C -1.512359788744E-01 -8.589842401511E-03 -1.788584233349E-04
H 1.614628446324E-02 -2.139042191055E+00 2.548893861617E-03
H 2.787144798369E-01 -2.142530852650E+00 1.435015837465E-03

**A3**

LATTICE PARAMETERS (ANGSTROMS AND DEGREES)

| A | B | C | ALPHA | BETA | GAMMA |
|---|---|---|---|---|---|
| 9.43907321 | 500.00000000 | 500.00000000 | 90.000000 | 90.000000 | 90.000000 |

*******************************************************************************

ATOMS IN THE ASYMMETRIC UNIT 50 - ATOMS IN THE UNIT CELL: 50

ATOM X/A Y(ANGSTROM) Z(ANGSTROM)

*******************************************************************************

C -9.308873694425E-03 -2.138885297421E-01 2.347323029239E-02
C 6.707350216770E-02 9.220923777800E-01 -3.807778986710E-01
H 8.830412508931E-03 1.803520912578E+00 -6.947773462818E-01
C 6.524974064744E-02 -1.375135543327E+00 4.371699278943E-01
C 2.133363900703E-01 9.232987004311E-01 -3.809330346334E-01
C 2.155328689350E-01 -1.373595702395E+00 4.374257598518E-01
C 2.899021532066E-01 -2.113489879464E-01 2.379006871945E-02
C 2.901874291557E-01 -2.521496984476E+00 8.475642639247E-01
H 2.714076528965E-01 1.805617519551E+00 -6.952764871588E-01
C 3.542329348589E-01 -3.521264120074E+00 1.200594758137E+00
C 4.392850659142E-01 -1.986275973606E-01 1.942352171978E-02
C 4.257120922383E-01 -4.625956521381E+00 1.590926399576E+00
C 4.903123579543E-01 -5.622151746090E+00 1.945821174540E+00
C 4.900432897069E-01 -7.923596726907E+00 2.771802573315E+00
C -4.314119122596E-01 -1.849210546814E-01 1.464828609016E-02
C -4.350139928959E-01 -6.770434362813E+00 2.347734591007E+00
C -4.348117269065E-01 -9.082738328042E+00 3.175546974587E+00
C -2.879871636280E-01 -1.856874291296E-01 1.458466969749E-02
C -2.848482098441E-01 -6.765453655889E+00 2.357302301283E+00
C -2.846425446667E-01 -9.078068432344E+00 3.184515770735E+00
C -2.096981123467E-01 -7.925129817965E+00 2.759952170664E+00
C -2.104967038589E-01 -5.619956846324E+00 1.941486852952E+00

C    -1.586887649373E-01 -2.022137869098E-01  1.969130297656E-02
C    -1.459130178620E-01 -4.622589848485E+00  1.589285804665E+00
C    -7.429619006216E-02 -3.519545065958E+00  1.196476845704E+00
C    -9.190339085525E-03 -2.525330031181E+00  8.446982249289E-01
C    -6.075871298043E-02 -7.922221943237E+00  2.766462052915E+00
C     6.859666314635E-02 -7.924788502257E+00  2.765358174276E+00
C     2.117578550339E-01 -7.924880211134E+00  2.767685930058E+00
C     3.411055282175E-01 -7.926814375336E+00  2.765517002088E+00
C     2.132736064664E-01 -1.675505057153E+01  5.957473066003E+00
C     6.700758282943E-02 -1.675548171351E+01  5.957080169885E+00
C    -4.315846827399E-01 -1.565008626736E+01  5.555608401119E+00
C     4.391128219107E-01 -1.563414418134E+01  5.550684727187E+00
C     2.897333553649E-01 -1.562198151480E+01  5.546452730888E+00
C    -9.486989067326E-03 -1.562294534157E+01  5.545180274041E+00
C    -1.588604511563E-01 -1.563523531786E+01  5.549206678379E+00
C    -2.881592468555E-01 -1.565060806887E+01  5.555170680436E+00
C     2.152409619868E-01 -1.446293047724E+01  5.125834028228E+00
C     6.497238603948E-02 -1.446370472169E+01  5.124548137080E+00
C     2.897156659801E-01 -1.331557833370E+01  4.711133216055E+00
C    -9.526563082916E-03 -1.331767630971E+01  4.706703906784E+00
C     4.908573565516E-01 -1.022684812709E+01  3.594936502467E+00
C     4.262016025559E-01 -1.122186533010E+01  3.952455630617E+00
C     3.545966710407E-01 -1.232262875885E+01  4.352136904728E+00
C    -7.465834821595E-02 -1.232729256736E+01  4.344724439197E+00
C    -2.101462970937E-01 -1.022674153893E+01  3.588634212943E+00
C    -1.460049112593E-01 -1.122384745535E+01  3.948569891348E+00
H     2.714479643607E-01 -1.763498059038E+01  6.276798031921E+00
H     8.866842754113E-03 -1.763551617170E+01  6.276714504841E+00

## A4

LATTICE PARAMETERS (ANGSTROMS AND DEGREES)

| A | B | C | ALPHA | BETA | GAMMA |
|---|---|---|---|---|---|
| 9.43895056 | 500.00000000 | 500.00000000 | 90.000000 | 90.000000 | 90.000000 |

*******************************************************************************

ATOMS IN THE ASYMMETRIC UNIT   68 - ATOMS IN THE UNIT CELL:   68

ATOM    X/A    Y(ANGSTROM)    Z(ANGSTROM)

*******************************************************************************

C    -4.484364735474E-01  2.470859598131E+01  8.297619404729E+00
C    -3.021656616272E-01  2.470843741747E+01  8.295715057842E+00
C     4.750297456446E-01  2.356840643126E+01  7.907976348164E+00
H     4.934663617475E-01  2.559428592243E+01  8.602176107584E+00
C    -2.257509196278E-01  2.356763228916E+01  7.905707280242E+00
H    -2.439411620731E-01  2.559395325959E+01  8.599187548842E+00
C    -4.505588714091E-01  2.240118846507E+01  7.509352835262E+00
C    -7.638025890143E-02  2.358005022719E+01  7.908131419933E+00
C     5.292181772888E-02  2.359630147145E+01  7.914285220079E+00
C    -3.002842864729E-01  2.240073618467E+01  7.507954664397E+00
C     4.748470373729E-01  2.124737034878E+01  7.115022350690E+00
C     1.963634279144E-01  2.359648253321E+01  7.914545443523E+00

| | | | |
|---|---|---|---|
| C | 3.256657367142E-01 | 2.358025930202E+01 | 7.913515424629E+00 |
| C | 4.103049669101E-01 | 2.024658303232E+01 | 6.772539787053E+00 |
| C | -2.258249870878E-01 | 2.124570407140E+01 | 7.114833746931E+00 |
| C | 3.389045949142E-01 | 1.913632276259E+01 | 6.396297732764E+00 |
| C | -1.607775014441E-01 | 2.024705833869E+01 | 6.774571505947E+00 |
| C | 2.747257401165E-01 | 1.813326669890E+01 | 6.053967490601E+00 |
| C | 4.974890848640E-02 | 1.698288998852E+01 | 5.666134301903E+00 |
| C | 1.999208340237E-01 | 1.697756748058E+01 | 5.676262941951E+00 |
| C | -8.915558290775E-02 | 1.913865617223E+01 | 6.396802372293E+00 |
| C | -2.549919336085E-02 | 1.581490289845E+01 | 5.289746961894E+00 |
| C | 2.748041614867E-01 | 1.581589526572E+01 | 5.274837140492E+00 |
| C | -2.454328357608E-02 | 1.813608246869E+01 | 6.060733796806E+00 |
| C | 4.937690600232E-02 | 1.465260900775E+01 | 4.889831699020E+00 |
| C | 4.237260738798E-01 | 1.581197498828E+01 | 5.281024387169E+00 |
| C | 1.995787925273E-01 | 1.464770662130E+01 | 4.899003682859E+00 |
| C | -4.469197203688E-01 | 1.581235436947E+01 | 5.280291107332E+00 |
| C | -2.518508075259E-02 | 1.349457428088E+01 | 4.516048274063E+00 |
| C | -3.037618618827E-01 | 1.581222203651E+01 | 5.283466700918E+00 |
| C | -1.744165630581E-01 | 1.581883687897E+01 | 5.283676652467E+00 |
| C | -8.977233760057E-02 | 1.249063374101E+01 | 4.182566661155E+00 |
| C | 2.739998286621E-01 | 1.349522153559E+01 | 4.506146189623E+00 |
| C | -1.612042505683E-01 | 1.137891527727E+01 | 3.814227824395E+00 |
| C | 3.391716645953E-01 | 1.249323019995E+01 | 4.176814152130E+00 |
| C | -2.252809929348E-01 | 1.037148307722E+01 | 3.482462720408E+00 |
| C | 4.109345613987E-01 | 1.138295720063E+01 | 3.810615859188E+00 |
| C | -4.502946640378E-01 | 9.217958339882E+00 | 3.106808938049E+00 |
| C | -3.000969697079E-01 | 9.219250685693E+00 | 3.095366786582E+00 |
| C | 4.744672900287E-01 | 8.055714733304E+00 | 2.712940661926E+00 |
| C | 4.753860804868E-01 | 1.037821181675E+01 | 3.477715015348E+00 |
| C | -2.252649108694E-01 | 8.047722400987E+00 | 2.722465315649E+00 |
| C | -7.634191067815E-02 | 8.048456172690E+00 | 2.714591246939E+00 |
| C | -4.506977321380E-01 | 6.884278037317E+00 | 2.340179499713E+00 |
| C | 5.301574389113E-02 | 8.051236669514E+00 | 2.717386931859E+00 |
| C | -3.005088953854E-01 | 6.885589745106E+00 | 2.328482399057E+00 |
| C | 1.962000069899E-01 | 8.053306410806E+00 | 2.716374567980E+00 |
| C | 3.255480833876E-01 | 8.055092103668E+00 | 2.720854962018E+00 |
| C | 4.745514239980E-01 | 5.731687486445E+00 | 1.952307865487E+00 |
| C | 4.104641955405E-01 | 4.724301897882E+00 | 1.621514990213E+00 |
| C | -2.261696604982E-01 | 5.725725504116E+00 | 1.955268382939E+00 |
| C | 3.389107053476E-01 | 3.612625176588E+00 | 1.252180236870E+00 |
| C | -1.615941667197E-01 | 4.721725293732E+00 | 1.622291349969E+00 |
| C | 2.745627711318E-01 | 2.607681146776E+00 | 9.188984497201E-01 |
| C | 4.956908032118E-02 | 1.450159914016E+00 | 5.408347319320E-01 |
| C | 1.998406637000E-01 | 1.449676457653E+00 | 5.394173873673E-01 |
| C | -8.986534113233E-02 | 3.609920590621E+00 | 1.256499390716E+00 |
| C | -2.495220532232E-02 | 2.792518751962E-01 | 1.546691269886E-01 |
| C | -2.494381248401E-02 | 2.607713904875E+00 | 9.251952887632E-01 |
| C | 2.742433431660E-01 | 2.783364312804E-01 | 1.525987029070E-01 |
| C | 5.145503976708E-02 | -8.653821227322E-01 | -2.238321906383E-01 |

C 4.236208966743E-01 2.659304066495E-01 1.486440680759E-01
C 1.977236364824E-01 -8.660855292267E-01 -2.243275561409E-01
C -4.470716957205E-01 2.507888209554E-01 1.440544578392E-01
C -3.036289637902E-01 2.513714565911E-01 1.449369686345E-01
C -1.743278013714E-01 2.671660693677E-01 1.503716018368E-01
H -6.760630950148E-03 -1.753964230564E+00 -5.175109545170E-01
H 2.558232289683E-01 -1.755305892457E+00 -5.180914193576E-01

**A5**

LATTICE PARAMETERS (ANGSTROMS AND DEGREES)

| A | B | C | ALPHA | BETA | GAMMA |
|---|---|---|---|---|---|
| 9.43958534 | 500.00000000 | 500.00000000 | 90.000000 | 90.000000 | 90.000000 |

*******************************************************************************

ATOMS IN THE ASYMMETRIC UNIT 86 - ATOMS IN THE UNIT CELL: 86

ATOM X/A Y(ANGSTROM) Z(ANGSTROM)

*******************************************************************************

C -7.295725578702E-03 -1.374838629765E+01 9.544323903213E+00
C 6.774280145206E-02 -1.274010423489E+01 8.842667219211E+00
C 6.782092039848E-02 -1.474186880218E+01 1.026602828210E+01
C 2.179479303095E-01 -1.274500162108E+01 8.835123506379E+00
C 2.180359264253E-01 -1.474695136589E+01 1.025818192688E+01
C 2.930687131090E-01 -1.373882801699E+01 9.556322985811E+00
C 2.925183538835E-01 -1.573232908345E+01 1.097252029512E+01
C 3.568989372267E-01 -1.659458972305E+01 1.158730099929E+01
C 4.419726907302E-01 -1.374288547151E+01 9.549754723029E+00
C 4.284357301733E-01 -1.754645125003E+01 1.226825591209E+01
C 4.932266265634E-01 -1.840569785678E+01 1.288352192512E+01
C -4.286719072086E-01 -1.374164949944E+01 9.551035572010E+00
C 4.927123152279E-01 -2.039001815138E+01 1.431193366477E+01
C -4.322279749057E-01 -1.940012138507E+01 1.358496257135E+01
C -4.322100584286E-01 -2.139289738749E+01 1.502049194357E+01
C -2.855526861424E-01 -1.374369690106E+01 9.548550893754E+00
C -2.820336872096E-01 -1.939465388064E+01 1.359230691734E+01
C -2.075449578228E-01 -1.841157617184E+01 1.287443091592E+01
C -2.820227091130E-01 -2.138672071931E+01 1.502895924039E+01
C -1.425789122063E-01 -1.755239408934E+01 1.226074990503E+01
C -2.069512168978E-01 -2.039800550970E+01 1.430055710596E+01
C -1.561979275332E-01 -1.374428000071E+01 9.550851404837E+00
C -7.099613033653E-02 -1.660041709425E+01 1.158038552276E+01
C -6.592275316372E-03 -1.573920309494E+01 1.096428783871E+01
C 2.160737139085E-01 -2.800730484494E+01 1.980039620678E+01
C 2.179316036323E-01 -2.602911978172E+01 1.837526140514E+01
C 2.924918775214E-01 -2.702889228276E+01 1.909631298731E+01
C -7.149518264378E-02 -2.418550626165E+01 1.704084579641E+01
C -6.749351189528E-03 -2.703132976096E+01 1.909594209483E+01
C -1.430404418425E-01 -2.323483683033E+01 1.635692799552E+01
C -2.076464337297E-01 -2.237852460406E+01 1.573631125224E+01
C 6.981120798358E-02 -2.800826372998E+01 1.980061805150E+01
C -6.945376906231E-03 -2.504250661965E+01 1.766067736451E+01

| | | | |
|---|---|---|---|
| C | 6.766957912949E-02 | -2.603065051043E+01 | 1.837480336082E+01 |
| H | 1.176673585286E-02 | -2.876863128749E+01 | 2.034766363074E+01 |
| H | 2.742420841851E-01 | -2.876656073638E+01 | 2.034769635911E+01 |
| C | -5.803674042584E-02 | -2.039351550886E+01 | 1.430612310040E+01 |
| C | 7.130471934786E-02 | -2.039175539445E+01 | 1.430273711954E+01 |
| C | 2.144558706611E-01 | -2.038997677112E+01 | 1.430482912206E+01 |
| C | 3.437943687077E-01 | -2.039427338946E+01 | 1.430579582130E+01 |
| C | 6.975390379720E-02 | 6.078645914188E-01 | -5.776139724315E-01 |
| C | -6.791977434202E-03 | -3.772126565336E-01 | 1.159795579925E-01 |
| C | 2.160160691009E-01 | 6.073445472807E-01 | -5.771976171294E-01 |
| C | 6.764554355039E-02 | -1.385493040932E+00 | 8.261129021906E-01 |
| C | 2.924525533469E-01 | -3.779984628636E-01 | 1.170726268300E-01 |
| C | 2.179082691485E-01 | -1.385638016764E+00 | 8.270893572832E-01 |
| C | 4.418214159207E-01 | -3.676178513971E-01 | 1.102161173043E-01 |
| C | 2.922849724473E-01 | -2.382181794217E+00 | 1.531631873602E+00 |
| C | -4.288833033348E-01 | -3.544083099983E-01 | 1.002623360507E-01 |
| C | -2.854592865116E-01 | -3.541906875206E-01 | 9.963899549480E-02 |
| C | 3.566409770910E-01 | -3.247756528667E+00 | 2.141234617078E+00 |
| C | -1.561613749000E-01 | -3.668216760370E-01 | 1.081396271452E-01 |
| C | 4.283484258693E-01 | -4.206019136313E+00 | 2.813208326865E+00 |
| C | -7.161559554360E-02 | -3.247504909457E+00 | 2.135398729171E+00 |
| C | -6.855219496566E-03 | -2.383391959843E+00 | 1.527512209263E+00 |
| C | -1.430988268731E-01 | -4.204555313400E+00 | 2.811086651311E+00 |
| C | 4.933441173836E-01 | -5.067889333687E+00 | 3.422286632533E+00 |
| C | -2.076517469392E-01 | -5.070229624155E+00 | 3.418935448759E+00 |
| C | 2.923431110400E-01 | -1.174609767945E+01 | 8.138988310420E+00 |
| H | 1.169150596208E-02 | 1.373662578840E+00 | -1.116925270364E+00 |
| C | -2.820600099683E-01 | -6.061162517371E+00 | 4.127008864673E+00 |
| C | -4.322426971903E-01 | -6.067491207402E+00 | 4.118090516362E+00 |
| C | 3.570574044670E-01 | -1.088389542864E+01 | 7.527239528333E+00 |
| C | -2.069646105140E-01 | -7.072245491630E+00 | 4.823705771098E+00 |
| C | -5.805387430369E-02 | -7.068039258846E+00 | 4.829521479687E+00 |
| C | 4.927021047116E-01 | -7.064229750920E+00 | 4.835592789044E+00 |
| C | 4.286363808259E-01 | -9.928302362870E+00 | 6.851829729144E+00 |
| C | -6.708716457986E-03 | -1.175313742479E+01 | 8.130255846506E+00 |
| C | 7.128679149440E-02 | -7.070215115552E+00 | 4.828576691998E+00 |
| H | 2.741772515427E-01 | 1.372729580224E+00 | -1.116172916944E+00 |
| C | 4.932940701673E-01 | -9.063679365332E+00 | 6.242942597411E+00 |
| C | -2.820168019551E-01 | -8.069643084407E+00 | 5.540430525947E+00 |
| C | 2.144473933397E-01 | -7.068698971205E+00 | 4.831114870667E+00 |
| C | -7.097259378676E-02 | -1.088814822106E+01 | 7.518222966508E+00 |
| C | -4.322055963874E-01 | -8.074826159222E+00 | 5.533179780027E+00 |
| C | 3.437900596335E-01 | -7.068981863311E+00 | 4.829791107234E+00 |
| C | -2.074828611532E-01 | -9.070688855271E+00 | 6.232559684734E+00 |
| C | -1.425064529631E-01 | -9.932649892057E+00 | 6.842183860541E+00 |
| C | 4.933842242019E-01 | -2.237589603852E+01 | 1.573969583511E+01 |
| C | 4.288651445585E-01 | -2.323430561599E+01 | 1.635815572929E+01 |
| C | 3.573429903166E-01 | -2.418280315126E+01 | 1.704545841028E+01 |
| C | 2.923294875578E-01 | -2.503873610716E+01 | 1.766235376713E+01 |
| C | -1.561117135474E-01 | -2.704190282219E+01 | 1.910368845597E+01 |
| C | -2.854133898984E-01 | -2.705343396447E+01 | 1.911212460482E+01 |
| C | -4.288387625300E-01 | -2.705261958900E+01 | 1.911232751782E+01 |
| C | 4.418652332222E-01 | -2.703896613516E+01 | 1.910335768437E+01 |

**Table S3:** DFT computed [PBE0/6-31G(d)] primiteve cell parameters and fractional coordinates of Z(n)-GDYNR of different widths. Space groups are all P1

## Z1

```
LATTICE PARAMETERS (ANGSTROMS AND DEGREES)
      A             B             C          ALPHA      BETA       GAMMA
   16.36146819   500.00000000   500.00000000    90.000000  90.000000  90.000000
*******************************************************************************
 ATOMS IN THE ASYMMETRIC UNIT   28 - ATOMS IN THE UNIT CELL:   28
    ATOM              X/A           Y(ANGSTROM)        Z(ANGSTROM)
*******************************************************************************
 C   -4.261216250627E-01 -5.308159610809E+00 -1.002590944289E+00
 C   -4.998120399951E-01 -4.632289595237E+00 -8.743345226672E-01
 C    7.387313789957E-02 -6.841465995049E-01 -1.294678883841E-01
 C   -2.767498846060E-01 -2.553898981741E+00 -4.836278158277E-01
 C   -4.087093752024E-04 -6.298682417201E-04 -5.166814214113E-04
 C   -3.524322197236E-01 -4.632074347445E+00 -8.753091443112E-01
 C   -2.121018733674E-01 -1.958678952062E+00 -3.717118574054E-01
 C    4.995910801751E-01 -3.251490461605E+00 -6.136834312481E-01
 C   -1.401119305652E-01 -1.293676306281E+00 -2.462398797615E-01
 C    4.245016627384E-01 -2.554355644389E+00 -4.816311239544E-01
 C    3.598512529245E-01 -1.959375438328E+00 -3.689653817318E-01
 C   -7.550854260742E-02 -6.973551093066E-01 -1.334457076090E-01
 C    2.878602977017E-01 -1.294410121744E+00 -2.432410157531E-01
 C    2.232537174692E-01 -6.981835405084E-01 -1.304746312136E-01
 C    1.481604172761E-01 -9.610795318600E-04  8.422105198642E-04
 C    1.894309412521E-04  1.380020272516E+00  2.610089695694E-01
 C   -3.518377880418E-01 -3.251271071488E+00 -6.146520600084E-01
 C   -4.261241441948E-01 -2.567961063701E+00 -4.850131459254E-01
 C    1.475693743039E-01  1.379670329076E+00  2.626087369791E-01
 C    7.387876812351E-02  2.055834025969E+00  3.898647823588E-01
 H   -4.261204188858E-01 -6.375760309897E+00 -1.204339570497E+00
 H    4.424704759539E-01 -5.158535016393E+00 -9.732004049514E-01
 H    7.386955570280E-02 -1.749927068431E+00 -3.316086442809E-01
 H   -5.753216936543E-02  1.906140867178E+00  3.600930021159E-01
 H   -4.261252272210E-01 -1.502059455135E+00 -2.835913825237E-01
 H    2.052889659862E-01  1.905553350019E+00  3.630260726598E-01
 H    7.388032249679E-02  3.123384972666E+00  5.919565958751E-01
 H   -2.947138866816E-01 -5.158154175642E+00 -9.749451363138E-01
```

## Z1.5

```
LATTICE PARAMETERS (ANGSTROMS AND DEGREES)
       A             B             C          ALPHA      BETA       GAMMA
   16.35142899   500.00000000   500.00000000    90.000000  90.000000  90.000000
*******************************************************************************
 ATOMS IN THE ASYMMETRIC UNIT   46 - ATOMS IN THE UNIT CELL:   46
    ATOM             X/A           Y(ANGSTROM)        Z(ANGSTROM)
*******************************************************************************
```

| | | | |
|---|---|---|---|
| C | -7.588331797034E-02 | -1.693963654703E+00 | 1.301173508778E+00 |
| C | -1.400757530615E-01 | -2.194492304413E+00 | 1.670410624436E+00 |
| C | -2.118418349880E-01 | -2.739825409022E+00 | 2.073153543314E+00 |
| C | -2.769135264859E-01 | -3.219743944576E+00 | 2.427236294929E+00 |
| C | -3.533648080872E-01 | -4.904532649100E+00 | 3.674967074809E+00 |
| C | -3.521742045948E-01 | -3.778646067269E+00 | 2.840828164458E+00 |
| C | -4.270209833359E-01 | -5.459765076838E+00 | 4.085319481558E+00 |
| C | -4.270570025783E-01 | -3.200654066469E+00 | 2.412458388890E+00 |
| C | 4.993030346758E-01 | -4.905691793665E+00 | 3.674280941373E+00 |
| C | 4.980760480162E-01 | -3.779504668374E+00 | 2.840599335050E+00 |
| H | 4.415141883786E-01 | -5.335134124980E+00 | 3.991914617876E+00 |
| C | 4.228030625814E-01 | -3.220723861238E+00 | 2.427444350948E+00 |
| C | 3.577739193531E-01 | -2.741119332423E+00 | 2.070859252888E+00 |
| C | 2.859648811424E-01 | 1.111056357088E+00 | -7.730642434166E-01 |
| C | 2.217932288423E-01 | 6.100692278728E-01 | -4.035160727842E-01 |
| C | 3.577180569229E-01 | 1.656956191177E+00 | -1.175545970825E+00 |
| C | -4.270812069571E-01 | -2.069171762603E+00 | 1.574342812464E+00 |
| C | 2.860183557783E-01 | -2.195478149176E+00 | 1.668004301106E+00 |
| C | -4.271148417420E-01 | -1.087303258248E+00 | 8.493240113345E-01 |
| C | 4.227937377978E-01 | 2.137206544213E+00 | -1.528937352270E+00 |
| C | -4.271077454502E-01 | 1.193512413598E-03 | 4.600852259454E-02 |
| C | 4.980683972892E-01 | 2.696850261691E+00 | -1.940850477198E+00 |
| C | -4.271089931399E-01 | 9.834109865640E-01 | -6.785378645577E-01 |
| C | 2.218933060030E-01 | -1.693678265368E+00 | 1.297342138707E+00 |
| C | 4.993038807782E-01 | 3.825386989603E+00 | -2.771347510661E+00 |
| C | -4.270721528329E-01 | 2.116985101940E+00 | -1.513835572728E+00 |
| C | 1.475599253187E-01 | 3.126877013578E-02 | 2.441928206912E-02 |
| C | -4.270148066535E-01 | 4.380949116768E+00 | -3.180099526243E+00 |
| C | 1.476093160920E-01 | -1.115212493657E+00 | 8.713114050473E-01 |
| C | -3.521835781960E-01 | 2.696255163229E+00 | -1.940192276637E+00 |
| C | -3.533635133806E-01 | 3.824757944267E+00 | -2.770784551522E+00 |
| C | 7.291334159046E-02 | 5.782894226976E-01 | -3.794229601971E-01 |
| C | -2.769313356935E-01 | 2.136113106056E+00 | -1.527889305987E+00 |
| C | -2.118738236051E-01 | 1.656490011856E+00 | -1.172750829780E+00 |
| C | -1.401215279909E-01 | 1.110543098709E+00 | -7.702760679141E-01 |
| C | 7.300594129861E-02 | -1.662948736065E+00 | 1.276454160631E+00 |
| C | -1.690378797442E-03 | 3.084362248206E-02 | 2.611593276921E-02 |
| C | -7.596477529902E-02 | 6.092188517744E-01 | -4.004754486681E-01 |
| C | -1.642069727354E-03 | -1.115571899147E+00 | 8.730910113214E-01 |
| H | 4.415185696692E-01 | 4.255689093280E+00 | -3.087983566361E+00 |
| H | 7.287544760975E-02 | 1.450266128471E+00 | -1.024570754706E+00 |
| H | -4.269907722819E-01 | 5.255680609764E+00 | -3.823985980205E+00 |
| H | -2.955559374881E-01 | 4.254199946893E+00 | -3.087487113344E+00 |
| H | 7.304433334054E-02 | -2.535546255026E+00 | 1.920757419223E+00 |
| H | -2.955607996497E-01 | -5.332796895551E+00 | 3.993460803683E+00 |
| H | -4.270052824913E-01 | -6.332555391036E+00 | 4.731826065746E+00 |

**Z2**

LATTICE PARAMETERS (ANGSTROMS AND DEGREES)

| A | B | C | ALPHA | BETA | GAMMA |
|---|---|---|---|---|---|
| 16.35211046 | 500.00000000 | 500.00000000 | 90.000000 | 90.000000 | 90.000000 |

*******************************************************************************

ATOMS IN THE ASYMMETRIC UNIT 64 - ATOMS IN THE UNIT CELL: 64

| ATOM | X/A | Y(ANGSTROM) | Z(ANGSTROM) |
|---|---|---|---|
| C | -1.805781448494E-03 | 6.176159581136E+00 | 7.296902219224E+00 |
| C | 7.274710565433E-02 | 6.610830317552E+00 | 7.822867409737E+00 |
| C | -2.420509817927E-03 | 5.272397485956E+00 | 6.198925520814E+00 |
| H | 7.275661400434E-02 | 7.300273411948E+00 | 8.659976338905E+00 |
| C | 1.472865039755E-01 | 6.177790036660E+00 | 7.295176565251E+00 |
| C | 2.216014051218E-01 | 6.635043800467E+00 | 7.848635984577E+00 |
| C | 7.272318705174E-02 | 4.824272465493E+00 | 5.652972643336E+00 |
| C | 2.857327847087E-01 | 7.031077733265E+00 | 8.330100765180E+00 |
| C | 1.478756132708E-01 | 5.273392490451E+00 | 6.197717729973E+00 |
| C | 3.575081218705E-01 | 7.459187940227E+00 | 8.855115022460E+00 |
| C | 7.271599808641E-02 | 3.928944166995E+00 | 4.565466998697E+00 |
| C | 4.990663290516E-01 | 9.164209549521E+00 | 1.093797634509E+01 |
| C | 4.226073415830E-01 | 7.835456575180E+00 | 9.316661439996E+00 |
| C | 7.268345342374E-02 | 3.153874779802E+00 | 3.622945027670E+00 |
| C | 2.222481987685E-01 | 4.825564581559E+00 | 5.653902710402E+00 |
| C | -4.272734742712E-01 | 9.601247932418E+00 | 1.147220234241E+01 |
| C | 4.978693232807E-01 | 8.274681191250E+00 | 9.855327766815E+00 |
| H | 4.412754834800E-01 | 9.504580739325E+00 | 1.134937450631E+01 |
| C | 7.265826708552E-02 | 2.295715914400E+00 | 2.576915419519E+00 |
| C | 2.869750033947E-01 | 4.438120666450E+00 | 5.183874872325E+00 |
| C | 7.264754103096E-02 | 1.521675773499E+00 | 1.633290315570E+00 |
| H | -4.272832071644E-01 | 1.029109277967E+01 | 1.231114458321E+01 |
| C | -3.536007357624E-01 | 9.162397627633E+00 | 1.093993418489E+01 |
| C | -4.272480078313E-01 | 7.816533301264E+00 | 9.300697987611E+00 |
| C | -2.168613965340E-03 | 1.739838558553E-01 | -1.382840663125E-02 |
| C | -3.523797923698E-01 | 8.272757715753E+00 | 9.857391964307E+00 |
| C | 7.268486547239E-02 | 6.302037943836E-01 | 5.434031360267E-01 |
| C | -4.272350670818E-01 | 6.922238403396E+00 | 8.213092054235E+00 |
| C | 3.584940416047E-01 | 4.006772984327E+00 | 4.661532615467E+00 |
| C | -9.139764833374E-04 | -7.128004615558E-01 | -1.098725134182E+00 |
| C | -4.272354707463E-01 | 6.146555658688E+00 | 7.270831530571E+00 |
| C | 4.232102228527E-01 | 3.619005429232E+00 | 4.191397950600E+00 |
| C | -4.272457457975E-01 | 5.287359936704E+00 | 6.225755928184E+00 |
| C | 1.475823204512E-01 | 1.753629722089E-01 | -1.335209127113E-02 |
| C | 7.277454515209E-02 | -1.148598033793E+00 | -1.632930586681E+00 |
| H | -5.868479580687E-02 | -1.051508176683E+00 | -1.512281752455E+00 |
| C | 2.228267536426E-01 | 6.151905530859E-01 | 5.254302963742E-01 |
| C | -4.272769460259E-01 | 4.512574456241E+00 | 5.283002374970E+00 |
| C | 2.879441942888E-01 | 9.911001056074E-01 | 9.865700499793E-01 |
| C | 1.464205797245E-01 | -7.114446207969E-01 | -1.098232968749E+00 |
| C | 4.975900448831E-01 | 3.170975598381E+00 | 3.648060636269E+00 |
| C | 3.597282230454E-01 | 1.418247546347E+00 | 1.512233737345E+00 |
| C | -4.272647566889E-01 | 3.618758326607E+00 | 4.194226287295E+00 |
| C | 4.238848910780E-01 | 1.813183718702E+00 | 1.993679756249E+00 |
| C | 4.981901644927E-01 | 2.268508891158E+00 | 2.548996640061E+00 |
| C | -3.521171386627E-01 | 3.170425693024E+00 | 3.648597312074E+00 |
| C | -4.272640064579E-01 | 1.835142894132E+00 | 2.021746861210E+00 |
| H | 7.281107735120E-02 | -1.836398819018E+00 | -2.473531694490E+00 |
| C | -3.527205751492E-01 | 2.268216649829E+00 | 2.549326086305E+00 |
| H | 2.042235896829E-01 | -1.049010407695E+00 | -1.511414692769E+00 |
| H | -4.272611062447E-01 | 1.146736812697E+00 | 1.183752029090E+00 |
| C | -2.784172910837E-01 | 1.813037281185E+00 | 1.993801392363E+00 |
| C | -2.143421981214E-01 | 1.417481577404E+00 | 1.510049005808E+00 |
| C | -2.777279080391E-01 | 3.617084908498E+00 | 4.192952627314E+00 |
| C | -1.425580952334E-01 | 9.891875680931E-01 | 9.854261941305E-01 |
| C | -2.130209331097E-01 | 4.004143688870E+00 | 4.664109738030E+00 |
| C | -1.415309138239E-01 | 4.434781543027E+00 | 5.187699777846E+00 |
| C | -7.679080868375E-02 | 4.824292534129E+00 | 5.655197604648E+00 |
| C | -7.745476900699E-02 | 6.127489522145E-01 | 5.241918708926E-01 |
| C | -7.610170433983E-02 | 6.631487339595E+00 | 7.852743169744E+00 |
| C | -1.401968461030E-01 | 7.026040618435E+00 | 8.336607287461E+00 |
| C | -2.119747203284E-01 | 7.454800346375E+00 | 8.860824491435E+00 |
| C | -2.771053590861E-01 | 7.831680546101E+00 | 9.320670962380E+00 |
| H | -2.958185338300E-01 | 9.501150777138E+00 | 1.135300125859E+01 |

**Z2.5**

LATTICE PARAMETERS (ANGSTROMS AND DEGREES)

| A | B | C | ALPHA | BETA | GAMMA |
|---|---|---|---|---|---|
| 16.35124189 | 500.00000000 | 500.00000000 | 90.000000 | 90.000000 | 90.000000 |

*******************************************************************************

ATOMS IN THE ASYMMETRIC UNIT 82 - ATOMS IN THE UNIT CELL: 82

| ATOM | X/A | Y(ANGSTROM) | Z(ANGSTROM) |
|---|---|---|---|
| C | -1.406594096661E-01 | -2.653532421269E+00 | -6.250926789379E-01 |
| C | -2.121953984745E-01 | -3.311928746303E+00 | -7.807424447385E-01 |
| C | -7.584767218769E-02 | -2.062858151468E+00 | -4.863296674642E-01 |
| C | -2.768060835083E-01 | -3.908138296604E+00 | -9.190092009416E-01 |
| C | -3.512398696540E-01 | -4.590903459121E+00 | -1.088555330613E+00 |
| C | -4.263007241750E-01 | -2.532815305427E+00 | -6.001455921397E-01 |
| C | -2.768098627780E-01 | -6.656270994625E+00 | -1.572878988941E+00 |
| C | -3.512451934828E-01 | -5.973620964269E+00 | -1.402586381811E+00 |
| C | -4.263056750444E-01 | -3.903315033119E+00 | -9.141906668148E-01 |
| C | -4.263125531531E-01 | -1.344812205366E+00 | -3.195594412641E-01 |
| C | -2.121863441670E-01 | -7.251915538625E+00 | -1.712125008975E+00 |
| C | -1.406457007711E-01 | -7.909466759936E+00 | -1.869240138001E+00 |
| C | -4.263130098365E-01 | -2.824765867512E-02 | -1.238160610377E-02 |
| C | -4.263145714331E-01 | -6.661098381007E+00 | -1.576714593533E+00 |
| C | 4.986244097699E-01 | -4.590843792880E+00 | -1.088118422092E+00 |
| C | -7.586976048431E-02 | -8.500927921275E+00 | -2.009071217289E+00 |
| C | -4.263225329916E-01 | 1.160158324134E+00 | 2.664964919457E-01 |
| C | 4.986196766463E-01 | -5.973585882889E+00 | -1.402111196152E+00 |
| C | -4.263313954746E-01 | -8.031375192833E+00 | -1.891691800323E+00 |
| C | 4.241810046098E-01 | -3.908339153723E+00 | -9.182724871282E-01 |
| C | -4.263136218740E-01 | 2.530169975866E+00 | 5.899554967406E-01 |
| C | -8.701480561372E-04 | -1.057008302036E+01 | -2.496848660278E+00 |
| C | -4.263421436336E-01 | -9.218905803400E+00 | -2.174297330132E+00 |
| C | 3.595482289439E-01 | -3.312880197022E+00 | -7.795351990801E-01 |
| C | 7.368208508876E-02 | -1.123439046639E+01 | -2.653114716697E+00 |
| C | -1.465901019209E-03 | -9.185989333033E+00 | -2.169709329414E+00 |
| C | 4.987983506599E-01 | 3.230417053505E+00 | 7.556760232116E-01 |
| C | -4.262863254393E-01 | -1.582303963899E+01 | -3.755526986718E+00 |
| C | -3.514116743572E-01 | 3.230207123115E+00 | 7.552229372673E-01 |
| C | -4.999791079956E-01 | -1.515284448962E+01 | -3.593555418250E+00 |
| C | 4.241819852296E-01 | -6.656192499475E+00 | -1.571779907264E+00 |
| C | -8.700699073068E-04 | 6.954807067446E-03 | -9.902402040718E-04 |
| C | -3.526226451852E-01 | -1.515186324905E+01 | -3.594007708028E+00 |
| C | -1.445241056181E-03 | -1.377565705008E+00 | -3.262663844035E-01 |
| C | 7.367045738489E-02 | 6.717009385134E-01 | 1.550617816041E-01 |
| C | 4.987751659691E-01 | -1.379073374079E+01 | -3.265165955111E+00 |
| C | -4.263638363111E-01 | -1.053473976309E+01 | -2.484672069041E+00 |
| C | 2.880132804581E-01 | -2.654695539464E+00 | -6.233561335447E-01 |
| C | 1.482399582719E-01 | -1.057033826292E+01 | -2.496153605613E+00 |
| C | 7.372141688062E-02 | -2.064755457365E+00 | -4.870560665376E-01 |
| C | -3.514342906776E-01 | -1.378968739192E+01 | -3.265858839968E+00 |
| C | 7.368857700769E-02 | -8.498901576223E+00 | -2.006886903702E+00 |
| C | -4.263444290855E-01 | -1.309081388493E+01 | -3.096896703111E+00 |
| C | -4.263766340935E-01 | -1.172213255388E+01 | -2.767860484236E+00 |
| C | 1.482394831203E-01 | 7.647384394468E-03 | -2.105164192085E-04 |
| C | 1.488625211689E-01 | -1.376825236932E+00 | -3.254195450674E-01 |
| C | 2.232720271820E-01 | -2.062000468553E+00 | -4.846810549123E-01 |
| C | 1.488416381353E-01 | -9.186295736301E+00 | -2.169101585819E+00 |
| C | -4.999721741602E-01 | 4.594020007139E+00 | 1.077779804342E+00 |
| C | 3.595496326253E-01 | -7.251499114871E+00 | -1.710955952205E+00 |
| C | -3.526160045473E-01 | 4.593815740928E+00 | 1.077356973168E+00 |
| C | 2.232426091823E-01 | -8.501345414603E+00 | -2.007967660270E+00 |
| C | 2.880005751848E-01 | -7.909391844316E+00 | -1.868225788560E+00 |
| C | -4.262886117988E-01 | 5.265405552824E+00 | 1.235943057249E+00 |
| H | 7.368034821249E-02 | -1.228974621766E+01 | -2.902714704194E+00 |
| H | -4.262631973262E-01 | -1.687877802073E+01 | -4.010499169258E+00 |
| H | 4.422489972163E-01 | -1.567260744146E+01 | -3.718802369062E+00 |
| H | -2.948280400250E-01 | -1.567090853976E+01 | -3.719396396489E+00 |
| H | 7.364480490834E-02 | 1.727482546062E+00 | 4.028644777405E-01 |
| H | 4.422490745014E-01 | 5.114021134651E+00 | 1.201162945840E+00 |
| H | -2.948280968049E-01 | 5.113626770497E+00 | 1.200404676651E+00 |
| H | -4.262805688018E-01 | 6.322379401394E+00 | 1.485754402263E+00 |
| C | -2.761867551230E-01 | -1.311331973597E+01 | -3.103710825839E+00 |
| C | -2.111398536111E-01 | -1.253251890246E+01 | -2.965071231436E+00 |
| C | -1.393842407872E-01 | -1.187304306265E+01 | -2.808060317814E+00 |
| C | -7.518399375112E-02 | -1.126863099997E+01 | -2.663004690368E+00 |
| C | -7.519092594172E-02 | 7.057894921615E-01 | 1.627155668057E-01 |
| C | -1.393345852538E-01 | 1.312511039959E+00 | 3.049348138431E-01 |
| C | -2.111063369414E-01 | 1.972167359862E+00 | 4.595896053363E-01 |
| C | -2.761571549181E-01 | 2.553383186796E+00 | 5.959354889660E-01 |
| C | 4.235029438338E-01 | -1.311520961970E+01 | -3.102721535192E+00 |
| C | 3.584638679625E-01 | -1.253429763600E+01 | -2.963524602269E+00 |
| C | 2.866950211242E-01 | -1.187520913574E+01 | -2.806128402679E+00 |
| C | 2.225495176922E-01 | -1.126908532247E+01 | -2.661676712052E+00 |
| C | 2.225378804990E-01 | 7.070826594902E-01 | 1.639747602834E-01 |
| C | 2.866893961115E-01 | 1.313556013319E+00 | 3.061509508278E-01 |
| C | 3.584632777214E-01 | 1.973054288134E+00 | 4.607251458438E-01 |
| C | 4.235292138197E-01 | 2.553904897287E+00 | 5.968326352058E-01 |
| C | 7.370305320820E-02 | -7.127922659006E+00 | -1.682303349199E+00 |
| C | 7.369290042685E-02 | -5.940554210291E+00 | -1.400922352534E+00 |
| C | 7.370215355265E-02 | -4.623804636315E+00 | -1.089915066270E+00 |
| C | 7.377983145508E-02 | -3.435943308976E+00 | -8.106687026300E-01 |

## Z3

LATTICE PARAMETERS (ANGSTROMS AND DEGREES)

| A | B | C | ALPHA | BETA | GAMMA |
|---|---|---|---|---|---|
| 16.35137871 | 500.00000000 | 500.00000000 | 90.000000 | 90.000000 | 90.000000 |

*******************************************************************************

ATOMS IN THE ASYMMETRIC UNIT 100 - ATOMS IN THE UNIT CELL: 100

| ATOM | X/A | Y(ANGSTROM) | Z(ANGSTROM) |
|---|---|---|---|
| C | -7.707384229787E-02 | 3.732038435583E+00 | 9.666559276206E+00 |
| C | -1.417123159904E-01 | 3.962874113423E+00 | 1.023263152980E+01 |
| C | -2.132710080900E-01 | 4.216745395777E+00 | 1.085896163793E+01 |
| C | -3.524537395790E-01 | 5.238179124834E+00 | 1.339182357162E+01 |
| C | -2.779304577191E-01 | 4.445142612869E+00 | 1.142534915762E+01 |
| C | -7.697059868189E-02 | 2.674374901245E+00 | 7.047396575729E+00 |
| C | -4.275541722775E-01 | 5.514955642870E+00 | 1.404397147949E+01 |
| C | -3.524018067016E-01 | 4.717693620079E+00 | 1.207271672413E+01 |
| C | -1.417478555454E-01 | 2.441685209046E+00 | 6.486394238571E+00 |
| C | 4.973932866803E-01 | 5.238321149640E+00 | 1.339021473230E+01 |
| C | -4.274565547356E-01 | 4.441038740908E+00 | 1.141905753628E+01 |
| C | 4.974405489256E-01 | 4.717659168100E+00 | 1.207126895740E+01 |
| C | -4.274177944738E-01 | 3.920994887094E+00 | 1.011260166940E+01 |
| C | -2.132884611315E-01 | 2.187229547080E+00 | 5.859574352034E+00 |
| C | -4.274943930727E-01 | 3.460387267654E+00 | 8.982172688165E+00 |
| C | -2.779751691441E-01 | 1.956782739721E+00 | 5.294393186441E+00 |
| C | -4.275279371359E-01 | 2.952529800348E+00 | 7.728936835765E+00 |
| C | -7.768562092173E-02 | 1.561436527906E-01 | 9.071582591238E-01 |
| C | -4.275729028573E-01 | 2.491873400606E+00 | 6.598758827602E+00 |
| C | -1.427316936495E-01 | 3.827339662032E-01 | 1.459568466642E+00 |
| C | -3.523568733231E-01 | 1.689468231973E+00 | 4.642923478523E+00 |
| C | -2.144905554891E-01 | 6.399050061704E-01 | 2.086608372805E+00 |
| C | -4.275414907611E-01 | 1.958087930633E+00 | 5.295436082360E+00 |
| C | -2.786238696040E-01 | 8.765661796840E-01 | 2.663323708354E+00 |
| C | -3.529352914853E-01 | 1.149460693445E+00 | 3.327340266243E+00 |
| C | 4.973060896461E-01 | 1.690068362101E+00 | 4.641677801563E+00 |
| C | -4.274796604044E-01 | 8.906328455263E-01 | 2.695579643863E+00 |
| C | 4.979449350550E-01 | 1.150055154956E+00 | 3.326104057619E+00 |
| H | -4.274529253456E-01 | 4.790827509522E-01 | 1.692259983732E+00 |
| C | 4.236590037743E-01 | 8.772866354889E-01 | 2.661259560450E+00 |
| C | 3.594835345771E-01 | 6.405838238176E-01 | 2.085801602376E+00 |
| C | 7.247711613063E-02 | 1.162224449705E+00 | 3.362611988347E+00 |
| C | 7.243514615978E-02 | 6.986971477187E-01 | 2.233311946032E+00 |
| C | 7.249140081503E-02 | 1.676832871475E+00 | 4.612706955895E+00 |
| C | 4.228997536783E-01 | 1.957522047571E+00 | 5.292336213108E+00 |
| C | 2.877116098360E-01 | 3.834781419108E-01 | 1.459252809395E+00 |
| C | 3.581977069262E-01 | 2.188451183427E+00 | 5.856761373883E+00 |
| C | 7.255403395449E-02 | 2.137882550561E+00 | 5.743016580010E+00 |
| C | 2.866646827894E-01 | 2.442994713702E+00 | 6.483933686850E+00 |
| C | 7.252278181998E-02 | 2.676317309706E+00 | 7.041819109220E+00 |
| C | 2.220274969852E-01 | 2.675250868678E+00 | 7.049428685218E+00 |
| C | 2.226141674434E-01 | 1.569959162683E-01 | 9.083961293654E-01 |
| C | -2.568366680823E-03 | 2.931993727454E+00 | 7.702825044063E+00 |
| C | 1.475804599890E-01 | 2.932519825338E+00 | 7.703574129009E+00 |
| C | 7.246069700048E-02 | 1.653335621441E-01 | 9.305152993105E-01 |
| C | -2.604747915140E-03 | 3.476706910717E+00 | 9.012104135948E+00 |
| C | 1.473665570355E-01 | -1.068237876628E-01 | 2.649400574204E-01 |
| C | 1.475491276543E-01 | 3.477022126375E+00 | 9.012944951901E+00 |
| C | 7.245858482639E-02 | 3.733306354432E+00 | 9.673626024466E+00 |
| C | -2.416493277696E-03 | -1.072749361381E-01 | 2.642748923303E-01 |
| C | 2.219970168392E-01 | 3.733016460417E+00 | 9.667824963038E+00 |
| C | 2.867154430615E-01 | 3.962830421502E+00 | 1.023185280726E+01 |
| C | 3.583080271117E-01 | 4.216626881599E+00 | 1.085741255201E+01 |
| C | 1.461810025607E-01 | -6.367834893650E-01 | -1.032101422108E+00 |
| C | -1.175362761952E-03 | -6.376239078445E-01 | -1.032593999632E+00 |
| C | 4.230088677843E-01 | 4.445126374605E+00 | 1.142256021368E+01 |
| C | 7.251685819004E-02 | -8.983092000859E-01 | -1.670942692502E+00 |
| H | 2.039824882339E-01 | -8.377340346000E-01 | -1.526627183724E+00 |
| H | -5.895155907594E-02 | -8.393543434319E-01 | -1.527489304933E+00 |
| H | 7.254113581083E-02 | -1.308246175530E+00 | -2.676588325438E+00 |
| C | -2.626705313269E-03 | 6.481120378973E+00 | 1.644996477951E+01 |
| C | -2.128464100604E-03 | 7.008454899939E+00 | 1.777070972937E+01 |
| C | 7.243378770475E-02 | 4.270343150162E+00 | 1.097315465682E+01 |
| C | 7.250174501969E-02 | 4.727859066328E+00 | 1.210488678663E+01 |
| C | 7.256046628084E-02 | 5.236747616337E+00 | 1.335750754185E+01 |
| C | 7.267378999649E-02 | 5.693929866082E+00 | 1.448911219222E+01 |
| C | 7.259098921793E-02 | 6.219219847394E+00 | 1.579590165467E+01 |
| C | -7.647641809714E-02 | 7.273852615983E+00 | 1.843664763949E+01 |
| C | 7.238245694014E-02 | 7.260883014279E+00 | 1.840616172285E+01 |
| C | 1.477031154897E-01 | 6.480971111687E+00 | 1.645317215746E+01 |
| C | 1.469897427570E-01 | 7.008369948947E+00 | 1.777387785931E+01 |
| C | -1.406021046819E-01 | 7.503758013682E+00 | 1.901628981429E+01 |
| C | -7.695534487997E-02 | 6.219844655035E+00 | 1.579453826728E+01 |
| C | 2.221493916402E-01 | 6.220071398493E+00 | 1.580116832931E+01 |
| C | 2.212520079527E-01 | 7.274166333268E+00 | 1.844219183989E+01 |
| C | -2.123722914539E-01 | 7.754040724365E+00 | 1.964607006883E+01 |
| C | 2.854413011228E-01 | 7.504514699046E+00 | 1.901975793589E+01 |
| C | -2.774163336545E-01 | 7.974088666996E+00 | 2.020115634837E+01 |
| C | 2.868233286080E-01 | 5.992412238497E+00 | 1.523449613210E+01 |
| C | -3.538230874344E-01 | 8.752005118892E+00 | 2.214840596701E+01 |
| C | -1.418351924713E-01 | 5.990904420878E+00 | 1.523466129869E+01 |
| C | -3.526525907165E-01 | 8.231554321487E+00 | 2.084748068956E+01 |
| C | -4.274756980984E-01 | 9.007480723601E+00 | 2.278997642718E+01 |
| C | 3.572227545459E-01 | 7.754763977422E+00 | 1.964916011076E+01 |
| C | 4.988244218048E-01 | 8.751219467286E+00 | 2.215010502382E+01 |
| C | -4.275672853510E-01 | 7.964514640080E+00 | 2.018014057720E+01 |
| C | 4.222915217866E-01 | 7.974987032709E+00 | 2.020342988549E+01 |
| C | 4.975656614519E-01 | 8.231686163454E+00 | 2.084886025153E+01 |
| C | 3.583371379150E-01 | 5.737803249526E+00 | 1.460688493379E+01 |
| C | -4.276194270422E-01 | 7.441968818660E+00 | 1.887300243381E+01 |
| C | -4.276229018645E-01 | 6.989168135165E+00 | 1.773937641292E+01 |
| C | 4.229467490024E-01 | 5.510175987966E+00 | 1.403863491256E+01 |
| C | -2.133777337590E-01 | 5.736393225192E+00 | 1.460831771419E+01 |
| C | -4.276103936807E-01 | 6.488402256451E+00 | 1.648366116872E+01 |
| C | -4.276061605062E-01 | 6.032580719954E+00 | 1.535122183847E+01 |
| C | -2.780571908326E-01 | 5.509223330827E+00 | 1.404209410016E+01 |
| H | 7.229864230049E-02 | 7.662035938095E+00 | 1.941360309024E+01 |
| H | -2.960202011924E-01 | 8.951329207628E+00 | 2.264348192503E+01 |
| H | -4.274389715285E-01 | 9.410467445407E+00 | 2.379852716907E+01 |
| H | 4.410554439560E-01 | 8.948956612558E+00 | 2.264687673406E+01 |

## Z3.5

LATTICE PARAMETERS (ANGSTROMS AND DEGREES)

| A | B | C | ALPHA | BETA | GAMMA |
|---|---|---|---|---|---|
| 16.35111230 | 500.00000000 | 500.00000000 | 90.000000 | 90.000000 | 90.000000 |

*******************************************************************************

ATOMS IN THE ASYMMETRIC UNIT 118 - ATOMS IN THE UNIT CELL: 118

| ATOM | X/A | Y(ANGSTROM) | Z(ANGSTROM) |
|---|---|---|---|
| C | 2.858718824578E-01 | -2.880638654097E+00 | -1.505331897179E-01 |
| C | 3.574120496751E-01 | -3.555929383213E+00 | -1.860931557746E-01 |
| C | 2.211259999488E-01 | -2.272879601762E+00 | -1.167915874239E-01 |
| C | 4.220716691177E-01 | -4.165639227693E+00 | -2.213734200000E-01 |
| C | 4.964974150077E-01 | -4.868554101925E+00 | -2.511427224690E-01 |
| C | 1.467301487117E-01 | -1.570224254565E+00 | -7.754067289898E-02 |
| C | -4.283987959287E-01 | -2.756725845481E+00 | -1.396846632426E-01 |
| C | 4.220150353293E-01 | -6.985683753861E+00 | -3.702719229515E-01 |
| C | 4.964763188263E-01 | -6.283728486558E+00 | -3.396104191204E-01 |
| C | -4.284156035946E-01 | -4.160035631205E+00 | -2.241842978202E-01 |
| C | -4.284103709406E-01 | -1.537721106583E+00 | -7.380075104261E-02 |
| C | 3.574057430556E-01 | -7.596923231221E+00 | -4.058200188895E-01 |
| C | 2.858334132993E-01 | -8.271749284542E+00 | -4.403803371392E-01 |
| C | -4.284182177251E-01 | -1.879145786701E-01 | 1.300889696312E-03 |
| C | 1.461293990880E-01 | -1.501765747763E-01 | -6.015591693228E-04 |
| C | -4.284445635841E-01 | -6.992631679309E+00 | -3.662107986599E-01 |
| C | -3.533439719334E-01 | -4.869041602034E+00 | -2.512075928535E-01 |
| C | 7.156200082508E-02 | -2.275227785261E+00 | -1.152906369842E-01 |
| C | 2.211472300627E-01 | -8.880818885178E+00 | -4.758462961630E-01 |
| C | -4.284409640421E-01 | 1.031094548824E+00 | 6.587816368829E-02 |
| C | -3.533493529880E-01 | -6.284224765468E+00 | -3.396877559989E-01 |
| C | -4.284645522038E-01 | -8.395755111194E+00 | -4.497182254660E-01 |
| C | -2.789071247878E-01 | -4.166473009678E+00 | -2.213554507057E-01 |
| C | 1.466707440880E-01 | -9.582382633857E+00 | -5.073152361280E-01 |
| C | -4.284374715952E-01 | 2.436899821655E+00 | 1.391002426458E-01 |
| C | -4.283928166460E-01 | -9.615030323533E+00 | -5.139732859689E-01 |
| C | -2.142715647252E-01 | -3.556094784844E+00 | -1.863042595288E-01 |
| C | 1.466367045555E-01 | -1.099785081192E+01 | -5.953828368843E-01 |
| C | 7.160743305540E-02 | -8.873264395696E+00 | -4.789989628722E-01 |
| C | -3.535430560549E-01 | 3.155350834242E+00 | 1.761642063154E-01 |
| C | 4.966711904346E-01 | 3.155438112610E+00 | 1.763220456403E-01 |
| C | -2.789034547769E-01 | -6.986627043387E+00 | -3.702106666748E-01 |
| C | -4.283708966812E-01 | -1.096411719025E+01 | -5.893265034378E-01 |
| C | -1.427398501950E-01 | -2.880467616501E+00 | -1.507369889430E-01 |
| C | -4.283671899626E-01 | -1.218341009233E+01 | -6.525880643742E-01 |
| C | -7.800276818052E-02 | -2.272494871517E+00 | -1.170147294905E-01 |
| C | 7.153711212689E-02 | -1.170592679453E+01 | -6.235038517436E-01 |
| C | -3.493678591738E-03 | -9.581357236929E+00 | -5.074102933083E-01 |
| C | -3.547538384520E-01 | 4.554617539513E+00 | 2.487176733783E-01 |
| C | -2.142645556473E-01 | -7.597008417053E+00 | -4.061920435314E-01 |
| C | 4.978891806871E-01 | 4.554701679425E+00 | 2.488043163623E-01 |
| C | -7.794130270460E-02 | -8.878947671014E+00 | -4.761467675903E-01 |
| C | -3.526992628153E-03 | -1.099681678787E+01 | -5.955246574636E-01 |
| C | -1.426790248935E-01 | -8.271329074029E+00 | -4.406602518632E-01 |
| C | -4.284305407492E-01 | 5.243645480215E+00 | 2.846912451375E-01 |
| C | 7.157131540090E-02 | 5.311801608879E-01 | 3.668418168834E-02 |
| C | -3.601559635311E-03 | -1.570023738521E+00 | -7.759446006836E-02 |
| C | -2.992531301280E-03 | -1.499819767228E-01 | -6.468501523611E-04 |
| H | -2.969639905665E-01 | 5.088039287505E+00 | 2.761910955757E-01 |
| H | 4.401023190649E-01 | 5.088210630182E+00 | 2.762423215998E-01 |
| H | -4.284277979668E-01 | 6.328256260944E+00 | 3.412253753372E-01 |
| H | 7.157751275949E-02 | 1.614034198162E+00 | 9.585814087794E-02 |
| C | 7.164552227743E-02 | -7.469615105685E+00 | -3.950546774289E-01 |
| C | 7.160487524130E-02 | -6.250563002369E+00 | -3.301317085871E-01 |
| C | 7.158204677359E-02 | -4.900473469876E+00 | -2.555153973275E-01 |
| C | 7.155250511072E-02 | -3.681717468909E+00 | -1.899693200566E-01 |
| C | 2.204316590469E-01 | 5.669497809188E-01 | 3.820818169688E-02 |
| C | 2.845925449928E-01 | 1.188712528459E+00 | 7.173072168947E-02 |
| C | 3.563585177586E-01 | 1.865245655177E+00 | 1.079170833288E-01 |
| C | 4.214092584265E-01 | 2.461383950951E+00 | 1.396616919521E-01 |
| C | -7.729583145941E-02 | 5.671035165177E-01 | 3.818066101761E-02 |
| C | -1.414423262734E-01 | 1.189249420136E+00 | 7.168218104706E-02 |
| C | -2.132162862068E-01 | 1.865601495749E+00 | 1.077947113482E-01 |
| C | -2.782848135863E-01 | 2.461219311676E+00 | 1.395025745484E-01 |
| C | 4.964818995729E-01 | -1.571030499607E+01 | -8.458405481071E-01 |
| C | -4.283995051829E-01 | -1.358666731636E+01 | -7.343013978162E-01 |
| C | -3.533602538107E-01 | -1.571110826669E+01 | -8.456791846575E-01 |
| C | -4.284794055317E-01 | -1.782302001849E+01 | -9.529618125835E-01 |
| C | 4.220587655885E-01 | -1.359236819055E+01 | -7.305342299616E-01 |
| C | -2.788560557003E-01 | -1.359393892713E+01 | -7.302414505180E-01 |
| C | -4.284393495441E-01 | -2.039214108164E+01 | -1.086873110380E+00 |
| C | 2.858057045523E-01 | -1.230837278876E+01 | -6.615725447437E-01 |
| C | 2.210814275073E-01 | -1.170035677688E+01 | -6.264753641225E-01 |
| C | -1.427123497142E-01 | -1.230689557640E+01 | -6.618188062642E-01 |
| C | -4.284353998873E-01 | -2.161129392401E+01 | -1.148607029693E+00 |
| C | -7.800309866282E-02 | -1.169840305438E+01 | -6.267473373089E-01 |
| C | -4.284437416990E-01 | -1.904214429452E+01 | -1.015421913365E+00 |
| C | -4.284301281681E-01 | -2.301728053728E+01 | -1.218237639042E+00 |
| C | -4.284236634936E-01 | -2.582440990769E+01 | -1.356355885333E+00 |
| C | 4.966788938868E-01 | -2.373592623862E+01 | -1.253530389269E+00 |
| C | -3.535354646058E-01 | -2.373582332101E+01 | -1.253454780148E+00 |
| C | 4.978962503042E-01 | -2.513538062092E+01 | -1.322318960407E+00 |
| C | -3.547470655250E-01 | -2.513527442411E+01 | -1.322349051527E+00 |
| C | 3.573948209090E-01 | -1.298267169979E+01 | -6.954229027740E-01 |
| C | 7.150083202009E-02 | -1.310964758897E+01 | -7.062185590572E-01 |
| C | -2.142793114512E-01 | -1.298176214791E+01 | -6.956438535939E-01 |
| C | -2.782762866903E-01 | -2.304162286021E+01 | -1.218515961429E+00 |
| C | 4.214178532176E-01 | -2.304179543074E+01 | -1.218652436737E+00 |
| C | 7.157108932250E-02 | -1.432880446912E+01 | -7.689641347769E-01 |
| C | 3.563547401964E-01 | -2.244594337169E+01 | -1.188587952541E+00 |
| C | -2.132255334924E-01 | -2.244540275905E+01 | -1.188384787381E+00 |
| C | 7.158651991409E-02 | -1.567907972668E+01 | -8.410713108937E-01 |
| C | 2.845836112462E-01 | -2.176943004517E+01 | -1.154184840912E+00 |
| C | 3.573959741447E-01 | -1.702297595934E+01 | -9.079015346095E-01 |
| C | -1.414564959518E-01 | -2.176885974164E+01 | -1.154011590304E+00 |
| C | -2.142747508725E-01 | -1.702389523192E+01 | -9.078115047134E-01 |
| C | 7.159611534242E-02 | -1.689795797663E+01 | -9.039282211675E-01 |
| C | 2.204319196841E-01 | -2.114733036455E+01 | -1.122167671535E+00 |
| C | -7.729743445366E-02 | -2.114699088715E+01 | -1.122167719779E+00 |
| C | 1.467427315384E-01 | -1.900990149528E+01 | -1.011707101932E+00 |
| C | 2.211414092042E-01 | -1.830729099438E+01 | -9.740550385471E-01 |
| C | 2.858668997072E-01 | -1.769886726296E+01 | -9.418059835288E-01 |
| C | 7.158226634858E-02 | -1.830459356840E+01 | -9.756208119971E-01 |
| C | 1.461309810780E-01 | -2.043010782664E+01 | -1.085140311632E+00 |
| C | 7.156666084855E-02 | -2.111135815019E+01 | -1.120729442513E+00 |
| C | -3.588239618094E-03 | -1.900964597480E+01 | -1.011744714091E+00 |
| C | -1.427335255525E-01 | -1.769921258396E+01 | -9.417526110384E-01 |
| C | -2.991076557121E-03 | -2.042988996906E+01 | -1.085158186427E+00 |
| C | -7.798354772885E-02 | -1.830690634448E+01 | -9.740876752880E-01 |
| C | 4.220404501259E-01 | -1.641276862114E+01 | -8.742907172269E-01 |
| C | -2.789426159137E-01 | -1.641433268260E+01 | -8.740026837328E-01 |
| C | 4.964998117231E-01 | -1.429496137210E+01 | -7.598798426211E-01 |

C -4.284525375756E-01 -1.641950632558E+01 -8.715125550876E-01
C -3.533254028209E-01 -1.429576209788E+01 -7.597503236823E-01
H -4.284208851778E-01 -2.690916048401E+01 -1.410007007371E+00
H 4.401089711260E-01 -2.566894276466E+01 -1.348464679748E+00
H -2.969572404646E-01 -2.566875701823E+01 -1.348635212266E+00
H 7.156067673996E-02 -2.219437109370E+01 -1.176909100476E+00

**Z4**

LATTICE PARAMETERS (ANGSTROMS AND DEGREES) - BOHR = 0.5291772083 ANGSTROM

PRIMITIVE CELL

| A | B | C | ALPHA | BETA | GAMMA |
|---|---|---|---|---|---|
| 16.35115399 | 500.00000000 | 500.00000000 | 90.000000 | 90.000000 | 90.000000 |

*******************************************************************************

ATOMS IN THE ASYMMETRIC UNIT 136 - ATOMS IN THE UNIT CELL: 136

| ATOM | X/A | Y(ANGSTROM) | Z(ANGSTROM) |
|---|---|---|---|
| C | -6.919558506351E-02 | 5.861620103489E+00 | 8.948618379242E+00 |
| C | -1.338924144952E-01 | 6.192314027576E+00 | 9.460840450717E+00 |
| C | -2.054728124441E-01 | 6.554777756193E+00 | 1.003098293856E+01 |
| C | -3.445252809338E-01 | 8.033950848225E+00 | 1.232867314097E+01 |
| C | -2.701227159783E-01 | 6.886581634329E+00 | 1.054397803430E+01 |
| C | -6.906729141135E-02 | 4.341238363593E+00 | 6.568796317600E+00 |
| C | -4.196087339978E-01 | 8.407612301151E+00 | 1.293203898721E+01 |
| C | -3.445582780126E-01 | 7.260811531602E+00 | 1.114000917722E+01 |
| C | -1.338242900967E-01 | 4.008711077742E+00 | 6.059829256773E+00 |
| C | -4.947232136020E-01 | 8.036042254777E+00 | 1.232860220357E+01 |
| C | -4.196680506185E-01 | 6.889608919272E+00 | 1.053633986671E+01 |
| C | -4.947413939408E-01 | 7.262013478639E+00 | 1.114048503042E+01 |
| C | -4.197040125855E-01 | 6.124822480783E+00 | 9.356498573501E+00 |
| C | -2.053594130041E-01 | 3.644590357462E+00 | 5.489789899567E+00 |
| C | -4.196402183006E-01 | 5.467189252916E+00 | 8.327905218403E+00 |
| C | -2.700779630569E-01 | 3.315216097040E+00 | 4.976965199998E+00 |
| C | -4.196402514407E-01 | 4.735517079924E+00 | 7.190850241622E+00 |
| C | -6.976787014262E-02 | 7.595136698847E-01 | 9.821140729792E-01 |
| C | -4.196397301164E-01 | 4.076563888789E+00 | 6.163565518287E+00 |
| C | -1.348238614804E-01 | 1.080674129782E+00 | 1.485183968789E+00 |
| C | -3.444494987139E-01 | 2.935157416213E+00 | 4.383954340081E+00 |
| C | -2.065853063836E-01 | 1.445249583173E+00 | 2.056345858153E+00 |
| C | -4.196305407036E-01 | 3.316901948428E+00 | 4.977529531272E+00 |
| C | -2.707138182227E-01 | 1.781296298849E+00 | 2.581638506328E+00 |
| C | -3.450264227975E-01 | 2.168340866511E+00 | 3.186284295709E+00 |
| C | -4.947856461470E-01 | 2.936114622697E+00 | 4.382635288581E+00 |
| C | -4.195801847506E-01 | 1.800394797594E+00 | 2.611422624092E+00 |
| C | -4.941543384947E-01 | 2.169027194425E+00 | 3.185098032719E+00 |
| H | -4.195607524190E-01 | 1.215865777439E+00 | 1.697950437420E+00 |
| C | 4.315547140556E-01 | 1.782365767928E+00 | 2.579576532266E+00 |
| C | 3.674018967509E-01 | 1.447017289739E+00 | 2.054774979898E+00 |
| C | 8.037988362678E-02 | 2.187459898798E+00 | 3.218226970902E+00 |
| C | 8.039098783516E-02 | 1.529626793239E+00 | 2.189831611824E+00 |
| C | 8.039440956757E-02 | 2.918033475115E+00 | 4.355881267740E+00 |
| C | 4.308077746391E-01 | 3.316971937145E+00 | 4.974011828722E+00 |
| C | 2.956213528307E-01 | 1.082250657024E+00 | 1.484165766975E+00 |
| C | 3.661039799956E-01 | 3.646756711797E+00 | 5.487168017035E+00 |
| C | 8.043610692917E-02 | 3.575419596855E+00 | 5.384531216713E+00 |
| C | 2.945858543995E-01 | 4.010940464137E+00 | 6.057418897879E+00 |
| C | 8.041208944496E-02 | 4.341494367355E+00 | 6.563214920187E+00 |
| C | 2.299023644168E-01 | 4.341767846509E+00 | 6.570035686039E+00 |
| C | 2.305486021985E-01 | 7.607825896279E-01 | 9.819586307004E-01 |
| C | 5.319039215632E-03 | 4.714013649550E+00 | 7.166355310403E+00 |
| C | 1.554799253121E-01 | 4.714618950510E+00 | 7.166583028613E+00 |
| C | 8.039312525846E-02 | 7.728492819865E-01 | 1.002883813414E+00 |
| C | 5.282851770883E-03 | 5.489586115821E+00 | 8.353436363917E+00 |
| C | 1.552906326888E-01 | 3.865359404235E-01 | 3.961479227548E-01 |
| C | 1.554562546010E-01 | 5.491268236636E+00 | 8.352872901375E+00 |
| C | 8.035906963084E-02 | 5.863070122849E+00 | 8.956427986474E+00 |
| C | 5.501339419369E-03 | 3.862139167057E-01 | 3.960269270809E-01 |
| C | 2.298946973219E-01 | 5.865064058254E+00 | 8.948269504992E+00 |
| C | 2.945373260711E-01 | 6.197306575372E+00 | 9.461449714296E+00 |
| C | 3.661124839119E-01 | 6.559407208895E+00 | 1.003175008396E+01 |
| C | 1.540860975969E-01 | -3.662224865030E-01 | -7.856122338381E-01 |
| C | 6.727496709086E-03 | -3.668396525780E-01 | -7.855351719693E-01 |
| C | 4.307820114088E-01 | 6.889498545090E+00 | 1.054534736539E+01 |
| C | 8.041129005675E-02 | -7.377936401032E-01 | -1.367023297975E+00 |
| H | 2.118766209831E-01 | -6.522458268254E-01 | -1.236627835615E+00 |
| H | -5.105478477961E-02 | -6.540114055501E-01 | -1.236089934314E+00 |
| H | 8.041952141770E-02 | -1.321803397336E+00 | -2.282696679046E+00 |
| C | 5.246072090964E-03 | 9.805740830766E+00 | 1.511663661312E+01 |
| C | 5.256599069145E-03 | 1.055722149604E+01 | 1.631930216813E+01 |
| C | 8.037304041002E-02 | 6.628919751988E+00 | 1.013540755422E+01 |
| C | 8.050465345200E-02 | 7.285366772834E+00 | 1.116476382815E+01 |
| C | 8.048735055639E-02 | 8.012229086029E+00 | 1.230393932797E+01 |
| C | 8.033539354684E-02 | 8.669419545225E+00 | 1.333289608082E+01 |
| C | 8.032753398192E-02 | 9.415536668262E+00 | 1.452427083382E+01 |
| C | -6.918481377078E-02 | 1.094007978062E+01 | 1.690867237698E+01 |
| C | 8.035367232723E-02 | 1.094663235665E+01 | 1.691156536196E+01 |
| C | 1.554373494551E-01 | 9.806142447583E+00 | 1.511578352990E+01 |
| C | 1.554366944720E-01 | 1.055643413537E+01 | 1.631915531381E+01 |
| C | -1.338724872829E-01 | 1.126506429906E+01 | 1.742503480447E+01 |
| C | -6.922546881626E-02 | 9.420995777030E+00 | 1.452907621047E+01 |
| C | 2.298788213893E-01 | 9.423005535865E+00 | 1.452623745475E+01 |
| C | 2.298774695147E-01 | 1.094029524373E+01 | 1.690833753960E+01 |
| C | -2.054594507523E-01 | 1.162912545203E+01 | 1.799364254586E+01 |
| C | 2.946471502102E-01 | 1.126406141968E+01 | 1.742242746526E+01 |
| C | -2.700966050246E-01 | 1.195487971396E+01 | 1.851114988368E+01 |
| C | 2.945892751152E-01 | 9.097687586143E+00 | 1.401089160634E+01 |
| C | -3.445978051124E-01 | 1.308313729226E+01 | 2.030642577172E+01 |
| C | -1.339944415742E-01 | 9.097170429375E+00 | 1.401474780415E+01 |
| C | -3.445506044013E-01 | 1.233692143249E+01 | 1.910064691841E+01 |
| C | -4.196910670597E-01 | 1.347050564772E+01 | 2.089998411870E+01 |
| C | 3.662328196606E-01 | 1.162822164515E+01 | 1.799104423738E+01 |
| C | -4.947591286089E-01 | 1.308290782968E+01 | 2.030552900779E+01 |
| C | -4.196238261740E-01 | 1.194907460783E+01 | 1.850650245089E+01 |
| C | 4.308404290771E-01 | 1.195485492033E+01 | 1.850891211113E+01 |
| C | -4.947302686299E-01 | 1.233670507043E+01 | 1.909982956634E+01 |
| C | 3.661566914874E-01 | 8.735534435493E+00 | 1.344034082138E+01 |
| C | -4.195832363337E-01 | 1.120610605791E+01 | 1.731315924862E+01 |
| C | -4.196392377145E-01 | 1.055104421921E+01 | 1.628269991299E+01 |
| C | 4.308401732724E-01 | 8.406807073682E+00 | 1.292635600286E+01 |
| C | -2.055256756978E-01 | 8.734563109352E+00 | 1.344380671007E+01 |
| C | -4.196477386312E-01 | 9.826565378275E+00 | 1.514238105090E+01 |
| C | -4.195641142758E-01 | 9.171860591265E+00 | 1.411177123833E+01 |
| C | -2.700783444958E-01 | 8.404024127881E+00 | 1.292656075016E+01 |
| C | 5.759271467011E-03 | 1.561078163568E+01 | 2.429891322787E+01 |
| C | 5.149460442376E-03 | 1.485250668117E+01 | 2.309569009976E+01 |
| C | 8.032082914646E-02 | 1.597435563230E+01 | 2.487599771799E+01 |
| C | 8.031303849729E-02 | 1.447533864688E+01 | 2.249852528560E+01 |
| C | 1.548849416228E-01 | 1.561068906103E+01 | 2.429840400579E+01 |
| C | 8.030045694882E-02 | 1.372253301611E+01 | 2.130801265775E+01 |
| C | 8.023474805409E-02 | 1.306959420835E+01 | 2.027682527222E+01 |
| C | 1.554886486875E-01 | 1.485170311977E+01 | 2.309563155788E+01 |
| C | 2.291900926216E-01 | 1.599368738280E+01 | 2.490569428605E+01 |
| C | 8.022192115078E-02 | 1.234776507442E+01 | 1.913346970956E+01 |
| C | 8.038965904268E-02 | 1.169332788129E+01 | 1.810290595527E+01 |
| C | 2.933670843972E-01 | 1.632649415714E+01 | 2.543149364319E+01 |
| C | 2.298729218105E-01 | 1.447699552614E+01 | 2.249946456271E+01 |
| C | -6.925583207432E-02 | 1.447808134476E+01 | 2.249970084689E+01 |
| C | -6.853980487322E-02 | 1.599245384274E+01 | 2.490698503417E+01 |
| C | 3.651298480447E-01 | 1.668722646285E+01 | 2.600495580808E+01 |
| H | 8.033465079233E-02 | 1.655113490195E+01 | 2.579435446284E+01 |
| C | -1.326852764473E-01 | 1.632454010078E+01 | 2.543418944876E+01 |
| C | 4.301895746559E-01 | 1.700499357232E+01 | 2.650989779751E+01 |
| C | -1.339992807012E-01 | 1.415319086435E+01 | 2.198498250610E+01 |
| C | -4.933816765941E-01 | 1.812156132493E+01 | 2.828464236671E+01 |
| C | 2.946551848434E-01 | 1.415342154568E+01 | 2.198518416853E+01 |
| C | -4.945718081940E-01 | 1.737474108145E+01 | 2.709908359161E+01 |
| C | -4.197139736554E-01 | 1.848929352019E+01 | 2.886875830186E+01 |
| C | -2.044540219890E-01 | 1.668548000712E+01 | 2.600766179096E+01 |
| C | -3.460210425207E-01 | 1.812124254214E+01 | 2.828558373940E+01 |
| C | -4.196660036428E-01 | 1.699045600539E+01 | 2.649124140580E+01 |
| C | -2.695165450773E-01 | 1.700334715250E+01 | 2.651263391523E+01 |
| C | -3.447870430962E-01 | 1.737412692349E+01 | 2.710024627527E+01 |
| C | -2.055162294004E-01 | 1.379098481867E+01 | 2.141332106330E+01 |
| C | -4.196386303274E-01 | 1.623862766080E+01 | 2.530110947232E+01 |
| C | -4.197101114900E-01 | 1.558684222868E+01 | 2.426877954222E+01 |
| C | -2.702001871038E-01 | 1.346678107530E+01 | 2.089687625201E+01 |
| C | 3.661884845087E-01 | 1.379089417399E+01 | 2.141366140099E+01 |
| C | -4.197204272680E-01 | 1.486674488498E+01 | 2.312479071658E+01 |
| C | -4.197146422093E-01 | 1.421377188978E+01 | 2.209324153932E+01 |
| C | 4.307958745682E-01 | 1.346554000389E+01 | 2.089541763334E+01 |
| H | 4.488229033205E-01 | 1.840576915735E+01 | 2.873660504173E+01 |
| H | -4.197363205774E-01 | 1.906858283856E+01 | 2.978748363806E+01 |
| H | -2.882387380440E-01 | 1.840593493816E+01 | 2.873786855008E+01 |

**Z4.5**

LATTICE PARAMETERS (ANGSTROMS AND DEGREES)

| A | B | C | ALPHA | BETA | GAMMA |
|---|---|---|---|---|---|
| 16.35104315 | 500.00000000 | 500.00000000 | 90.000000 | 90.000000 | 90.000000 |

*******************************************************************************

ATOMS IN THE ASYMMETRIC UNIT 154 - ATOMS IN THE UNIT CELL: 154

| ATOM | X/A | Y(ANGSTROM) | Z(ANGSTROM) |
|---|---|---|---|
| C | -1.421624232146E-01 | 2.960667712918E+00 | 4.239699973213E-01 |
| C | -2.136940892729E-01 | 3.628915530090E+00 | 5.287882962947E-01 |
| C | -7.743051364865E-02 | 2.359196626777E+00 | 3.279998800889E-01 |
| C | -2.783579496728E-01 | 4.231601887575E+00 | 6.264992675019E-01 |
| C | -3.527761562783E-01 | 4.927861749316E+00 | 7.286344862960E-01 |
| C | -4.278889593665E-01 | 2.839018305655E+00 | 3.995500432602E-01 |
| C | -2.782760178136E-01 | 7.021335045222E+00 | 1.064542694826E+00 |
| C | -3.527414058283E-01 | 6.326117653082E+00 | 9.640020399343E-01 |
| C | -4.278690596514E-01 | 4.225930537663E+00 | 6.293940375904E-01 |
| C | -4.278584388903E-01 | 1.633461635064E+00 | 2.074309117726E-01 |
| C | -2.136889530179E-01 | 7.626336583358E+00 | 1.161459979361E+00 |
| C | -1.421128934009E-01 | 8.293652165500E+00 | 1.267287440267E+00 |
| C | -4.278622562902E-01 | 2.988327306367E-01 | -8.049694480676E-03 |
| C | -4.278153563457E-01 | 7.028508746191E+00 | 1.064263519115E+00 |
| C | 4.970628051362E-01 | 4.928571885806E+00 | 7.290428159098E-01 |
| C | -7.734403691787E-02 | 8.893956841473E+00 | 1.361949515773E+00 |
| C | -4.278445061113E-01 | -9.066946487685E-01 | -1.999660156022E-01 |
| C | 4.970861304081E-01 | 6.326829784448E+00 | 9.644372225304E-01 |
| C | -4.277899469665E-01 | 8.414963235407E+00 | 1.296038525210E+00 |
| C | 4.226263613389E-01 | 4.232874254078E+00 | 6.270951365692E-01 |
| C | -4.278489625470E-01 | -2.297263119436E+00 | -4.192394698710E-01 |
| C | -2.942050112634E-03 | 1.098919428476E+01 | 1.693735174692E+00 |
| C | -4.277965103649E-01 | 9.620673935823E+00 | 1.488170745478E+00 |
| C | 3.580097859200E-01 | 3.628801570136E+00 | 5.293807027623E-01 |
| C | 7.213330261324E-02 | 1.168804108579E+01 | 1.816224881486E+00 |
| C | -2.906494881448E-03 | 9.587019875667E+00 | 1.481568475635E+00 |
| C | 4.972578155119E-01 | -3.007928679149E+00 | -5.309619902088E-01 |
| C | -3.529555276523E-01 | -3.007932936810E+00 | -5.313051181545E-01 |
| C | 4.226585277338E-01 | 7.023031845760E+00 | 1.065764871390E+00 |
| C | -2.404312852690E-03 | 2.606753015167E-01 | -9.772574072537E-03 |
| C | -3.022946439876E-03 | 1.664692093815E+00 | 2.163296899726E-01 |
| C | 7.216595597729E-02 | -4.126565290981E-01 | -1.183988954651E-01 |
| C | -4.277753613180E-01 | 1.095496019175E+01 | 1.701304259415E+00 |
| C | 2.864687330693E-01 | 2.960462402396E+00 | 4.246767333663E-01 |
| C | 1.472457696968E-01 | 1.098999378447E+01 | 1.694315084278E+00 |
| C | 7.213595074564E-02 | 2.361953072523E+00 | 3.288482381001E-01 |
| C | 7.220201779760E-02 | 8.888887747401E+00 | 1.360134174296E+00 |
| C | -4.277080612710E-01 | 1.216059052467E+01 | 1.894127895772E+00 |
| C | 1.467198483449E-01 | 2.610920325108E-01 | -9.478256490962E-03 |
| C | 1.473097470896E-01 | 1.665124542988E+00 | 2.166806947541E-01 |
| C | 2.217085905806E-01 | 2.359773471666E+00 | 3.288817727604E-01 |
| C | 1.472725966577E-01 | 9.587884577468E+00 | 1.482257684112E+00 |
| C | 4.984696055930E-01 | -4.391963932597E+00 | -7.491986211819E-01 |
| C | 3.580449156104E-01 | 7.627286790583E+00 | 1.163143365972E+00 |
| C | -3.541698060953E-01 | -4.391972766517E+00 | -7.494901636435E-01 |
| C | 2.217489533951E-01 | 8.895795572061E+00 | 1.363432825047E+00 |
| C | 2.864798845181E-01 | 8.294386051312E+00 | 1.269061565194E+00 |
| C | -4.278511831665E-01 | -5.073471959825E+00 | -8.565371332518E-01 |
| H | 7.217948378124E-02 | -1.483316209439E+00 | -2.908664945051E-01 |
| H | 4.406872716919E-01 | -4.919789572272E+00 | -8.322976955391E-01 |
| H | -2.963890512441E-01 | -4.919812548332E+00 | -8.327953589512E-01 |
| H | -4.278529441151E-01 | -6.146380386076E+00 | -1.025110448602E+00 |
| C | -7.669986187158E-02 | -4.485251487246E-01 | -1.238508085360E-01 |
| C | -1.408900007637E-01 | -1.062467177646E+00 | -2.225749814133E-01 |
| C | -2.126540254746E-01 | -1.731543472985E+00 | -3.292208377709E-01 |
| C | -2.776981849880E-01 | -2.321326171151E+00 | -4.227639920956E-01 |
| C | 2.210279491096E-01 | -4.477920282491E-01 | -1.232599108046E-01 |
| C | 2.851691945827E-01 | -1.063182531653E+00 | -2.215290286753E-01 |
| C | 3.569395535846E-01 | -1.732187403608E+00 | -3.281619589605E-01 |
| C | 4.220009476405E-01 | -2.321395657784E+00 | -4.220256152272E-01 |
| C | 7.224324867270E-02 | 7.499048380194E+00 | 1.147183932894E+00 |
| C | 7.205265515345E-02 | 6.293537970667E+00 | 9.548538219407E-01 |
| C | 7.204597008117E-02 | 4.957919153045E+00 | 7.439441346739E-01 |
| C | 7.210524775401E-02 | 3.752609425469E+00 | 5.520507362874E-01 |
| C | -3.527695900112E-01 | 1.565158729692E+01 | 2.440300934728E+00 |
| C | -4.277711715852E-01 | 1.355005624520E+01 | 2.106106091214E+00 |
| C | 4.970342474806E-01 | 1.564964648852E+01 | 2.440382267994E+00 |
| C | -4.279713148554E-01 | 1.773854507580E+01 | 2.774991413596E+00 |
| C | -2.782305528935E-01 | 1.355734906032E+01 | 2.110966580535E+00 |
| C | 4.226831936635E-01 | 1.355368420613E+01 | 2.110896701274E+00 |
| C | -4.277798595447E-01 | 2.027783443790E+01 | 3.184328225552E+00 |
| C | -1.421236520562E-01 | 1.228365433834E+01 | 1.906147717950E+00 |
| C | -7.741477401212E-02 | 1.168153151544E+01 | 1.811642864375E+00 |
| C | 2.864121587556E-01 | 1.228481859056E+01 | 1.907192059029E+00 |
| C | -4.279585830356E-01 | 2.148307898596E+01 | 3.379508745468E+00 |
| C | 2.216816621653E-01 | 1.168329979444E+01 | 1.812691859531E+00 |
| C | -4.277883724332E-01 | 1.894404342139E+01 | 2.968736371297E+00 |
| C | -4.278983561130E-01 | 2.286885487900E+01 | 3.615262676475E+00 |
| C | -4.277793448817E-01 | 2.566954322210E+01 | 4.062694967522E+00 |
| C | -3.527842907449E-01 | 2.356954373044E+01 | 3.718364327391E+00 |
| C | 4.970492571166E-01 | 2.357126510288E+01 | 3.719150093506E+00 |
| C | -3.527294111122E-01 | 2.496690672515E+01 | 3.959172557949E+00 |
| C | 4.971141975447E-01 | 2.496861720302E+01 | 3.959972698158E+00 |
| C | -2.136842871981E-01 | 1.295075169390E+01 | 2.015628298200E+00 |
| C | 3.579893865229E-01 | 1.295118690571E+01 | 2.016488272715E+00 |
| C | 4.225657149593E-01 | 2.287738717445E+01 | 3.614145846064E+00 |
| C | -2.783591412100E-01 | 2.287381957239E+01 | 3.612911766881E+00 |
| C | 7.210431854369E-02 | 1.307757544868E+01 | 2.029231573795E+00 |
| C | -2.137139813067E-01 | 2.227094855859E+01 | 3.512299958577E+00 |
| C | 3.579712170965E-01 | 2.227312768502E+01 | 3.513535543151E+00 |
| C | 7.224390838603E-02 | 1.428307919774E+01 | 2.223215582850E+00 |
| C | -1.421236594813E-01 | 2.160441062697E+01 | 3.406620318844E+00 |
| C | -2.137275247680E-01 | 1.694950245953E+01 | 2.655858862204E+00 |
| C | 2.864082698373E-01 | 2.160549920781E+01 | 3.407697719326E+00 |
| C | 3.579201715570E-01 | 1.694500765918E+01 | 2.656572228132E+00 |
| C | -7.736155387635E-02 | 2.100461111768E+01 | 3.307376254413E+00 |
| C | 7.225494509244E-02 | 1.561736771455E+01 | 2.437320724533E+00 |
| C | 2.217224459436E-01 | 2.100366350445E+01 | 3.308040085039E+00 |
| C | -7.739363023254E-02 | 1.821619038291E+01 | 2.861838963526E+00 |
| C | -1.421582452100E-01 | 1.761635571099E+01 | 2.763176530277E+00 |
| C | 7.216549670859E-02 | 1.682310451068E+01 | 2.629824193576E+00 |
| C | 1.472681853692E-01 | 2.030889278299E+01 | 3.202561737426E+00 |
| C | 7.218515409057E-02 | 2.101082475707E+01 | 3.304683833212E+00 |
| C | 7.215752666475E-02 | 1.820947009226E+01 | 2.862567242891E+00 |
| C | 2.863641148351E-01 | 1.761254057609E+01 | 2.764079213872E+00 |
| C | 1.472614655319E-01 | 1.891087942363E+01 | 2.965382155576E+00 |
| C | -2.929855216993E-03 | 1.891126774607E+01 | 2.964904524783E+00 |
| C | -2.912311187487E-03 | 2.030941067547E+01 | 3.202219902902E+00 |
| C | 2.216980770989E-01 | 1.821524377454E+01 | 2.862330747609E+00 |
| C | -2.783554223758E-01 | 1.634522841980E+01 | 2.560371588976E+00 |
| C | 4.225504766235E-01 | 1.634101115190E+01 | 2.560485398984E+00 |
| C | -3.527032515001E-01 | 1.424938646596E+01 | 2.229578432659E+00 |
| C | -4.279034085857E-01 | 1.634900645420E+01 | 2.563313282069E+00 |
| C | 4.971016588839E-01 | 1.424736012897E+01 | 2.229696868823E+00 |
| C | -4.278231119343E-01 | 3.496564656955E+01 | 5.569115825511E+00 |
| C | 4.984852989869E-01 | 3.428464349201E+01 | 5.460904958776E+00 |
| C | -3.541554364017E-01 | 3.428397494123E+01 | 5.460788933146E+00 |
| C | 4.972470291290E-01 | 3.290091234589E+01 | 5.240854745489E+00 |
| C | -3.529658951501E-01 | 3.290026299820E+01 | 5.240481280245E+00 |
| C | 4.219762937162E-01 | 3.221492059719E+01 | 5.131046152137E+00 |
| C | -2.777130011839E-01 | 3.221379897858E+01 | 5.130229693929E+00 |
| C | 3.569692117670E-01 | 3.162423437397E+01 | 5.036433022175E+00 |
| C | -4.278723389879E-01 | 3.219012093751E+01 | 5.127468530210E+00 |
| C | -2.126286400632E-01 | 3.162541106294E+01 | 5.035378887958E+00 |
| C | 7.221946405735E-02 | 2.239758582242E+01 | 3.536911870156E+00 |
| C | 7.219260898335E-02 | 3.030606419362E+01 | 4.823077465373E+00 |
| C | 2.852131303071E-01 | 3.095509255438E+01 | 4.928491517847E+00 |
| C | -1.408435129365E-01 | 3.095709054343E+01 | 4.927691905539E+00 |
| C | 2.210572954180E-01 | 3.034037845711E+01 | 4.828622648432E+00 |
| C | -2.389279800913E-03 | 2.963326513562E+01 | 4.713368190919E+00 |
| C | -7.667602317909E-02 | 3.034252612025E+01 | 4.828706600657E+00 |
| C | 1.467373049885E-01 | 2.963207900639E+01 | 4.713314593565E+00 |
| C | -4.278989737175E-01 | 3.079994249547E+01 | 4.905811414430E+00 |
| C | 7.213192093474E-02 | 2.360261778274E+01 | 3.732097439516E+00 |
| C | -3.035937805046E-03 | 2.822952007853E+01 | 4.485406822481E+00 |
| C | 1.472991931422E-01 | 2.822835086014E+01 | 4.485392142749E+00 |
| C | 7.211021122516E-02 | 2.753204341859E+01 | 4.372044273288E+00 |
| C | -4.278666023519E-01 | 2.959488321454E+01 | 4.710939647655E+00 |
| C | 7.212301521346E-02 | 2.493698240056E+01 | 3.951159491150E+00 |
| C | 7.207549972279E-02 | 2.614172068431E+01 | 4.146713567675E+00 |
| C | 2.216802682968E-01 | 2.753351621091E+01 | 4.370953204973E+00 |
| C | -7.745692022920E-02 | 2.753590430736E+01 | 4.370693842956E+00 |
| C | 2.864619207851E-01 | 2.693377721602E+01 | 4.273217370429E+00 |
| C | -1.421248235601E-01 | 2.693302222380E+01 | 4.272290938973E+00 |
| C | -4.278479596401E-01 | 2.826069313278E+01 | 4.492568657411E+00 |
| C | 3.580235628938E-01 | 2.626645291329E+01 | 4.165756921313E+00 |
| C | -2.136506831616E-01 | 2.626501918232E+01 | 4.164511913333E+00 |
| C | 4.227102091610E-01 | 2.566475474617E+01 | 4.065600782388E+00 |
| C | -4.277211388075E-01 | 2.705570333357E+01 | 4.296977649905E+00 |
| C | -2.782688504124E-01 | 2.566141356936E+01 | 4.064218112526E+00 |
| H | -4.278039174556E-01 | 3.603826022646E+01 | 5.739532085766E+00 |
| H | 4.407137739464E-01 | 3.481265938323E+01 | 5.544780124448E+00 |
| H | -2.963641917023E-01 | 3.481142291279E+01 | 5.544619989989E+00 |
| H | 7.222674881868E-02 | 3.137647396250E+01 | 4.997147772760E+00 |

**Z5**

LATTICE PARAMETERS (ANGSTROMS AND DEGREES)

| A | B | C | ALPHA | BETA | GAMMA |
|---|---|---|---|---|---|
| 16.35099588 | 500.00000000 | 500.00000000 | 90.000000 | 90.000000 | 90.000000 |

*******************************************************************************

ATOMS IN THE ASYMMETRIC UNIT 172 - ATOMS IN THE UNIT CELL: 172

| ATOM | X/A | Y(ANGSTROM) | Z(ANGSTROM) |
|---|---|---|---|

*******************************************************************************

| | | | |
|---|---|---|---|
| C | -7.070973266264E-02 | 3.767146179747E+00 | 9.882192570777E+00 |
| C | -1.353312724182E-01 | 4.002491644764E+00 | 1.044689760811E+01 |
| C | -2.068996686875E-01 | 4.261226988210E+00 | 1.107123342395E+01 |
| C | -3.460665541260E-01 | 5.304515439786E+00 | 1.359454255337E+01 |
| C | -2.715858522546E-01 | 4.493829180607E+00 | 1.163505733422E+01 |
| C | -7.061768785908E-02 | 2.695407958371E+00 | 7.269227519905E+00 |
| C | -4.211794461388E-01 | 5.586985665025E+00 | 1.424447414675E+01 |
| C | -3.460452333986E-01 | 4.771809469307E+00 | 1.228035468288E+01 |
| C | -1.353162581763E-01 | 2.460535271320E+00 | 6.706575279707E+00 |
| C | -4.962616024242E-01 | 5.304759528160E+00 | 1.359357467824E+01 |
| C | -4.211198423624E-01 | 4.489978594836E+00 | 1.162926461027E+01 |
| C | -4.962232365461E-01 | 4.771861915899E+00 | 1.227941568229E+01 |
| C | -4.211002971110E-01 | 3.959859678329E+00 | 1.032706668330E+01 |
| C | -2.068566465801E-01 | 2.203594345685E+00 | 6.081128606672E+00 |
| C | -4.211267354777E-01 | 3.492333233098E+00 | 9.199254923820E+00 |
| C | -2.715883701931E-01 | 1.970962615993E+00 | 5.518152887538E+00 |
| C | -4.211357261791E-01 | 2.977518596864E+00 | 7.949140601214E+00 |
| C | -7.131475219669E-02 | 1.532421909320E-01 | 1.138618245204E+00 |
| C | -4.211490466783E-01 | 2.511141471782E+00 | 6.821165631596E+00 |
| C | -1.363735776599E-01 | 3.814651442657E-01 | 1.690024154637E+00 |
| C | -3.459782030746E-01 | 1.700798314557E+00 | 4.868237401620E+00 |
| C | -2.081368420000E-01 | 6.407848965432E-01 | 2.316100578701E+00 |
| C | -4.211448399032E-01 | 1.971660331325E+00 | 5.520182347263E+00 |
| C | -2.722805575273E-01 | 8.795247121465E-01 | 2.891714221725E+00 |
| C | -3.465805660088E-01 | 1.155146981926E+00 | 3.554940746871E+00 |
| C | -4.963098069370E-01 | 1.700652047344E+00 | 4.868219879095E+00 |
| C | -4.211431444343E-01 | 8.927679260171E-01 | 2.925142012994E+00 |
| C | -4.957064804718E-01 | 1.154833420759E+00 | 3.555038066786E+00 |
| H | -4.211419182705E-01 | 4.763308208825E-01 | 1.923802973194E+00 |
| C | 4.299944292916E-01 | 8.792196609593E-01 | 2.891900348678E+00 |
| C | 3.658235485799E-01 | 6.404412966043E-01 | 2.317097554732E+00 |
| C | 7.884280215190E-02 | 1.166250050681E+00 | 3.590986183610E+00 |
| C | 7.882064425984E-02 | 6.990064634732E-01 | 2.463241410131E+00 |
| C | 7.886755804183E-02 | 1.685506512016E+00 | 4.839284316807E+00 |
| C | 4.292992658167E-01 | 1.970757918185E+00 | 5.518021313477E+00 |
| C | 2.940575543954E-01 | 3.810563925465E-01 | 1.691160935899E+00 |
| C | 3.645822061342E-01 | 2.203816835309E+00 | 6.081233206966E+00 |
| C | 7.892616123445E-02 | 2.151285820939E+00 | 5.967613326872E+00 |
| C | 2.930459911440E-01 | 2.460589918724E+00 | 6.707145255417E+00 |
| C | 7.889177027457E-02 | 2.696162817669E+00 | 7.263621601290E+00 |
| C | 2.284020383282E-01 | 2.695213498584E+00 | 7.271531996633E+00 |
| C | 2.289880351898E-01 | 1.528979151088E-01 | 1.139989621490E+00 |
| C | 3.790817324836E-03 | 2.955571519703E+00 | 7.923228764668E+00 |
| C | 1.539553789612E-01 | 2.955368467941E+00 | 7.924381565439E+00 |
| C | 7.883532606645E-02 | 1.617846967720E-01 | 1.161982025117E+00 |
| C | 3.752522536800E-03 | 3.507634612300E+00 | 9.229276739549E+00 |
| C | 1.537386385648E-01 | -1.127241435052E-01 | 4.973087590899E-01 |
| C | 1.539237668133E-01 | 3.507762365660E+00 | 9.230338456741E+00 |
| C | 7.882160397273E-02 | 3.767371403007E+00 | 9.889595881183E+00 |
| C | 3.951875279629E-03 | -1.123801448247E-01 | 4.964847968856E-01 |
| C | 2.283491147903E-01 | 3.767492330534E+00 | 9.884163542929E+00 |
| C | 2.930306216585E-01 | 4.002282241440E+00 | 1.044726684636E+01 |
| C | 3.646086132189E-01 | 4.260778688842E+00 | 1.107141482336E+01 |
| C | 1.525444382926E-01 | -6.469085730934E-01 | -7.980367785602E-01 |
| C | 5.183296157091E-03 | -6.467908439500E-01 | -7.987413517926E-01 |
| C | 4.293589879817E-01 | 4.493636550366E+00 | 1.163313676496E+01 |
| C | 7.887279504759E-02 | -9.100024792967E-01 | -1.436135403860E+00 |
| H | 2.103387536414E-01 | -8.508215348232E-01 | -1.291566687191E+00 |
| H | -5.259663805375E-02 | -8.502525097895E-01 | -1.292928681299E+00 |
| H | 7.888805420333E-02 | -1.324406372495E+00 | -2.440049004727E+00 |
| C | 3.770999891888E-03 | 6.566030141721E+00 | 1.664554562148E+01 |
| C | 3.737882673535E-03 | 7.116096526294E+00 | 1.795242036951E+01 |
| C | 7.879250412298E-02 | 4.312487984620E+00 | 1.118522899800E+01 |
| C | 7.888046067173E-02 | 4.778807010230E+00 | 1.231358267118E+01 |
| C | 7.890135890045E-02 | 5.298834097152E+00 | 1.356090786609E+01 |
| C | 7.892565217765E-02 | 5.764166148446E+00 | 1.468966418704E+01 |
| C | 7.889317107050E-02 | 6.307877373710E+00 | 1.598571022377E+01 |
| C | -7.071036758799E-02 | 7.374908950076E+00 | 1.860557197969E+01 |
| C | 7.881446944008E-02 | 7.374605889545E+00 | 1.861331169447E+01 |
| C | 1.539781670276E-01 | 6.566483231021E+00 | 1.664656059858E+01 |
| C | 1.539322573787E-01 | 7.116396751115E+00 | 1.795356310352E+01 |
| C | -1.353507349332E-01 | 7.609495118815E+00 | 1.917027640156E+01 |
| C | -7.066784656066E-02 | 6.306300535176E+00 | 1.599197379742E+01 |
| C | 2.284424754282E-01 | 6.307686707839E+00 | 1.599366863832E+01 |
| C | 2.283396608512E-01 | 7.375831835893E+00 | 1.860782810849E+01 |
| C | -2.069064965791E-01 | 7.867683951209E+00 | 1.979491602458E+01 |
| C | 2.930469889757E-01 | 7.610475039588E+00 | 1.917040358836E+01 |
| C | -2.715646537663E-01 | 8.100042604630E+00 | 2.035982401858E+01 |
| C | 2.930418787982E-01 | 6.072825648204E+00 | 1.542790970113E+01 |
| C | -3.460693314612E-01 | 8.907651573493E+00 | 2.232023753995E+01 |
| C | -1.353646065341E-01 | 6.072162014983E+00 | 1.542894660496E+01 |
| C | -3.460266316819E-01 | 8.377096776346E+00 | 2.100521175426E+01 |
| C | -4.211897368286E-01 | 9.189188580957E+00 | 2.297056700483E+01 |
| C | 3.646222704163E-01 | 7.868449862828E+00 | 1.979488100200E+01 |
| C | -4.962690743544E-01 | 8.907519812158E+00 | 2.231930321671E+01 |
| C | -4.211044513380E-01 | 8.094991927634E+00 | 2.035412119294E+01 |
| C | 4.293660127097E-01 | 8.100061387687E+00 | 2.035742463159E+01 |
| C | -4.962173128054E-01 | 8.376989266264E+00 | 2.100425724020E+01 |
| C | 3.645995545385E-01 | 5.814552395811E+00 | 1.480317997282E+01 |
| C | -4.210752968808E-01 | 7.565585463517E+00 | 1.905203004235E+01 |
| C | -4.210865000338E-01 | 7.098157812094E+00 | 1.792399043543E+01 |
| C | 4.292783833021E-01 | 5.582354762131E+00 | 1.423893604041E+01 |
| C | -2.069509506369E-01 | 5.813893630229E+00 | 1.480515229371E+01 |
| C | -4.211173688615E-01 | 6.585069382803E+00 | 1.667416653104E+01 |
| C | -4.212070917698E-01 | 6.117390487516E+00 | 1.554626291226E+01 |
| C | -2.716275908093E-01 | 5.581916411715E+00 | 1.424066534258E+01 |
| C | -4.211305989046E-01 | 1.018194104140E+01 | 2.540231725664E+01 |
| C | -4.211186701021E-01 | 1.069590279769E+01 | 2.665165010729E+01 |
| C | -4.211794214813E-01 | 9.716082327427E+00 | 2.427364193040E+01 |
| C | -4.210903612125E-01 | 1.115893089274E+01 | 2.778145442895E+01 |
| C | -2.716400475747E-01 | 9.184697644436E+00 | 2.296657469984E+01 |
| C | -4.211421168278E-01 | 1.170302890566E+01 | 2.907768906399E+01 |
| C | -2.069771907107E-01 | 9.415728771864E+00 | 2.353201427448E+01 |
| C | 4.292393032377E-01 | 9.184001883018E+00 | 2.296415019610E+01 |
| C | -3.460685851959E-01 | 1.196253190845E+01 | 2.973803336235E+01 |
| C | -4.962322741290E-01 | 1.196289756452E+01 | 2.973706851296E+01 |
| C | -2.716049822550E-01 | 1.170147096153E+01 | 2.908582626928E+01 |
| C | 4.293094953164E-01 | 1.170260818771E+01 | 2.908419919491E+01 |
| C | 3.645873015367E-01 | 9.415817851409E+00 | 2.352965240603E+01 |
| C | 3.646242397952E-01 | 1.146924663245E+01 | 2.852053904004E+01 |
| C | -2.069631248209E-01 | 1.146630028452E+01 | 2.852154149555E+01 |
| C | -1.353875165327E-01 | 9.673151597815E+00 | 2.415614652397E+01 |
| C | -3.460878470649E-01 | 1.251449226249E+01 | 3.104420281221E+01 |
| C | -4.962483644405E-01 | 1.251468920638E+01 | 3.104330201309E+01 |
| C | 2.930348095995E-01 | 9.673100851112E+00 | 2.415468267303E+01 |
| C | -4.211811242713E-01 | 1.277388483994E+01 | 3.170402309624E+01 |
| C | 2.930694635231E-01 | 1.121380401776E+01 | 2.789503585899E+01 |
| C | -1.353950656392E-01 | 1.121109384148E+01 | 2.789581067006E+01 |
| C | -4.211503848029E-01 | 1.636043374726E+01 | 4.041149380585E+01 |
| C | 4.293138802165E-01 | 1.277512782184E+01 | 3.169609852666E+01 |
| C | -2.716557396832E-01 | 1.277531666172E+01 | 3.169736810242E+01 |
| C | -4.948341552220E-01 | 1.609925184707E+01 | 3.977306505669E+01 |
| C | 2.283991019525E-01 | 9.906827477673E+00 | 2.471971344888E+01 |
| C | -7.069567404837E-02 | 9.907281676286E+00 | 2.471940128365E+01 |
| C | -4.212178240858E-01 | 1.331741796321E+01 | 3.300051471498E+01 |
| C | -3.474737897834E-01 | 1.609876109843E+01 | 3.977304848521E+01 |
| C | 2.283516819907E-01 | 1.097743943564E+01 | 2.733369633893E+01 |
| C | -4.960545458784E-01 | 1.556833203554E+01 | 3.847640513973E+01 |
| C | 3.646547307604E-01 | 1.300942563643E+01 | 3.226024715732E+01 |
| C | -7.071243828121E-02 | 1.097659029868E+01 | 2.733238464319E+01 |
| C | -4.212151116198E-01 | 1.378092290357E+01 | 3.412983715920E+01 |
| C | -2.069954194492E-01 | 1.300925479518E+01 | 3.226146469248E+01 |
| C | -3.462660609750E-01 | 1.556777466260E+01 | 3.847643790020E+01 |
| C | -4.211630569573E-01 | 1.529519059141E+01 | 3.781074235524E+01 |
| C | 4.286829636846E-01 | 1.530454687530E+01 | 3.783341803319E+01 |
| C | -4.212072333938E-01 | 1.429731962533E+01 | 3.537905248994E+01 |
| C | -4.211724296108E-01 | 1.476147753203E+01 | 3.650813804213E+01 |
| C | 1.539517856544E-01 | 1.016569944920E+01 | 2.537353153172E+01 |
| C | 3.636352097046E-01 | 1.507753619129E+01 | 3.728116750961E+01 |
| C | 1.539285286442E-01 | 1.071632524660E+01 | 2.668027106214E+01 |
| C | -2.710074757079E-01 | 1.530384896973E+01 | 3.783340377039E+01 |
| C | 2.931212993639E-01 | 1.326587536808E+01 | 3.288612846306E+01 |
| C | -2.059396374638E-01 | 1.507700723227E+01 | 3.728175600830E+01 |
| C | -1.354651945076E-01 | 1.326602079680E+01 | 3.288670496055E+01 |
| C | 7.885766388368E-02 | 9.906541711527E+00 | 2.471329045489E+01 |
| C | 2.918703994142E-01 | 1.481987812820E+01 | 3.665463615123E+01 |
| C | 7.886897155923E-02 | 9.363360098686E+00 | 2.341690229782E+01 |
| C | 7.883151000574E-02 | 8.899254870309E+00 | 2.228756317402E+01 |
| C | 7.882233079940E-02 | 1.097522160432E+01 | 2.734010871381E+01 |
| C | 2.283924974910E-01 | 1.349835337520E+01 | 3.344929842986E+01 |
| C | 2.277192022585E-01 | 1.458281101391E+01 | 3.607856134066E+01 |
| C | -1.341700456114E-01 | 1.481932993934E+01 | 3.665531231390E+01 |
| C | 7.880683539221E-02 | 1.151791664883E+01 | 2.863700234272E+01 |
| C | 7.883061571075E-02 | 8.381151845430E+00 | 2.103933415963E+01 |
| C | -7.073041018678E-02 | 1.349860310284E+01 | 3.344950913801E+01 |
| C | 3.753919166392E-03 | 1.016578081286E+01 | 2.537293199392E+01 |
| H | -4.211447993727E-01 | 1.677177388277E+01 | 4.141666676453E+01 |
| C | 3.741143923784E-03 | 1.071605141709E+01 | 2.667976771210E+01 |
| C | 7.884991841210E-02 | 1.198184477856E+01 | 2.976620276786E+01 |
| H | 4.473811017195E-01 | 1.630136397332E+01 | 4.026766352459E+01 |
| C | 7.877290400246E-02 | 7.916873142784E+00 | 1.991000812998E+01 |
| C | 1.540099547431E-01 | 1.376786132618E+01 | 3.409962833039E+01 |
| C | 1.534133427870E-01 | 1.430940716041E+01 | 3.541454984294E+01 |
| C | -7.002195903538E-02 | 1.458230616624E+01 | 3.607913372096E+01 |
| H | -2.896833153723E-01 | 1.630065704575E+01 | 4.026756232277E+01 |
| C | 7.886151057489E-02 | 1.249873830602E+01 | 3.101571373441E+01 |
| C | 7.884955888963E-02 | 1.296376161253E+01 | 3.214418443098E+01 |
| C | 7.883721099944E-02 | 1.349914325580E+01 | 3.344679006541E+01 |
| C | 7.884684421411E-02 | 1.456919490784E+01 | 3.604544926079E+01 |
| C | 3.668258203000E-03 | 1.376781059092E+01 | 3.409969065480E+01 |
| C | 4.277901815042E-03 | 1.430906387881E+01 | 3.541482043473E+01 |
| H | 7.885042277224E-02 | 1.498219215904E+01 | 3.704821169087E+01 |

**Table S4: B**and gaps of A(n)- and Z(n)-GDYNRs of different widths computed at PBE0/6-31G(d) and HSE06/6-31G(d) level of theory

| NR | Band Gap PBE0 (eV) | Band Gap HSE06 (eV) |
|---|---|---|
| **A1** | 2.836 | 2.2248 |
| **A2** | 2.281 | 1.7159 |
| **A3** | 2.014 | 1.4797 |
| **A4** | 1.883 | 1.3588 |
| **A5** | 1.804 | 1.2882 |
| **Z1** | 3.933 | 3.2814 |
| **Z1.5** | 2.866 | 2.2919 |
| **Z2** | 2.410 | 1.8702 |
| **Z2.5** | 2.173 | 1.6492 |
| **Z3** | 2.030 | 1.5146 |
| **Z3.5** | 1.937 | 1.4259 |
| **Z4** | 1.873 | 1.3636 |
| **Z4.5** | 1.826 | 1.3181 |
| **Z5** | 1.791 | / |

**Table S5:** DFT computed **[**PBE0/6-31G(d)] values of vibrational frequency ($cm^{-1}$), IR intensity (km/mol) and Raman activity ($A^4$/amu) of 2D-GDY. It should be noted that there are two modes with large imaginary frequency values (-388 and -96 $cm^{-1}$) in this Table. The occurrence of imaginary frequencies associated to out-of-plane displacements of carbon atoms in arenes such as the GDY systems investigated in this work is a spurious computational effect related with the incompleteness of the atomic orbital basis set [H. Torii et al. J. Mol. Struct. - Theochem 500, 311 (2000), D. Moran et al. J. Am. Chem. Soc. 128, 9342 (2006), D. Asturiol et al. :J. Chem. Phys. 128, 144108 (2008)]. It is not expected to affect the in-plane vibrational structure.

| **Wavenumber ($cm^{-1}$)** | **Symmetry species** | **IR intensity [km/mol]** | **Raman activity [$A^4$/amu]** |
|---|---|---|---|
| -388 | $B_{2g}$ | 0 | 0 |
| -96 | $B_{2u}$ | 0 | 0 |
| -10 | $A_{2u}$ | 0 | 0 |
| -3 | $E_{1u}$ | 0 | 0 |
| 17 | $E_{1u}$ | 0 | 0 |
| 87 | $A_{2g}$ | 0 | 0 |
| 96 | $E_{2u}$ | 0 | 0 |
| 96 | $E_{2u}$ | 0 | 0 |
| 137 | $E_{1g}$ | 0 | 5 |
| 137 | $E_{1g}$ | 0 | 5 |
| 172 | $E_{1u}$ | 0.3 | 0 |
| 172 | $E_{1u}$ | 0.3 | 0 |
| 183 | $A_{2u}$ | 0.1 | 0 |
| 307 | $B_{2g}$ | 0 | 0 |
| 392 | $E_{2u}$ | 0 | 0 |
| 392 | $E_{2u}$ | 0 | 0 |
| 396 | $E_{2g}$ | 0 | 973 |
| 396 | $E_{2g}$ | 0 | 973 |
| 410 | $B_{2u}$ | 0 | 0 |
| 418 | $E_{1u}$ | 6 | 0 |
| 418 | $E_{1u}$ | 6 | 0 |
| 474 | $E_{1g}$ | 0 | 40 |
| 474 | $E_{1g}$ | 0 | 40 |
| 485 | $A_{2u}$ | 0.2 | 0 |
| 485 | $A_{2g}$ | 0 | 0 |
| 504 | $B_{1u}$ | 0 | 0 |
| 573 | $E_{2g}$ | 0 | 740 |
| 573 | $E_{2g}$ | 0 | 740 |
| 583 | $E_{2u}$ | 0 | 0 |
| 583 | $E_{2u}$ | 0 | 0 |
| 640 | $B_{2g}$ | 0 | 0 |
| 652 | $E_{1g}$ | 0 | 393 |
| 652 | $E_{1g}$ | 0 | 393 |
| 696 | $A_{2g}$ | 0 | 0 |
| 790 | $E_{2g}$ | 0 | 2673 |
| 790 | $E_{2g}$ | 0 | 2673 |
| 893 | $E_{1u}$ | 6.7 | 0 |
| 893 | $E_{1u}$ | 6.7 | 0 |
| 983 | $A_{1g}$ | 0 | 231213 |
| 1317 | $B_{2u}$ | 0 | 0 |
| 1342 | $B_{1u}$ | 0 | 0 |
| 1382 | $E_{2g}$ | 0 | 6285 |
| 1382 | $E_{2g}$ | 0 | 6285 |
| 1429 | $E_{1u}$ | 522 | 0 |
| 1429 | $E_{1u}$ | 522 | 0 |
| 1490 | $A_{1g}$ | 0 | 869870 |
| 1573 | $E_{2g}$ | 0 | 248915 |
| 1573 | $E_{2g}$ | 0 | 248915 |
| 2267 | $E_{1u}$ | 240 | 0 |
| 2267 | $E_{1u}$ | 240 | 0 |
| 2276 | $A_{1g}$ | 0 | 3707443 |
| 2282 | $B_{1u}$ | 0 | 0 |
| 2335 | $E_{2g}$ | 0 | 638183 |
| 2335 | $E_{2g}$ | 0 | 638183 |

**Table S6:** Sketches of DFT computed [PBE0/6-31G(d)] Raman and IR active normal modes of 2D-GDY.
Red arrows indicate atomic displacements while green and blue colors indicate the active bond stretching vibrations in each normal mode. The different colors indicate different phases of the vibration, while the line width is proportional to its amplitude.

| Normal Mode | Frequency ($cm^{-1}$) | Symmety species | IR intensity (km/mol) | Raman activity ($A^4$/amu) |
|---|---|---|---|---|
| **R’** 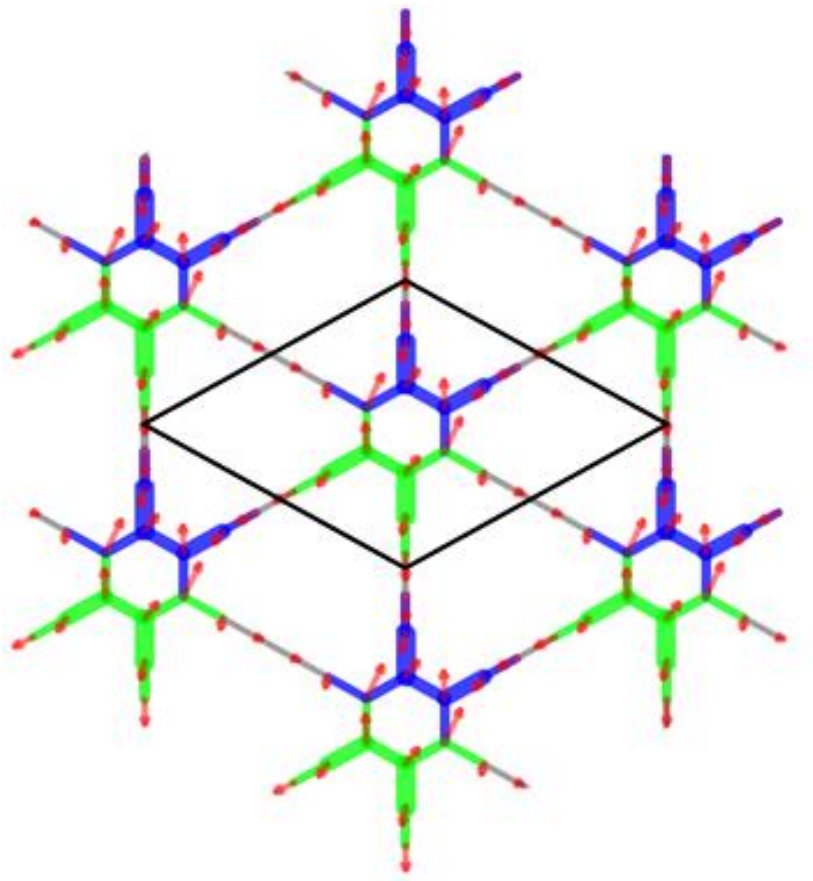 | 893 | $E_{1u}$ | 7 | 0 |
| 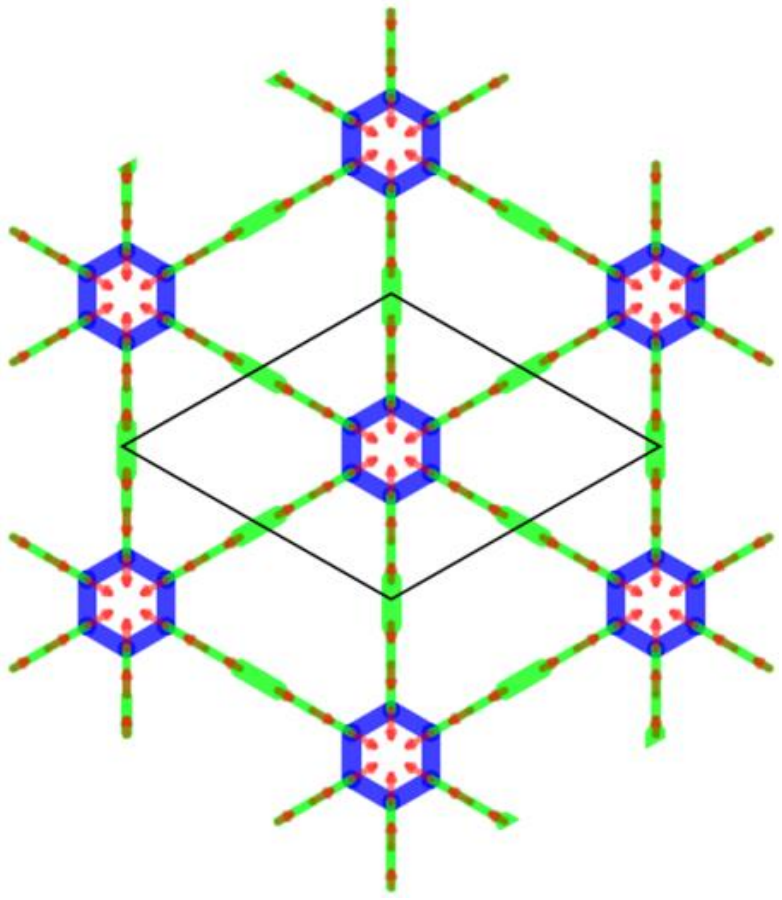 **B** | 983 | $A_{1g}$ | 0 | 231213 |
| 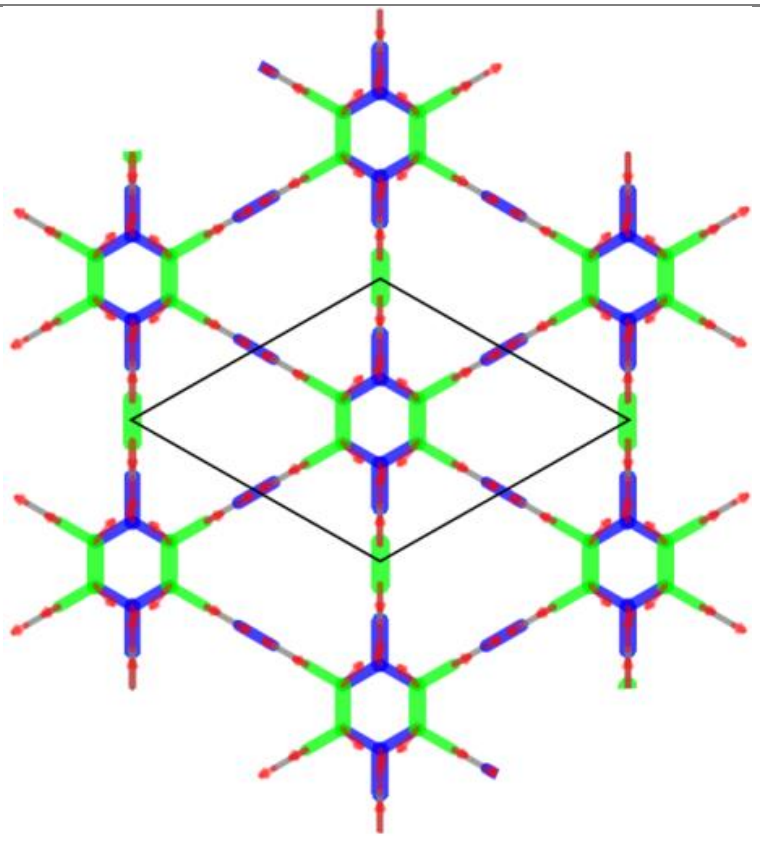 **G’** | 1382 | $E_{2g}$ | 0 | 6285 |
| 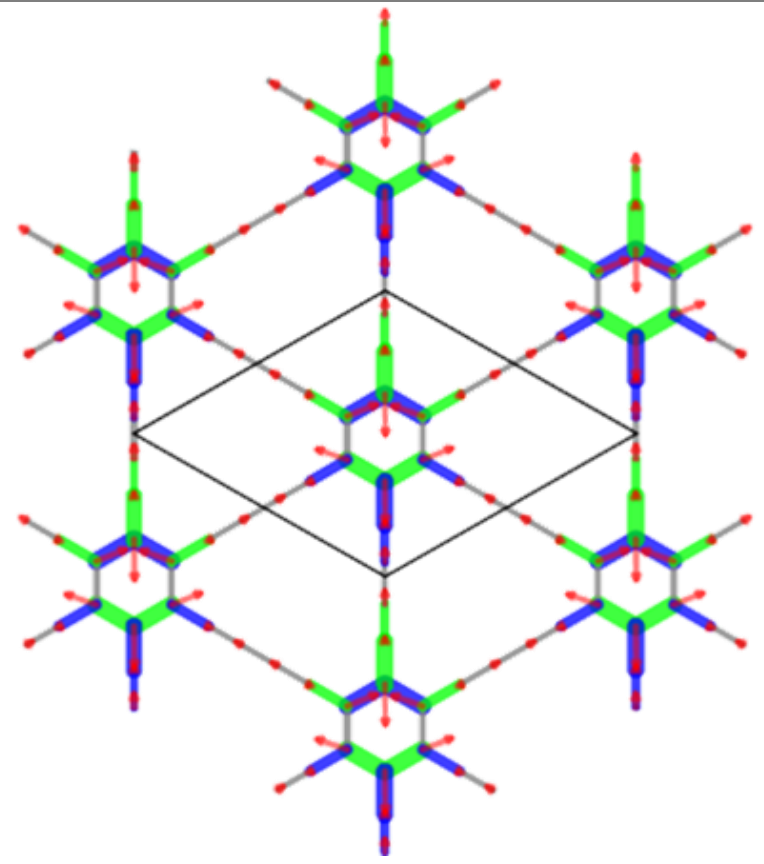 R | 1429 | $E_{1u}$ | 522 | 0 |
| 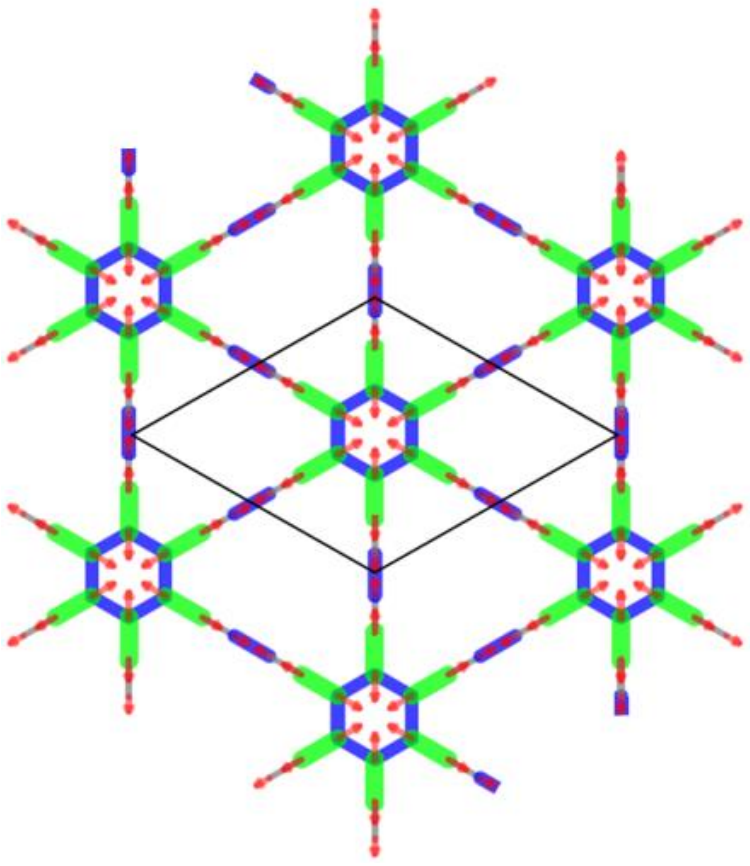 D | 1490 | $A_{1g}$ | 0 | 869870 |
| 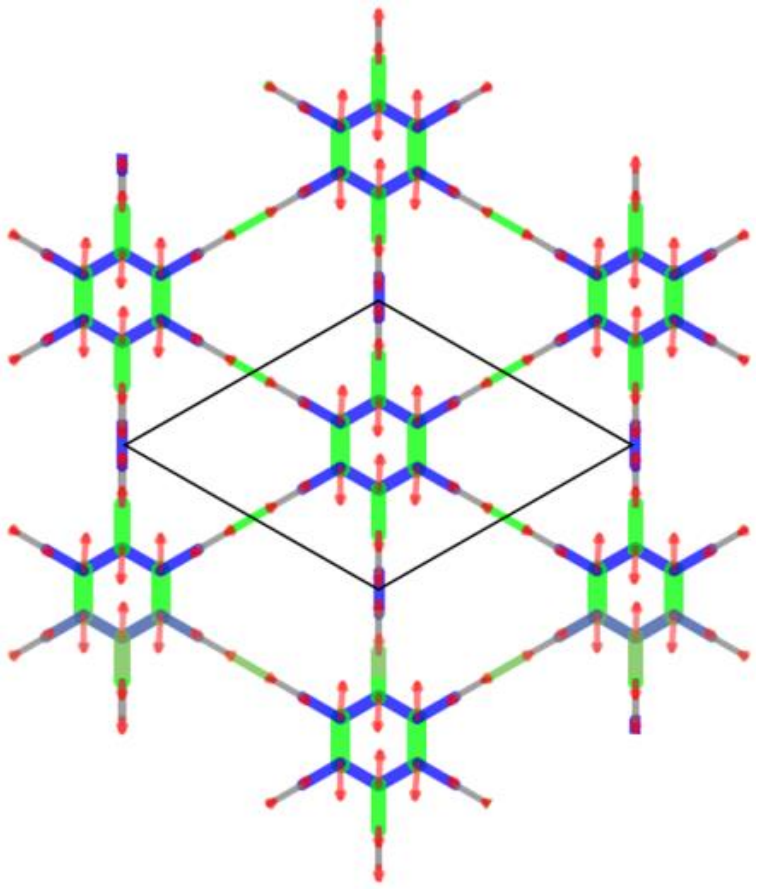 G | 1573 | $E_{2g}$ | 0 | 248915 |
| 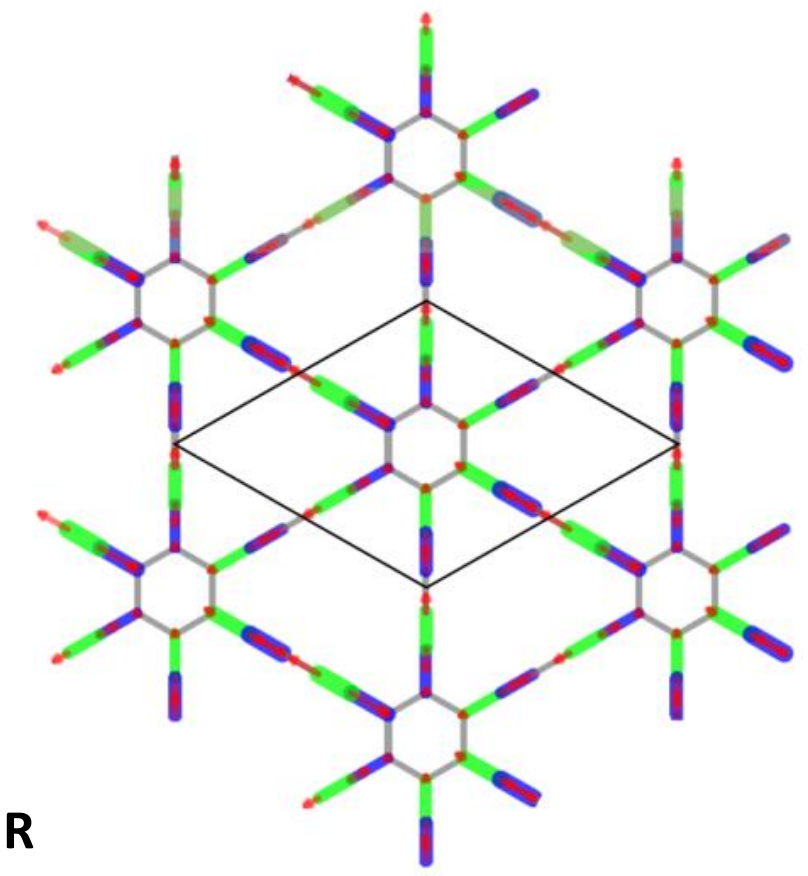 **Y^IR** | 2267 | $E_{1u}$ | 240 | 0 |
| 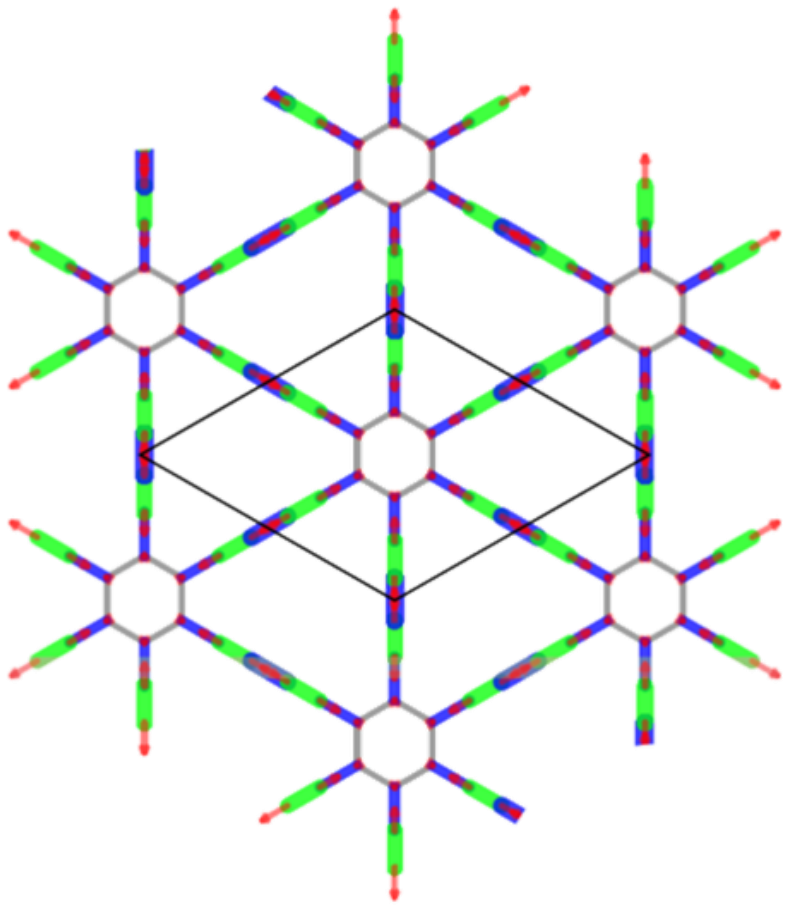 **Y** | 2276 | $A_{1g}$ | 0 | 3707443 |
| 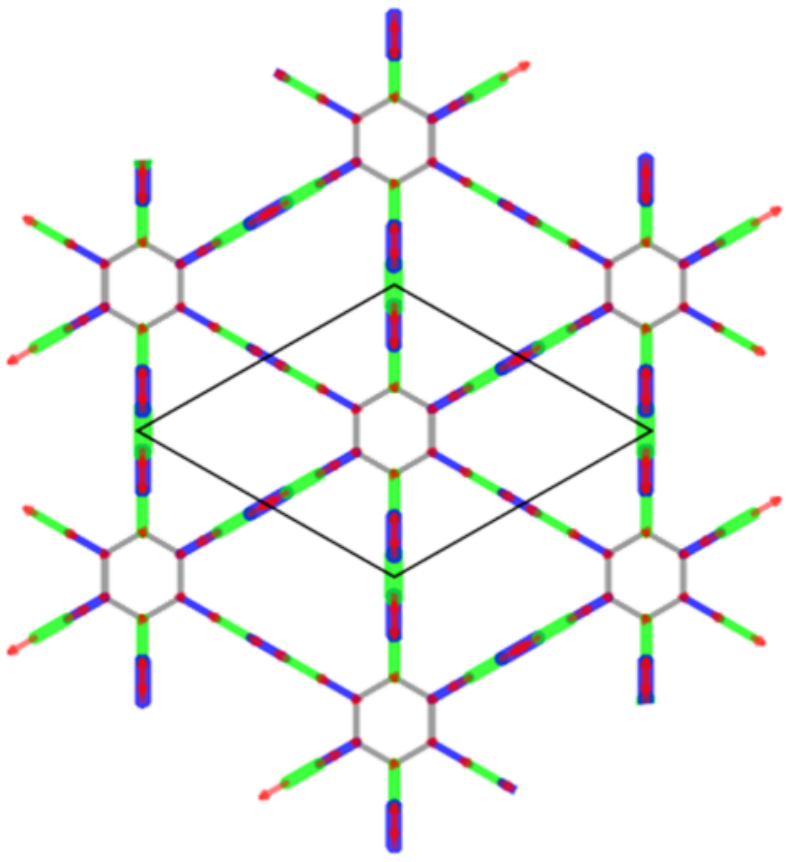 **Y'** | 2335 | $E_{2g}$ | 0 | 638183 |

**Table S7:** Sketches of DFT computed [PBE0/6-31G(d)] Raman active normal modes of A(n)- and (Z(n)- GDYNR.
Red arrows indicate atomic displacements while green and blue colors indicate the active bond stretching vibrations in each normal mode. The different colors indicate different phases of the vibration, while the line width is proportional to its amplitude.

| Normal modes A3 | Frequency [$cm^{-1}$] | Raman activity [$A^4$/amu] | IR Intensity [km/mol] |
|---|---|---|---|
| 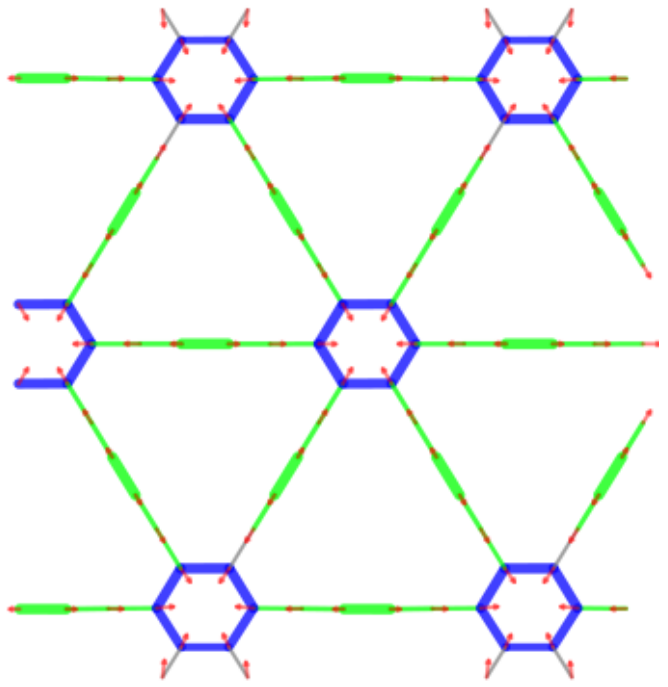 | 990 | 179315 | 0 |
| 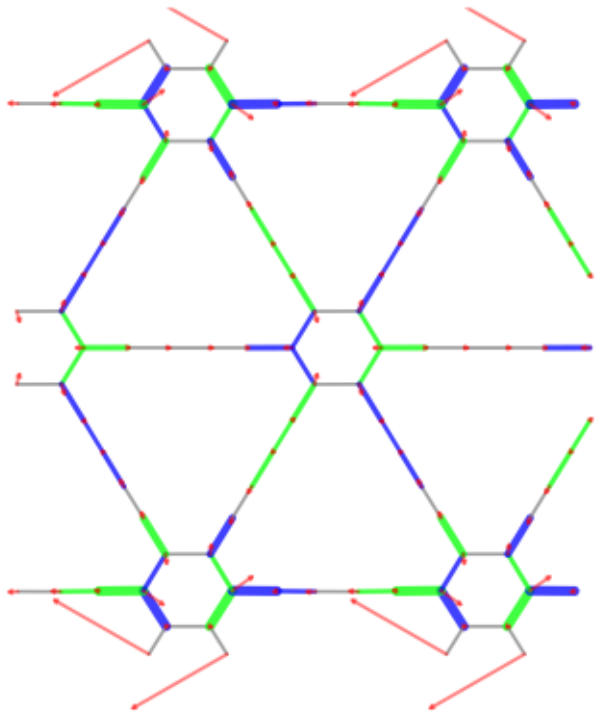 | 1266 | 1 | 86 |
| 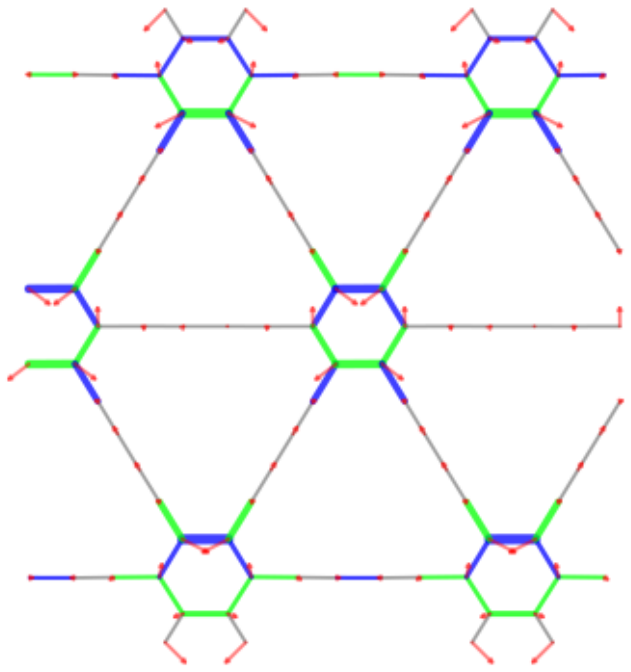 | 1433 | 0 | 172 |
| 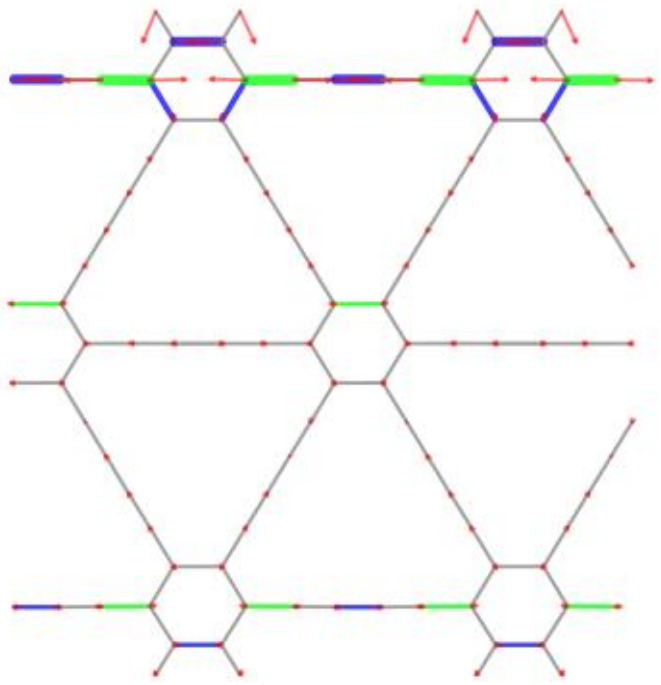 | 1452 | 157137 | 1 |
| 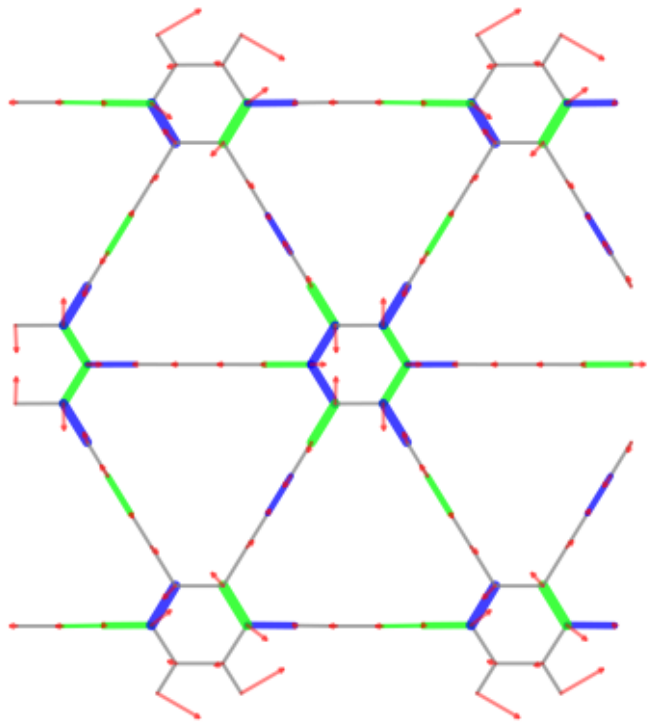 | 1469 | 9 | 796 |
| 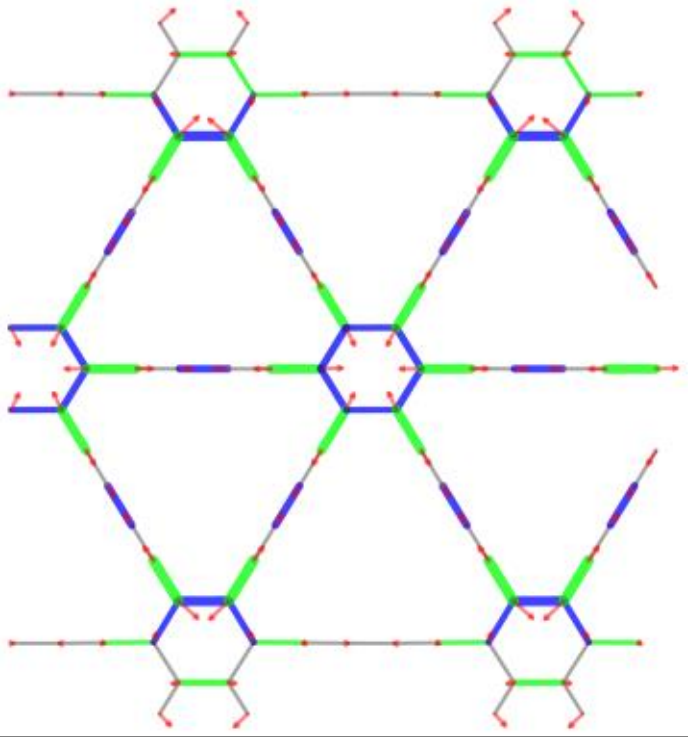 | 1499 | 373617 | 0 |
| 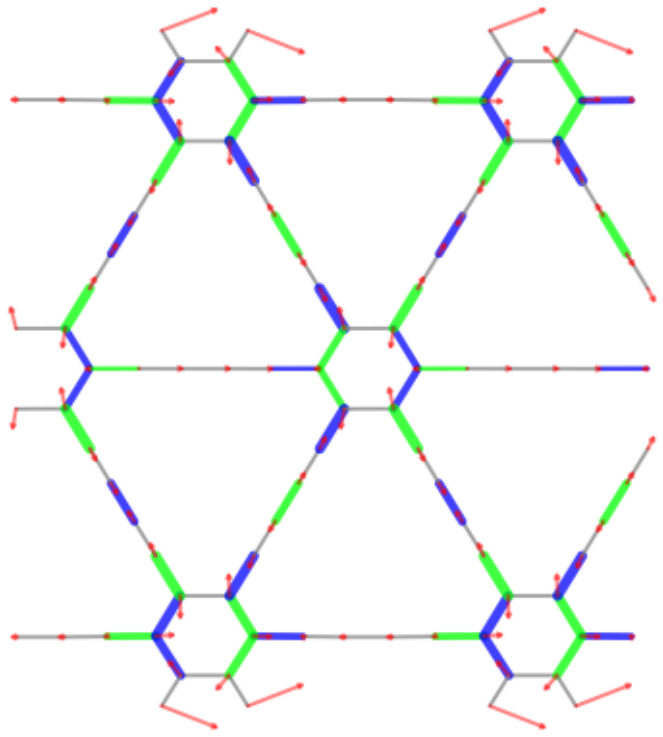 | 1519 | 40 | 113 |
| 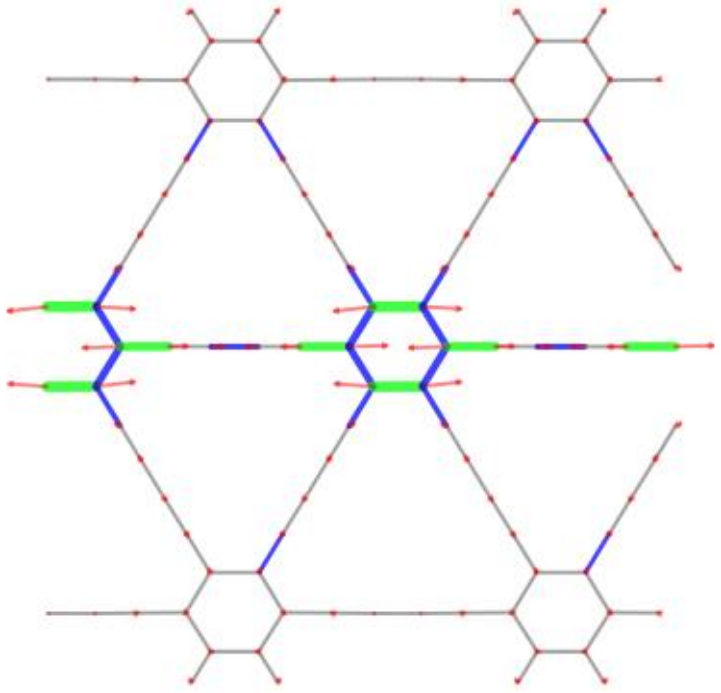 | 1571 | 212736 | 0 |

| Normal modes Z3 | Frequency [cm$^{-1}$] | Raman activity [A$^4$/amu] | IR Intensity [km/mol] |
|---|---|---|---|
| 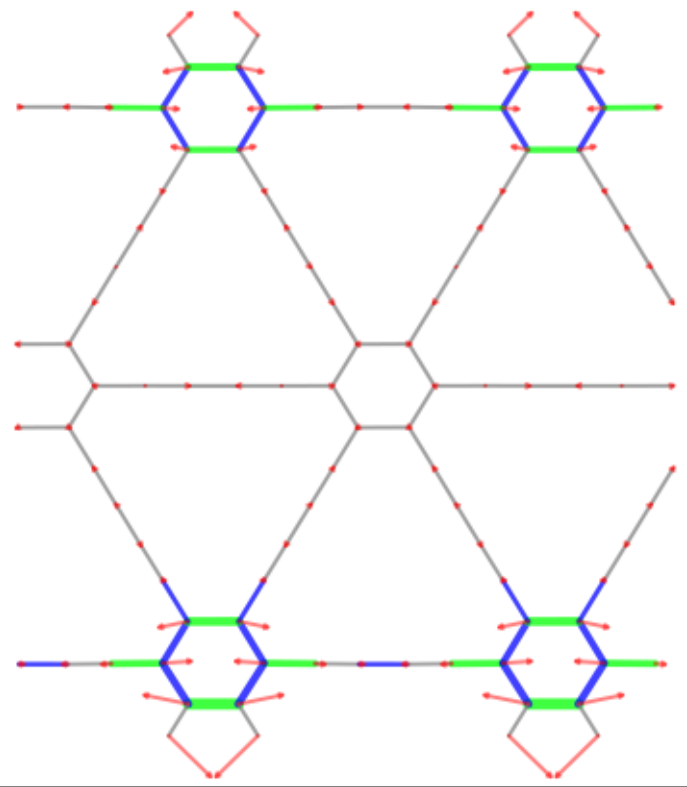 | 1630 | 609858 | 0 |
| 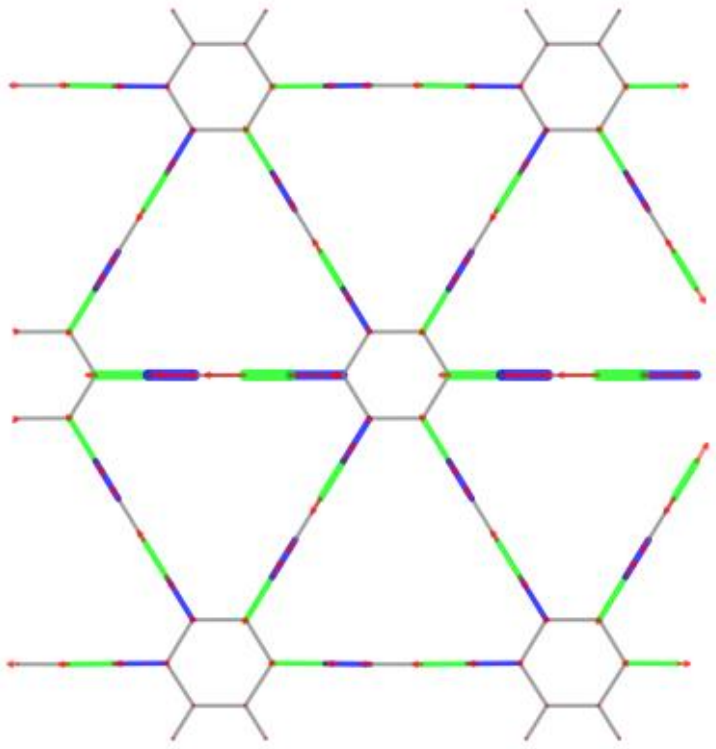 | 2270 | 7 | 322 |
| 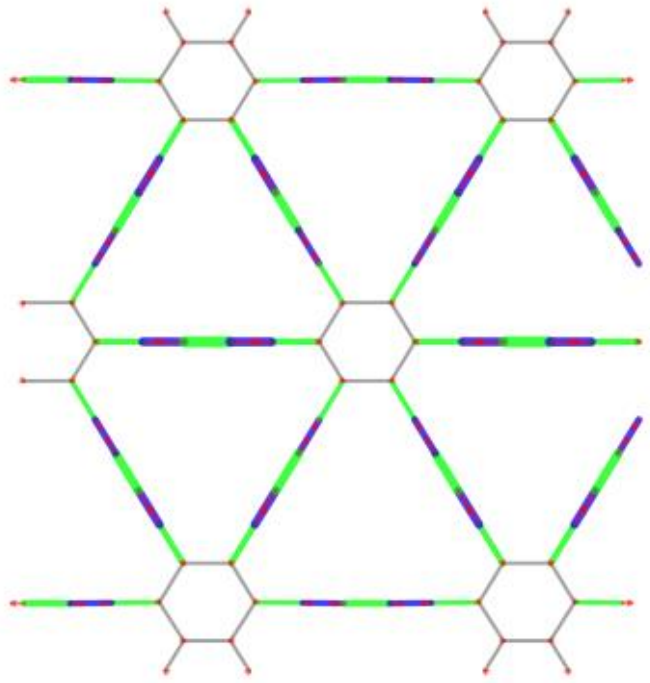 | 2293 | 2763591 | 0 |
| 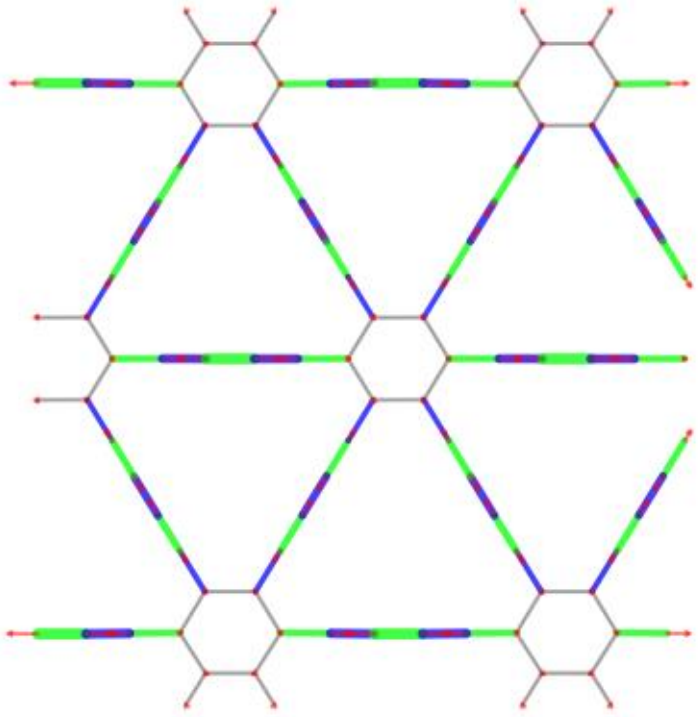 | 2342 | 1592954 | 0 |
| 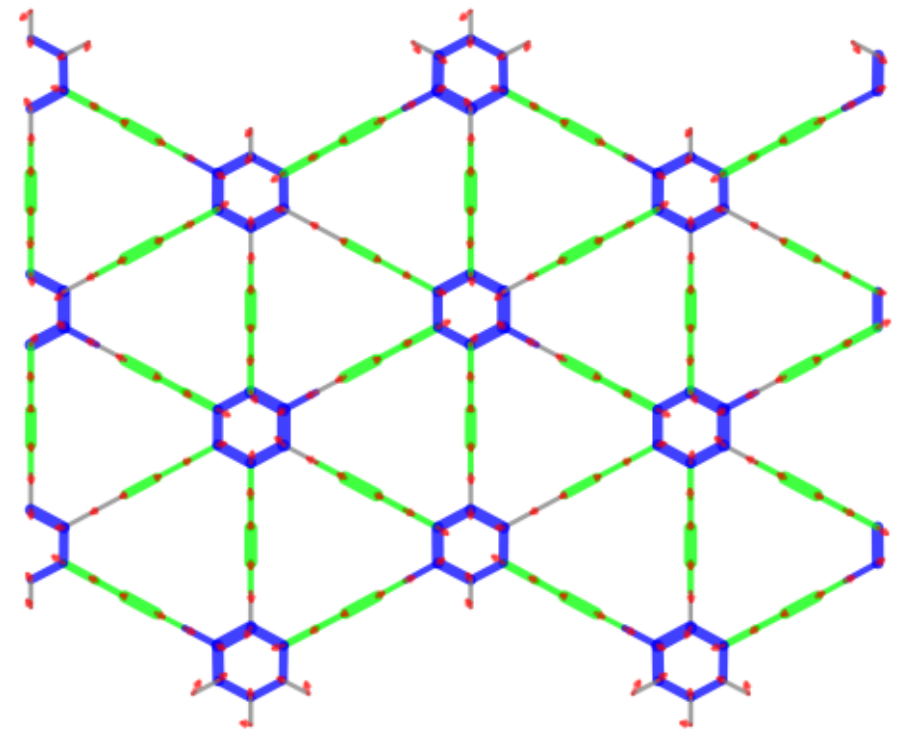 | 990 | 265254 | 1 |
| 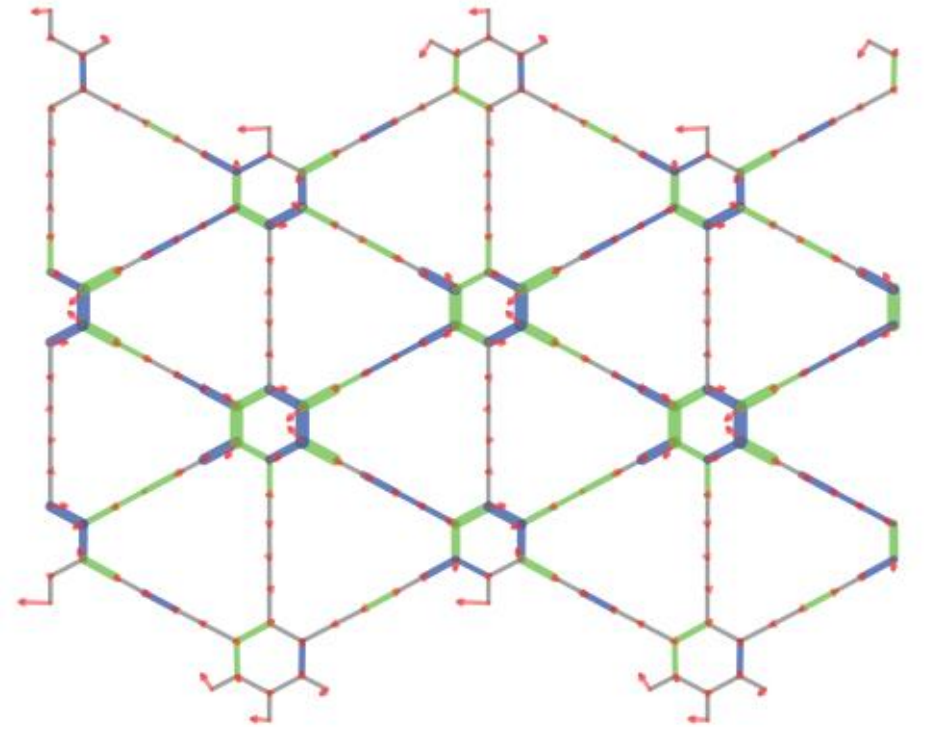 | 1448 | 5119 | 1569 |
| 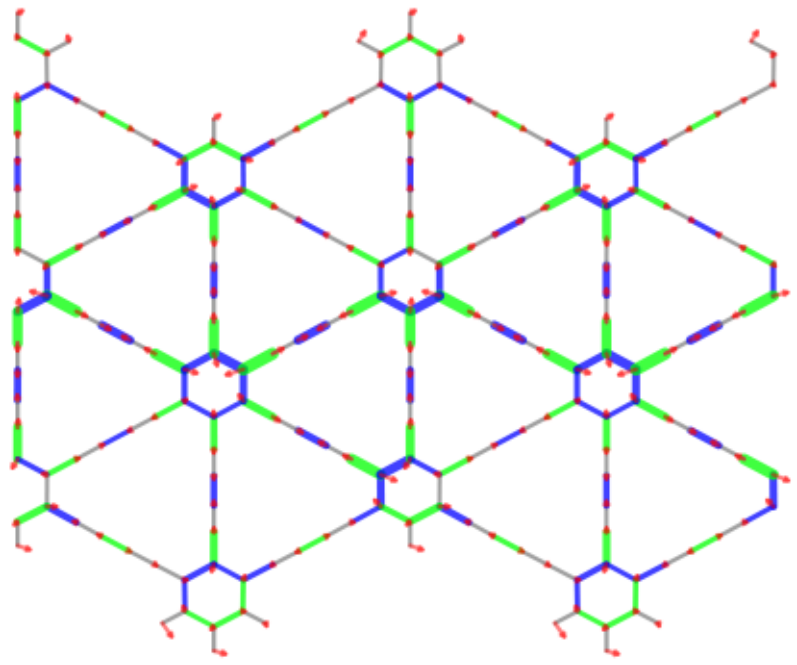 | 1498 | 631709 | 22 |
| 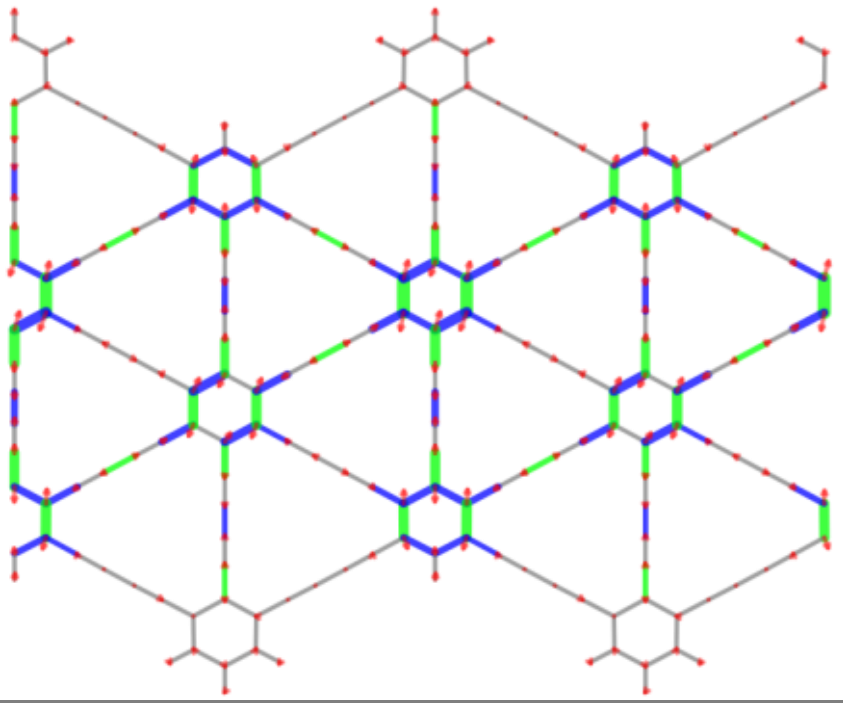 | 1571 | 391266 | 0 |
| 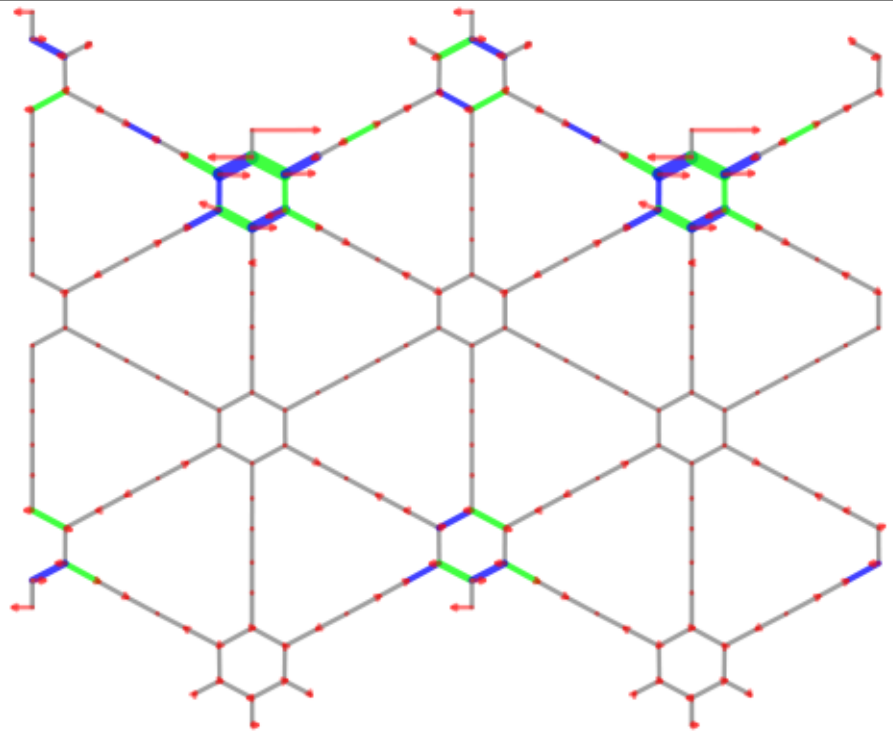 | 1619 | 63719 | 35 |
| 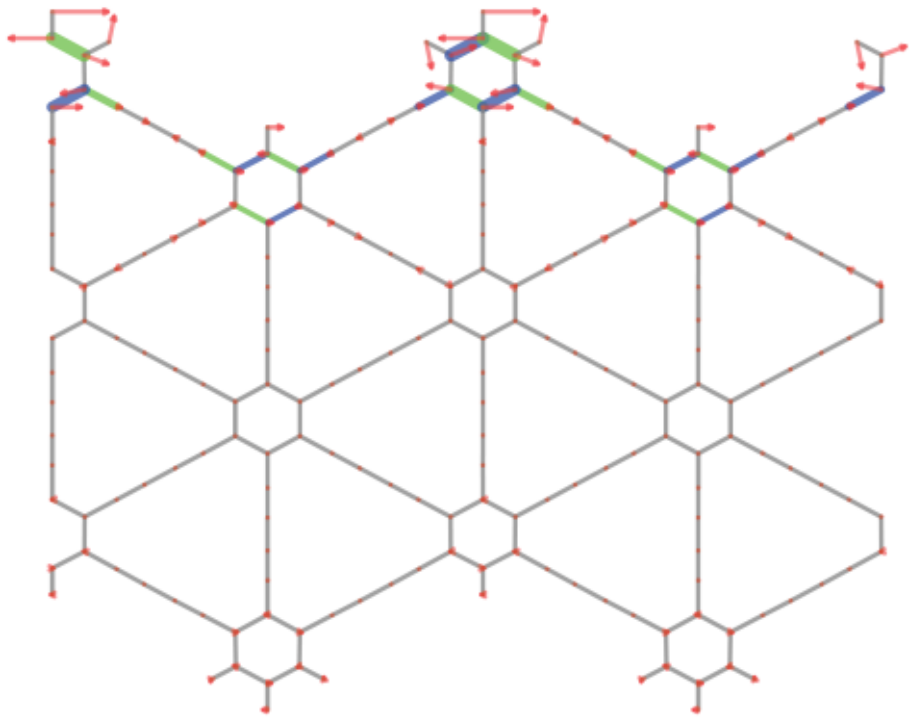 | 1627 | 9775 | 45 |
| 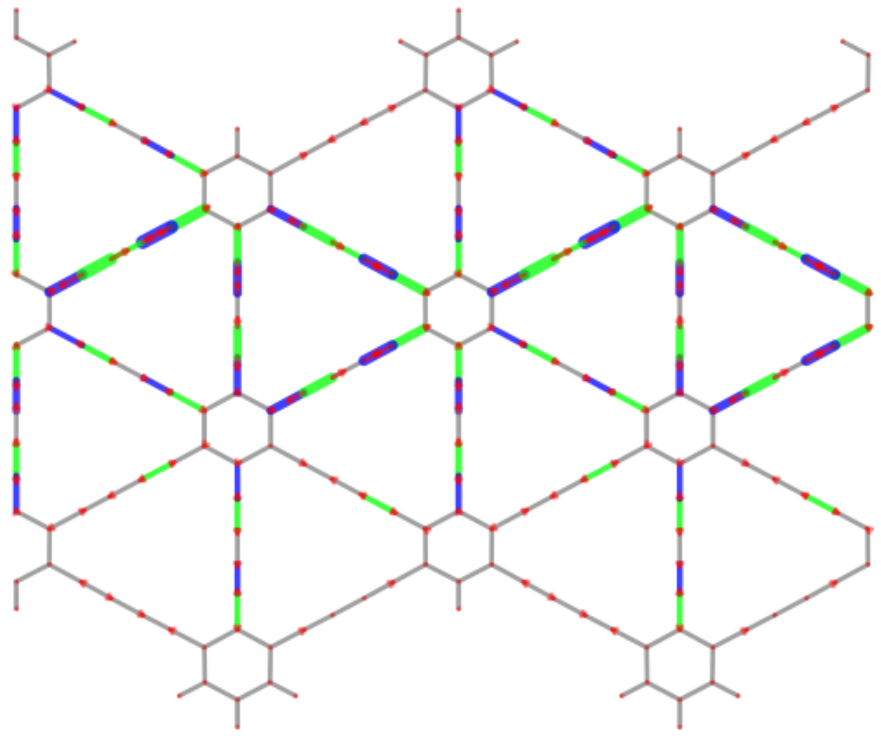 | 2269 | 13163 | 266 |
| 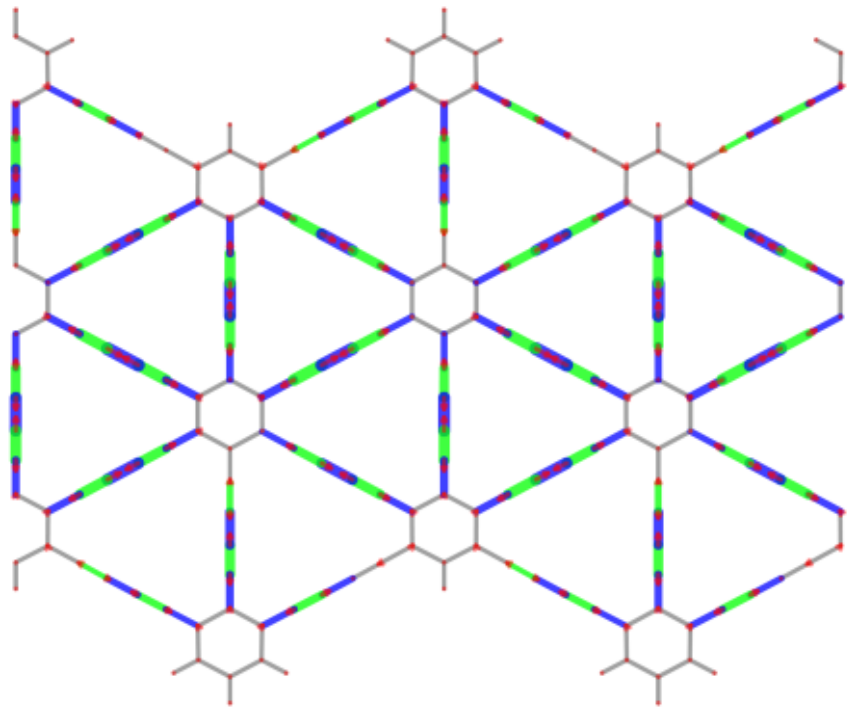 | 2292 | 4984467 | 1 |
| 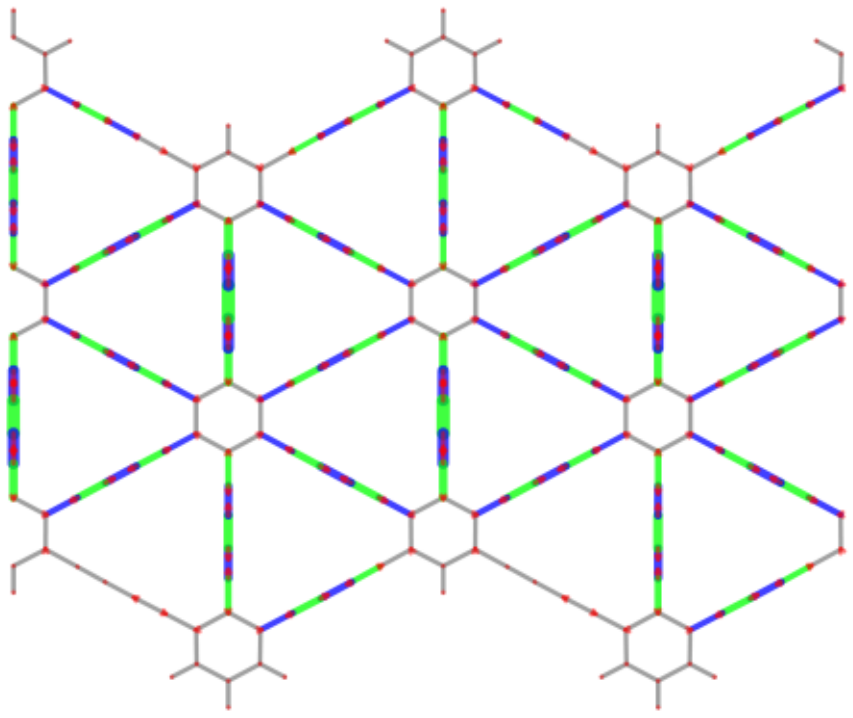 | 2347 | 988883 | 0 |
| 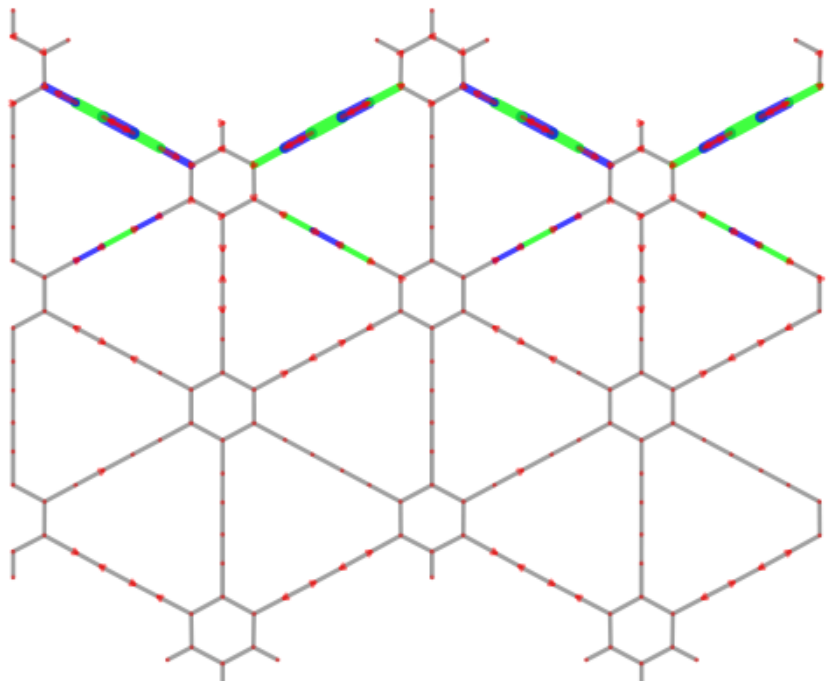 | 2350 | 110847 | 97 |

**Table S8:** DFT computed **[**PBE0/6-31G(d)] values of vibrational frequency ($cm^{-1}$), IR intensity (km/mol) and Raman activity ($A^4$/amu) of the diphenyl-polyyne with 4 sp-carbon atoms.

| Frequency ($cm^{-1}$) | Symmetry species | IR intensity [km/mol] | Raman activity [$A^4$/amu] |
|---|---|---|---|
| 7 | $A_u$ | 0 | 0 |
| 29 | $B_{1u}$ | 0.56 | 0 |
| 34 | $B_{2u}$ | 0.8 | 0 |
| 79 | $B_{2g}$ | 0 | 5.77 |
| 96 | $B_{1g}$ | 0 | 4.28 |
| 190 | $B_{2u}$ | 0.0 | 0 |
| 208 | $B_{1u}$ | 0.04 | 0 |
| 233 | $A_g$ | 0 | 0.13 |
| 315 | $B_{1g}$ | 0 | 15.11 |
| 402 | $B_{2u}$ | 0.43 | 0 |
| 412 | $A_u$ | 0 | 0 |
| 413 | $B_{3g}$ | 0 | 0.01 |
| 463 | $B_{2g}$ | 0 | 0.21 |
| 475 | $B_{3u}$ | 18.79 | 0 |
| 511 | $B_{1u}$ | 1.75 | 0 |
| 520 | $B_{1g}$ | 0 | 3.12 |
| 553 | $B_{2u}$ | 7.29 | 0 |
| 560 | $B_{2g}$ | 0 | 33.5 |
| 633 | $B_{1u}$ | 0.0 | 0 |
| 634 | $B_{2g}$ | 0 | 4.71 |
| 663 | $B_{1g}$ | 0 | 106.41 |
| 677 | $A_g$ | 0 | 43.99 |
| 707 | $B_{1g}$ | 0 | 8.74 |
| 708 | $B_{2u}$ | 32.88 | 0 |
| 776 | $B_{2u}$ | 62.03 | 0 |
| 777 | $B_{1g}$ | 0 | 2.78 |
| 788 | $B_{3u}$ | 8.21 | 0 |
| 864 | $A_u$ | 0 | 0 |
| 864 | $B_{3g}$ | 0 | 15.47 |
| 938 | $B_{2u}$ | 4.71 | 0 |
| 938 | $B_{1g}$ | 0 | 12.85 |
| 985 | $A_u$ | 0 | 0 |
| 985 | $B_{3g}$ | 0 | 0.12 |
| 1010 | $B_{1g}$ | 0 | 4.55 |
| 1010 | $B_{2u}$ | 0.07 | 0 |
| 1011 | $A_g$ | 0 | 69.03 |
| 1018 | $B_{3u}$ | 3.41 | 0 |
| 1024 | $A_g$ | 0 | 1012.84 |
| 1067 | $B_{3u}$ | 5.34 | 0 |
| 1071 | $A_g$ | 0 | 27.77 |
| 1116 | $B_{2g}$ | 0 | 0.39 |
| 1116 | $B_{1u}$ | 6.28 | 0 |
| 1188 | $B_{1u}$ | 0.01 | 0 |
| 1188 | $B_{2g}$ | 0 | 34.48 |
| 1206 | $B_{3u}$ | 1.9 | 0 |
| 1208 | $A_g$ | 0 | 207.7 |
| 1254 | $B_{3u}$ | 7.68 | 0 |
| 1336 | $B_{1u}$ | 0.18 | 0 |
| 1336 | $B_{2g}$ | 0 | 0.52 |
| 1384 | $B_{2g}$ | 0 | 54.58 |
| 1384 | $B_{1u}$ | 0.51 | 0 |
| 1408 | $A_g$ | 0 | 728.46 |
| 1495 | $B_{2g}$ | 0 | 50.06 |
| 1495 | $B_{1u}$ | 10.55 | 0 |
| 1544 | $B_{3u}$ | 42.17 | 0 |
| 1561 | $A_g$ | 0 | 968.32 |
| 1654 | $B_{1u}$ | 1.47 | 0 |
| 1654 | $B_{2g}$ | 0 | 0.26 |
| 1687 | $B_{3u}$ | 18.24 | 0 |
| 1688 | $A_g$ | 0 | 4480.41 |
| 2282 | $B_{3u}$ | 0.15 | 0 |
| 2369 | $A_g$ | 0 | 29522.21 |
| 3203 | $B_{3u}$ | 6.15 | 0 |
| 3203 | $A_g$ | 0 | 57.54 |
| 3212 | $B_{2g}$ | 0 | 356.23 |
| 3212 | $B_{1u}$ | 13.21 | 0 |
| 3224 | $B_{3u}$ | 45.33 | 0 |
| 3224 | $A_g$ | 0 | 479.01 |
| 3232 | $B_{2g}$ | 0 | 19.34 |
| 3232 | $B_{1u}$ | 41.15 | 0 |
| 3236 | $B_{3u}$ | 30.96 | 0 |
| 3236 | $A_g$ | 0 | 668.23 |

**Table S9:** Sketches of DFT computed [PBE0/6-31G(d)] relevant normal modes of the diphenyl-polyyne with 4 sp-carbon atoms ; the two different colors indicate stretching of the bonds with different phases

| Normal modes | Frequency [$cm^{-1}$] | Symmetry species | IR intensity [km/mol] | Raman activity [$A^4$/amu] |
|---|---|---|---|---|
| 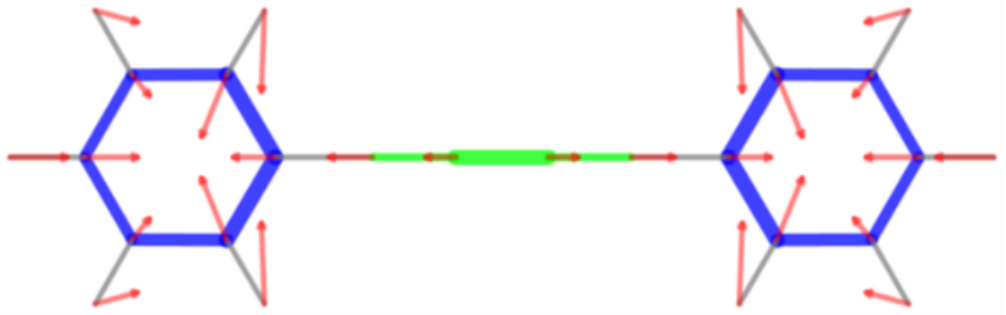 | 1024 | $A_g$ | 0 | 1013 |
| 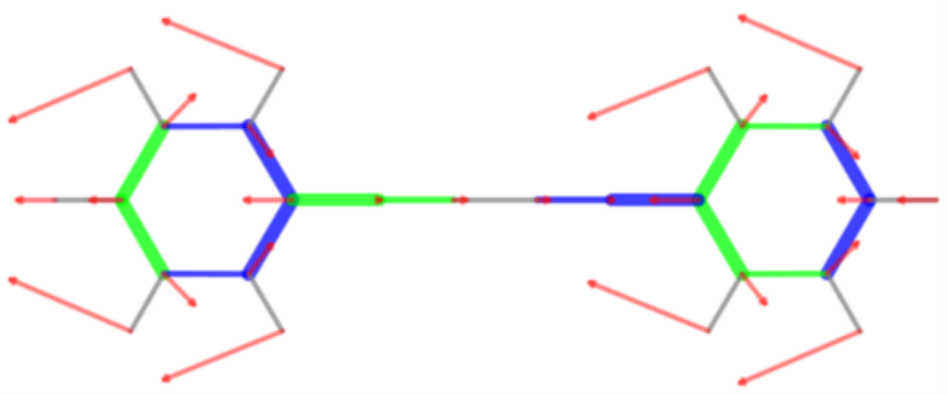 | 1561 | $A_g$ | 0 | 968 |
| 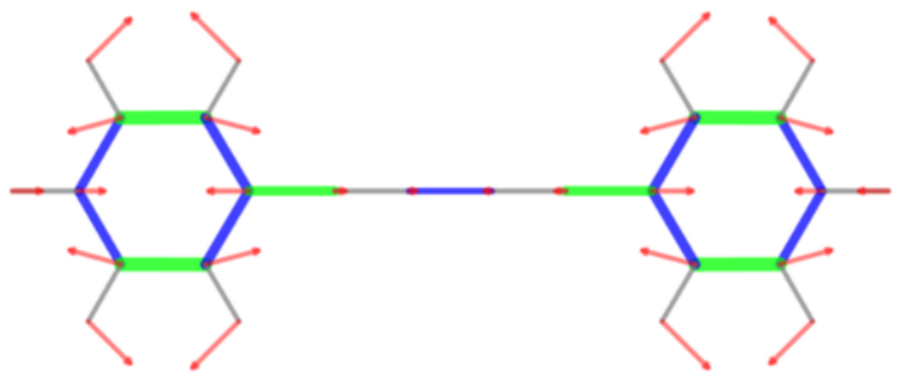 | 1688 | $A_g$ | 0 | 4480 |
| 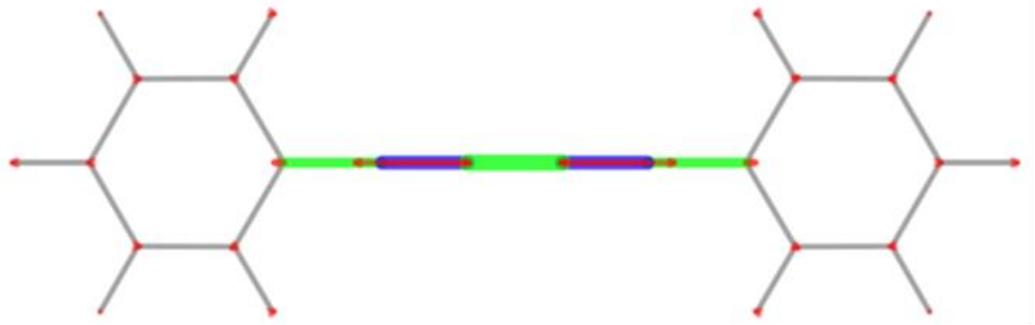 | 2369 | $A_g$ | 0 | 29522 |

**Table S10:** DFT computed **[**PBE0/6-31G(d)] Raman polarizability tensors associated to the most relevant Raman transitions of 2D-GDY

| Symmetry species | Frequency ($cm^{-1}$) | Raman activity ($A^4/amu$) | $d\alpha_{xx}/dq$ ($A^2/\sqrt{amu}$) | $d\alpha_{xy}/dq$ ($A^2/\sqrt{amu}$) | $d\alpha_{yy}/dq$ ($A^2/\sqrt{amu}$) | $d\alpha_{xz}/dq$ ($A^2/\sqrt{amu}$) | $d\alpha_{yz}/dq$ ($A^2/\sqrt{amu}$) | $d\alpha_{zz}/dq$ ($A^2/\sqrt{amu}$) |
|---|---|---|---|---|---|---|---|---|
| A1g | 983 | 231213 | 93 | 0 | 93 | 0 | 0 | 0.0 |
| E2g | 1382 | 12571 | -17 | 0 | 17 | 0 | 0 | 0.0 |
| | | | 0 | 17 | 0 | 0 | 0 | 0.0 |
| A1g | 1490 | 869870 | 179 | 0 | 179 | 0 | 0 | 0.0 |
| E2g | 1573 | 497834 | 109 | 0 | -109 | 0 | 0 | 0.0 |
| | | | 0 | -109 | 0 | 0 | 0 | 0.0 |
| A1g | 2276 | 3707443 | 371 | 0 | 371 | 0 | 0 | 0.0 |
| E2g | 2335 | 1276368 | 174 | 0 | -174 | 0 | 0 | 0.0 |
| | | | 0 | -174 | 0 | 0 | 0 | 0.0 |

**Table S11:** DFT computed **[**PBE0/6-31G(d)] dipole derivatives associated to the most relevant IR transitions of 2D-GDY

| Symmetry Species | Frequency ($cm^{-1}$) | IR Intensity (km/mol) | $d\mu_x/dq$ (debye/$(A*\sqrt{amu})$) | $d\mu_y/dq$ (debye/$(A*\sqrt{amu})$) | $d\mu_z/dq$ (debye/$(A*\sqrt{amu})$) |
|---|---|---|---|---|---|
| E1u | 893 | 13 | 0 | 0.4 | 0 |
| E1u | 1429 | 1044 | 0 | -3.5 | 0 |
| E1u | 2267 | 480 | -2.4 | 0 | 0 |

**Table S12:** DFT computed **[**PBE0/6-31G(d)] Raman polarizability tensors associated to the most relevant Raman transitions of A3-GDYNR. The phonons are classified according to the $D_{2h}$ symmetry group. In the last column, the ratio between yy and xx components highlights the strong anisotropy of the Raman tensors: one only element (xx), where x is the crystal axis direction, is relevant.

| Symmetry | Frequency ($cm^{-1}$) | Raman activity ($A^4/amu$) | $d\alpha_{xx}/dq$ ($A^2/\sqrt{amu}$) | $d\alpha_{xy}/dq$ ($A^2/\sqrt{amu}$) | $d\alpha_{yy}/dq$ ($A^2/\sqrt{amu}$) | $d\alpha_{xz}/dq$ ($A^2/\sqrt{amu}$) | $d\alpha_{yz}/dq$ ($A^2/\sqrt{amu}$) | $d\alpha_{zz}/dq$ ($A^2/\sqrt{amu}$) | $\frac{(d\alpha_{yy}/dq)}{(d\alpha_{xx}/dq)}$ |
|---|---|---|---|---|---|---|---|---|---|
| Ag | 990 | 179315 | -119 | 0 | -17 | 0 | 0 | 0.0 | 0.14 |
| Ag | 1452 | 157137 | 91 | 0 | 4 | 0 | 0 | 0.0 | 0.04 |
| Ag | 1499 | 373617 | 170 | 0 | 31 | 0 | 0 | 0.0 | 0.18 |
| Ag | 1571 | 212736 | 134 | 0 | -17 | 0 | 0 | 0.0 | -0.13 |
| Ag | 1630 | 609858 | -217 | 0 | 3 | 0 | 0 | 0.0 | -0.01 |
| Ag | 2293 | 2763591 | 468 | 0 | 64 | 0 | 0 | -0.1 | 0.14 |
| Ag | 2342 | 1592954 | 367 | 0 | -33 | 0 | 0 | 0.0 | -0.09 |

**Table S13:** DFT computed **[**PBE0/6-31G(d)] dipole derivatives associated to the most relevant IR transitions of A3-GDYNR. The phonons are classified according to the $D_{2h}$ symmetry group.

| Symmetry | Frequency ($cm^{-1}$) | IR Intensity ($km/mol$) | $d\mu_x/dq$ (debye/($A*\sqrt{amu}$)) | $d\mu_y/dq$ (debye/($A*\sqrt{amu}$)) | $d\mu_z/dq$ (debye/($A*\sqrt{amu}$)) |
|---|---|---|---|---|---|
| B3u | 1266 | 86 | -1,4 | 0.0 | 0.0 |
| B2u | 1433 | 172 | 0,0 | -2.0 | 0.0 |
| B3u | 1469 | 796 | 4,3 | 0.0 | 0.0 |
| B3u | 1519 | 113 | 1,6 | 0.0 | 0.0 |
| B3u | 2270 | 322 | 2,8 | 0.1 | 0.0 |

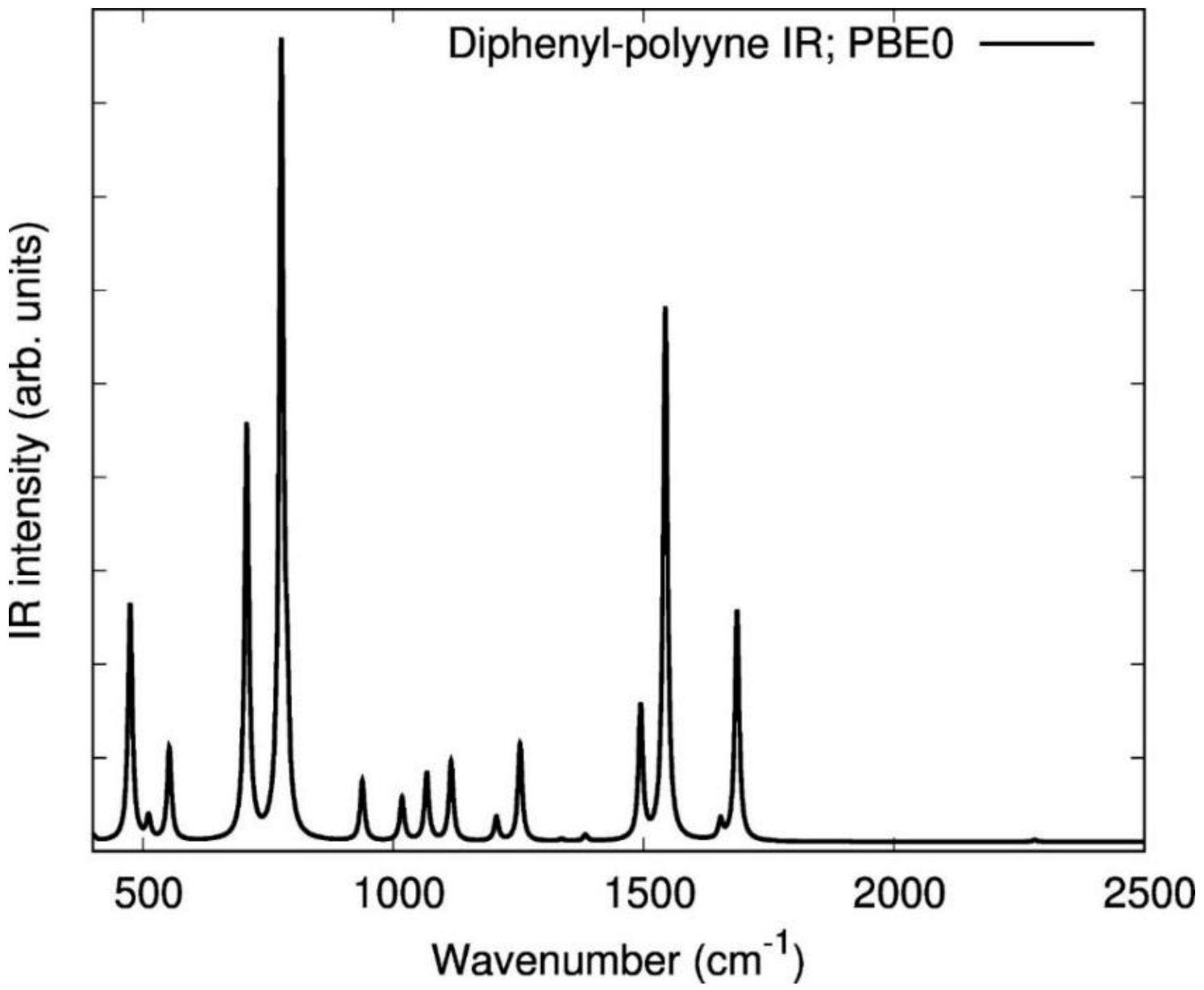

**Figure S1:** DFT computed **[**PBE0/6-31G(d)] IR spectrum of the diphenyl-polyyne with 4 sp-carbon atoms

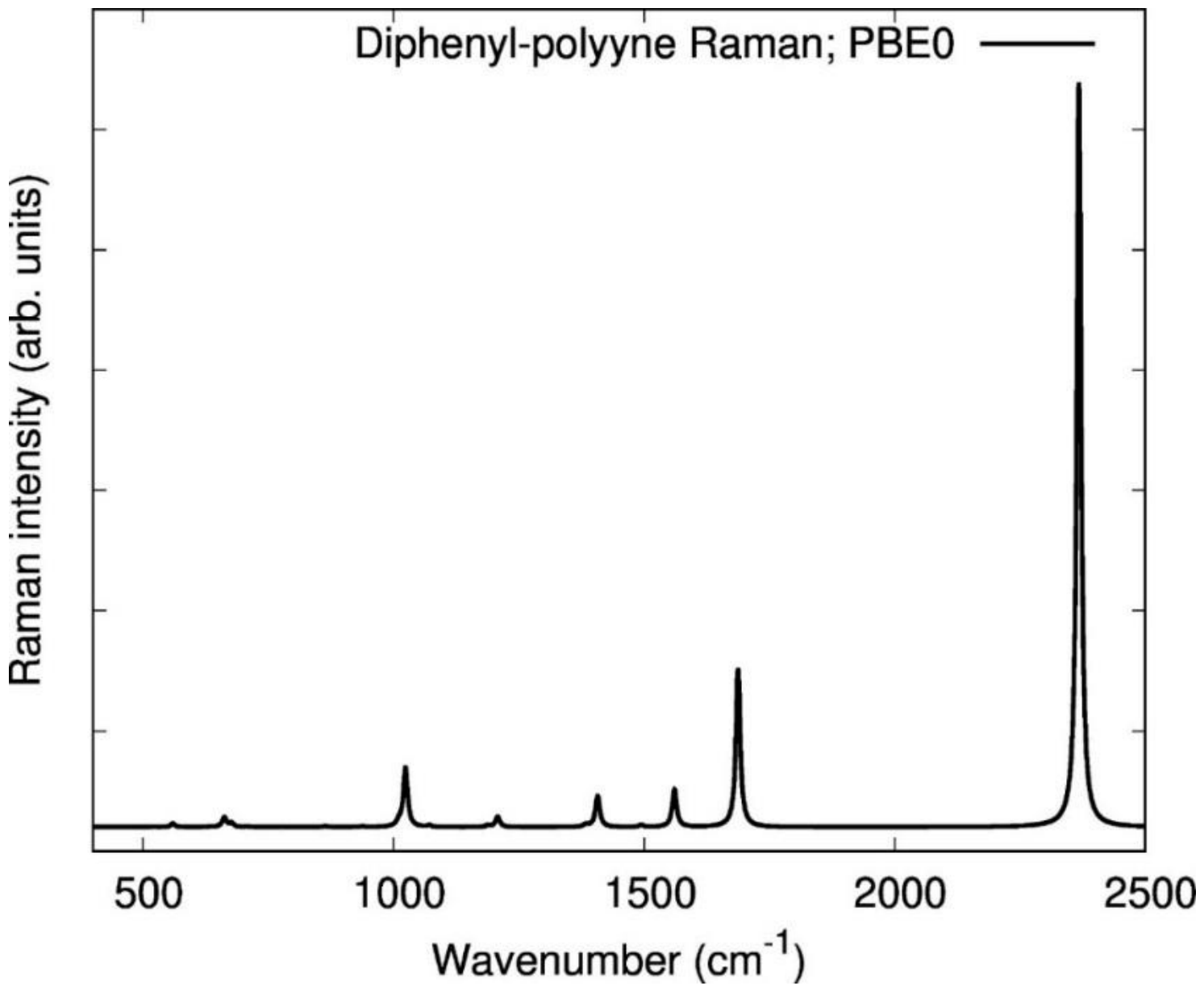

**Figure S2:** DFT computed **[**PBE0/6-31G(d)] Raman spectrum of the diphenyl-polyyne with 4 sp-carbon atoms

## Spectroscopic marker bands of armchair and zigzag GDYNR.

In Figure S3 we directly compare the Raman and IR spectra of A(3)-GDYNR and Z(*n*)-GDYNR (*n = 2.5, 3*) with those of 2D-GDY in order to reveal the peculiar marker bands of A- and Z- GDYNR.

As mentioned in the paper, while in the Raman spectra Z-edges do not show intense marker bands, A(3)-GDYNR presents a relevant marker bands at 1630 $cm^{-1}$, which can be considered not only as a marker of confinement, but also a reliable marker of the presence A edges.

By comparing the Raman spectra of integer and half integer Z(*n*)-GDYNRs, we can notice small but clear differences. For instance, Z(3) shows one only band originated by the superposition of two normal modes located at 1490 $cm^{-1}$ and at 1498 $cm^{-1}$ while for Z(2.5), in the same spectral range, two individual components are computed, located at 1482 $cm^{-1}$ and at 1503 $cm^{-1}$ respectively, giving rise to a doublet of peaks . However, these bands find a clear correspondence with D line of 2D crystal and cannot be taken as marker band of the ribbons.

Considering the IR spectrum, a reliable marker of Z(*n*)-GDYNRs can be found in the extra band predicted at about 2350 $cm^{-1}$, which does not appear in the case of for A(*n*)-GDYNRs. In addition, another marker of Z edge is the band at about 1620 $cm^{-1}$. On the other hand, A(*n*)-GDNRs display several marker bands at about 1270 and 1405 $cm^{-1}$.

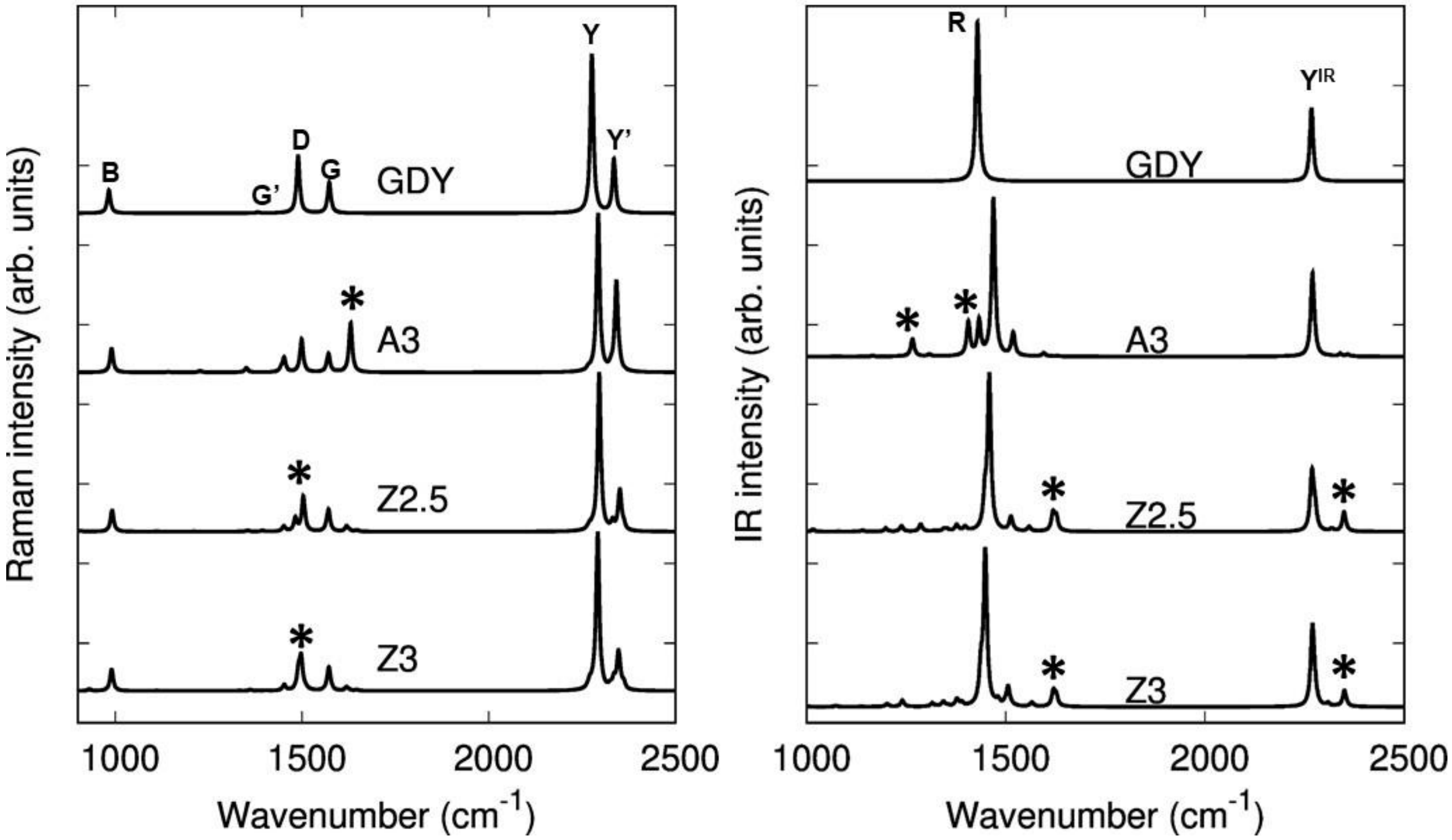

**Figure S3:** Comparison among the DFT computed (PBE0/6-31G(d)) Raman and IR spectra of 2D-GDY, with GDYNRs having different edges and the same width. The wavenumber values are not scaled. The bands discussed in the text are marked by *.